\documentstyle[epsfig,12pt]{article}
\begin{document}
\renewcommand{\baselinestretch}{1.5}

\newcommand\beq{\begin{equation}}
\newcommand\eeq{\end{equation}}
\newcommand\bea{\begin{eqnarray}}
\newcommand\eea{\end{eqnarray}}

\newcommand\tphi{{\tilde \phi}}
\newcommand\tPi{{\tilde \Pi}}
\newcommand\sqpi{{\sqrt\pi}} 
\newcommand\bo{{\bar\omega}}
\newcommand\sqfp{2 \sqrt{\pi}}
\newcommand\sqk{\sqrt{K}}
\newcommand\bphi{{\bar\phi}}
\newcommand\bPi{{\bar\Pi}}
\newcommand\ua{\uparrow}
\newcommand\da{\downarrow}
\newcommand\td{{\tilde d}}
\newcommand\sumi{\sum_{i=1}^{N}}
\newcommand\rhr{\rho_R }
\newcommand\rhl{\rho_L }
%\hfill MRI-PHY/P20000533
%\hfill cond-mat/0005492

\begin{center}
{\Large An Introduction To Bosonization And Some Of Its Applications}
\end{center}

\vskip .8 true cm
\centerline{\bf Sumathi Rao \footnote{{\it E-mail address}: 
sumathi@mri.ernet.in}}
\centerline{\it Mehta Research Institute, Chhatnag Road, Jhunsi,}
\centerline{\it Allahabad 211019, India.}

\vskip .2 true cm
\centerline{\it and}
\vskip .2 true cm

\centerline{\bf Diptiman Sen \footnote{{\it E-mail address}: 
diptiman@cts.iisc.ernet.in}} 
\centerline{\it Centre for Theoretical Studies,} 
\centerline{\it Indian Institute of Science, Bangalore 560012, India.} 
\vskip 1.5 true cm

\centerline{\bf Lectures given in the SERC School at MRI on }
\centerline{\bf "Field Theories in Condensed Matter Systems"}
\centerline{\bf February - March, 2000}
\vskip 1.6 true cm

\begin{abstract}

We discuss the technique of bosonization for studying systems of interacting
fermions in one dimension. After briefly reviewing the low-energy properties 
of Fermi and Luttinger liquids, we present some of the relations between 
bosonic and fermionic operators in one dimension. We use these relations
to calculate the correlation functions and the renormalization group 
properties of various operators for a system of spinless fermions. We then 
apply the methods of bosonization to study the Heisenberg antiferromagnetic 
spin 1/2 chain, the Hubbard model in one dimension, and transport 
in clean quantum wires and in the presence of isolated impurities. 

%\noindent PACS numbers: 73.20.Dx, 73.40.Gk, 71.10.Pm

\end{abstract}
\vskip 1.6 true cm

\newpage
\noindent {\large {\bf List of topics:}}
\vskip .4 true cm

\noindent 1. Introduction to Fermi and Luttinger liquids

\noindent 2. Bosonization for spinless fermions

\noindent 3. Calculation of correlation functions using bosonization

\noindent 4. Renormalization group analysis of perturbed models

\noindent 5. Heisenberg antiferromagnetic spin-$1/2$ chain

\noindent 6. Hubbard model

\noindent 7. Transport in a Luttinger liquid without impurities

\noindent 8. Transport in the presence of isolated impurities

%\newpage

\section{Fermi and Luttinger liquids}

In two and three dimensions, many systems of interacting fermions at low 
temperatures are described by the Fermi liquid theory developed by Landau 
(see Ref. 1 for a brief review). According to this theory, at zero 
temperature, the ground state of each species of fermions has a Fermi surface 
in momentum space located at an energy called the Fermi energy $E_F$, such 
that all the states within that surface ($i.e.$, with energies less than 
$E_F$) are occupied while all the states outside it are unoccupied. An 
elementary low-energy excitation is one in which 
a particle is added (annihilated) in a state just outside (inside) the Fermi 
surface; these are called particle and hole excitations respectively. In 
an interacting system, these one-particle excitations are accompanied by 
a cloud of particle-hole pairs, and they are more commonly called 
quasiparticles; these carry the same charge as a single particle (or hole).
If the particle number is held fixed, the 
low-energy excitations of the system consist of particle-hole pairs in which 
a certain number of particles are excited from states within the Fermi 
surface to states outside it. A few of these excitations have both low
wave numbers and low energies with the energy being proportional to
the wave number; such excitations can be thought of as sound waves. But most 
of the particle-hole excitations do not have such a linear relationship 
between energy and wave number; in fact, for most such excitations, a given
energy can correspond to many possible momenta. 

Another interesting property of a Fermi liquid in two and three dimensions 
is that the one-particle momentum distribution function $n({\vec k})$, 
obtained by Fourier transforming the one-particle equal-time correlation 
function, has a finite discontinuity at the Fermi surface as shown in Fig. 1 
(a). This discontinuity is called the quasiparticle renormalization factor
$z_{\vec k}$; it is also equal to the residue of the pole in the one-particle 
propagator. For non-interacting fermions, $z_{\vec k} =1$; but for 
interacting fermions, 
$0 < z_{\vec k} <1$ because a quasiparticle is a superposition of many states, 
only some of which are one-particle excitations. To compute $z_{\vec k}$, we
consider the one-particle Green's function $G({\vec x},t)$ defined as the
expectation value of the time-ordered product of the fermion operator 
$\psi ({\vec x},t)$ in the ground state $\vert 0 \rangle$, namely,
\beq
G ({\vec x},t) ~=~ \langle 0 \vert ~T \psi ({\vec x},t) \psi^\dagger 
({\vec 0},0) ~\vert 0 \rangle ~.
\eeq
(We will ignore the spin label here). The Fourier transform of this function 
can be written as
\beq
{\cal G} ({\vec k}, \omega ) ~=~ \frac{i}{\omega - \epsilon_{\vec k} - 
\Sigma ({\vec k}, \omega )} ~,
\label{green}
\eeq
where $\epsilon_{\vec k}$ is the dispersion relation for the non-interacting 
theory; we absorb the chemical potential $\mu$ in the definition of 
$\epsilon_{\vec k}$ so that $\epsilon_{\vec k} = 0$ for ${\vec k}$ lying on 
the Fermi surface. (We will set $\hbar =1$). The self-energy 
$\Sigma ({\vec k}, \omega )$ contains the effects of all the interactions as 
well as any prescription necessary to shift the pole slightly off the real 
axis in $\omega$. For a Fermi liquid, ${\cal G} ({\vec k}, \omega)$ has a pole
near the real axis of $\omega$ for any value of $\vec k$ on the Fermi
surface. In addition, $\Sigma$ is sufficiently analytic at all such points 
so that the derivative $\partial \Sigma /\partial \omega$ has a finite value. 
The quasiparticle renormalization factor $z_{\vec k}$ is then given 
by the residue at the pole, $i.e.$,
\beq
z _{\vec k}~=~ \Bigl( ~1~ -~ \frac{\partial \Sigma}{\partial \omega} ~
\Bigr)^{-1} ~.
\eeq
This gives the discontinuity in $n({\vec k})$ at the Fermi surface.

Finally, in a Fermi liquid,
the various correlation functions decay asymptotically at long distances as
power laws, with the exponents being independent of the strength of the
interactions. Thus non-interacting and interacting systems have the same
exponents and there is a universality.

The discussion above does not apply if the ground state of the system 
spontaneously breaks some symmetry, for instance, if it is superconducting, 
or forms a crystal or develops charge or spin density ordering.

In contrast to a Fermi liquid, interacting fermion systems in one dimension 
behave quite differently \cite{hald,schu,gogo}; we will assume again that the 
ground state breaks no symmetry. Such systems are called Luttinger liquids 
and they have the following general properties. First of all, there are no 
single particle or quasiparticle excitations. Thus {\it all} the low-energy 
excitations can be thought of as particle-hole excitations; further, all of 
these take the form of sound waves with a linear dispersion relation. 
(As we will see below, there are also excitations of another kind possible 
which correspond to adding a small number of particles $N_R$ and $N_L$ to 
the right and left Fermi points. However, these correspond to only two 
oscillator degrees of freedom, and therefore do not 
contribute to thermodynamic properties like the specific heat). Secondly, 
there is no discontinuity in the momentum distribution function at the Fermi 
momentum, as indicated in Fig. 1 (b). Rather, there is a cusp there whose 
form is determined by a certain exponent. Finally, this exponent depends on 
the strength of the interactions in a non-universal manner, and it also 
governs the power-law fall-offs of the correlation functions at large 
space-time distances \cite{kriv}.

Let us be more specific about the nature of the low-energy excitations
in a one-dimensional system of fermions. Assume that we have a system of 
length $L$ with a boundary condition to be specified later. The translation 
invariance and the finite length make the 
one-particle momenta discrete. Suppose that the system has $N_0$ particles 
with a ground state energy $E_0 (N_0)$ and a ground state momentum $P_0 =0$;
we are assuming that the system conserves parity. We will be interested
in the thermodynamic limit $N_0 , L \rightarrow \infty$ keeping the particle 
density $\rho_0 = N_0 /L$ fixed. Let us first consider a single species of 
non-interacting fermions which have two possible directions of motion, 
right-moving with $d \epsilon_k /dk = v_F$ and left-moving with $d 
\epsilon_k /dk = -v_F$. Here $\epsilon_k$ is the energy of a low-lying 
one-particle excitation, $k$ is its momentum 
measured with respect to a right Fermi momentum $k_F$ and 
a left Fermi momentum $- k_F$ respectively, and $v_F$ is called the Fermi 
velocity. (See Fig. 2 for a typical picture of the momentum states of a 
lattice model). The values of $k_F$ and $v_F$ are defined for the 
non-interacting system; hence they depend on the density $\rho_0$ but not on 
the strength of the interaction. Then a low-lying excitation consists of 
two pieces \cite{hald},

\noindent (i) a set of bosonic excitations each of which can have either 
positive momentum $q$ or negative momentum $-q$ with an energy $\epsilon_q 
= v_F q$, where $0 < q << k_F$, and

\noindent (ii) a certain number of particles $N_R$ and $N_L$ added to the
right and left Fermi points respectively, where $N_R , N_L << N_0$. (Note
that $N_R$ and $N_L$ can be positive, negative or zero. It is convenient
to assume that $N_R \pm N_L$ are even integers; then the total number of
particles $N_0 + N_R + N_L$ is always even or always odd. We can choose
the boundary condition (periodic or antiperiodic) to ensure that the
ground state is always non-degenerate).

It turns out that the Hamiltonian and momentum operators for a 
one-dimensional system (which may have interactions) have the general form 
\bea
H ~= && E_0 (N_0) ~+~ \sum_{q>0} ~vq~ {[} ~{\tilde b}_{R,q}^\dagger 
{\tilde b}_{R,q} ~+~ 
{\tilde b}_{L,q}^\dagger {\tilde b}_{L,q} ~{]} ~\nonumber \\
&& +~ \mu (N_R + N_L ) ~+~ \frac{\pi v}{2LK} (N_R + N_L )^2 ~+~ \frac{\pi 
vK}{2L} ~(N_R - N_L )^2 ~, \nonumber \\
P ~= && \sum_{q>0} ~q~ {[} ~{\tilde b}_{R,q}^\dagger {\tilde b}_{R,q} ~-~ 
{\tilde b}_{L,q}^\dagger {\tilde b}_{L,q} ~{]} ~+~ [~ k_F ~+~ 
\frac{\pi}{L} (N_R + N_L ) ~]~ (N_R - N_L ) ~,
\label{ham1}
\eea
where $v$ is the sound velocity, $q$ is the momentum of the low-energy 
bosonic excitations created and annihilated by ${\tilde b}_q^\dagger$ and 
${\tilde b}_q$,
$K$ is a positive dimensionless number, and $\mu$ is the chemical potential
of the system. We will see later that $v$ and $K$ are the two important 
parameters which determine all the low-energy properties of a system. Their 
values generally depend on both the strength of the interactions
and the density. If the fermions are non-interacting, we have
\beq
v ~=~ v_F \quad {\rm and} \quad K ~=~ 1 ~.
\label{vk}
\eeq
Note that one can numerically find the values of $v$ and $K$ by studying the 
$1/L$ dependence of the low-energy excitations of finite size systems.

It is interesting that the expression for the momentum operator in Eq. 
(\ref{ham1}) is independent of the interaction strength. We can understand 
the last term in the momentum as follows. For a continuum system, the Fermi 
momentum $k_F (N)$ is related to the density by the relation
\beq
L ~\int_{-k_F (N)}^{k_F (N)} ~\frac{dk}{2\pi} ~=~ N ~.
\eeq
Thus a system of $N_0$ particles has a Fermi momentum 
\beq
k_F ~= ~\frac{\pi N_0}{L} ~=~ \pi \rho_0 ~,
\eeq
while a system of $N =N_0 + N_R + N_L$ particles has a Fermi momentum 
equal to $k_F + (\pi /L)(N_R + N_L)$. If the $N$ particles occupy the
momenta states symmetrically about zero momentum, the total momentum of that
state is zero; in this state, both the right and left Fermi points have 
$(N_R + N_L)/2$ particles more than the original ground state. Now let us 
shift $(N_R - N_L)/2$ particles from the left Fermi point to the right 
Fermi point, so that the right Fermi point has $N_R$ particles more and 
the left Fermi point has $N_L$ particles more than the original
system. We then see that the total momentum has changed from zero to
$[k_F + (\pi /L) (N_R + N_L) ](N_R - N_L )$; this is the last term in the
expression for the momentum operator.

The form of the parameterization of the last two terms in the Hamiltonian in 
Eq. (\ref{ham1}) can be understood as follows. (Note that these two terms 
vanish in the thermodynamic limit and do not contribute to the specific heat. 
However they are required for the completeness of the theory up to 
terms of order $1/L$, and for a comparison with conformal field theory). 
Specifically, we will prove that if the coefficients of $(\pi /2L) 
(N_R + N_L )^2$ and of $(\pi /2L) (N_R - N_L )^2$ in Eq. (\ref{ham1})
are denoted by $A$ and $B$ respectively, then 
\beq
A B ~=~ v^2 ~.
\eeq
It will then follow that if $A$ is equal to $v/K$, $B$ must be equal to $vK$. 
Although the expressions in Eq. 
(\ref{ham1}) are valid for lattice models also,
let us for simplicity consider a continuum model which is invariant under 
Galilean transformations. First, let us set $N_R = N_L$, so that we have
added $\Delta N = 2N_R$ particles to the system. The sound velocity $v$ of a 
one-dimensional system is related to the density of particles $\rho = N/L$ 
(where $N= N_0 + N_R + N_L$), the particle mass $m$, and the pressure $\cal 
P$ as
\beq
m \rho v^2 ~=~ - ~L~ \Bigl( ~\frac{\partial {\cal P}}{\partial L} ~
\Bigr)_N ~.
\eeq 
The pressure is related to the ground state energy by ${\cal P} = 
- ( \partial E_0 /\partial L )_N$. Hence
\beq
m \rho v^2 ~=~ -~ L ~\Bigl( ~\frac{\partial^2 E_0}{\partial L^2} ~\Bigr)_N ~
=~ \frac{N^2}{L} ~\Bigl( ~\frac{\partial^2 E_0}{\partial N^2} ~\Bigr)_L ~,
\label{eqna}
\eeq
where the second equality follows from the first because
$E_0$ depends on $N$ and $L$ only through the combination $N/L$. Comparing
Eqs. (\ref{ham1}) and (\ref{eqna}), we see that the coefficient of $(\pi /
2L)(\Delta N)^2$ is given by
\beq
A ~=~ \frac{mv^2}{\pi \rho_0} ~.
\label{defa}
\eeq
(In certain expressions such as Eq. (\ref{defa}), we have ignored the 
difference between $\rho$ and $\rho_0$ since $\Delta N \ll N_0$).
Next, let us take $N_L = - N_R$; this corresponds to moving $N_R$ particles
from the left Fermi point $-k_F$ to the right Fermi point $k_F$
keeping the total number 
of particles equal to $N_0$. The change in momentum is therefore given by
$\Delta P = 2\pi \rho_0 N_R$. Since we can also view 
such an excitation as a center of mass excitation with momentum $\Delta P$,
the change in energy is given by $\Delta E = (\Delta P)^2 /(2mN)$ since the
total mass of the system is $mN$. It follows from this that the coefficient
of $(\pi /2L)(N_R - N_L)^2$ satisfies
\beq
B ~=~ \frac{\pi \rho_0}{m} ~.
\eeq
We thus see that $AB =v^2$ independently of the nature of the interactions 
between the particles.

We now consider the other important property of a Luttinger liquid, 
namely, the absence of a discontinuity in $n(k)$ at the Fermi 
momenta or, equivalently, the absence of a pole in the one-particle
propagator. Thus the effect of interactions is so drastic in one dimension
that the self-energy $\Sigma$ in Eq. (\ref{green}) becomes non-analytic at 
the Fermi points. As a result, $n(k)$ becomes continuous at $k= \pm k_F$
with the form 
\beq
n (k) ~= ~ n (k_F ) ~+~~ {\rm constant} ~\cdot ~{\rm sign} (k-k_F ) ~\vert 
k - k_F \vert^\beta ~,
\eeq
where ${\rm sign} (z) \equiv 1$ if $z>0$, $-1$ if $z<0$ and $0$ if $z=0$. 
The exponent $\beta$ is a positive number 
whose value depends on the strength of the
interactions; for a non-interacting system, $\beta =0$ and we recover the
discontinuity in $n(k)$. Similarly, the density of states (DOS) 
is obtained by integrating Eq. (\ref{green}) over all momenta; near zero 
energy it vanishes with a power-law form
\beq
{\tilde n} (\omega ) ~\sim ~\vert \omega \vert^{\beta} ~,
\eeq
which signals the absence of one-particle states in the low-energy spectrum.
We will see later how the exponent $\beta$ can be calculated in an 
interacting system called the Tomonaga-Luttinger model.

\section{Bosonization}

The basic idea of bosonization is that there are certain objects which can 
be calculated either in a fermionic theory or in a bosonic theory, and the
two calculations give the same answer \cite{hald,schu,gogo,aff1,tomo,vond}. 
Further, a particular quantity may seem very difficult to compute in one 
theory and may be easily calculable in the other theory. Bosonization works 
best in two space-time dimensions although there have been some attempts to 
extend it to higher dimensions. 

In two dimensions, bosonization can be 
studied in either real time (Minkowski space) or in imaginary time (Euclidean
space). In both cases, there is a one-to-one correspondence between the 
correlation functions of some fermionic and bosonic operators. We will work 
in real time here because bosonization has an added advantage in that
case, namely, that there is a direct relationship between the creation and 
annihilation operators for a boson in terms of the corresponding operators
for a fermion \cite{vond}. To show this, we just need to consider a bosonic 
and a fermionic Fock space. A Hamiltonian is {\it not} needed at this stage;
we need to introduce a Hamiltonian only when discussing interactions and
time-dependent correlation functions.

\subsection{Bosonization of a fermion with one chirality}

Let us begin by considering just one component, say, right-moving, of a
single species of fermions on a circle of length $L$ with the following
boundary condition on the one-particle wave functions ${\tilde \psi} (x)$,
\beq
{\tilde \psi} (L) ~=~ e^{-i \pi \sigma} ~{\tilde \psi} (0) ~.
\eeq
Thus $\sigma =0$ and $1$ correspond to periodic and antiperiodic boundary
conditions, but any value of $\sigma$ lying in the range $0 \le \sigma <2$
is allowed in principle. (If we assume that the particles are charged, 
then $\pi \sigma$ can be identified with an Aharonov-Bohm phase and can be 
varied by changing the magnetic flux through the circle). The normalized 
one-particle wave functions are then given by 
\bea
{\tilde \psi}_{n_k} ~&=&~ \frac{1}{\sqrt L} ~e^{ikx} ~, \nonumber \\
k ~&=&~ \frac{2\pi}{L} ~(n_k ~-~ \frac{\sigma}{2} ) ~,
\label{wave}
\eea
where $n_k = 0, \pm 1, \pm 2, ...$ is an integer. We now introduce a second
quantized Fermi field
\beq
\psi_R (x) ~=~ \frac{1}{\sqrt L} ~\sum_{k=-\infty}^{\infty} ~c_{R,k} ~
e^{ikx} ~,
\label{psi1}
\eeq
where the subscript $R$ stands for right-moving, and
\beq
\{ c_{R,k} , c_{R,k^\prime} \} ~=~ 0 ~, \quad {\rm and} \quad \{ c_{R,k} , 
c_{R,k^\prime}^\dagger \} ~=~ \delta_{k k^\prime} ~.
\label{anti}
\eeq
Using the identity
\beq
\sum_{n=-\infty}^{\infty} ~e^{iny} ~=~ 2\pi ~\sum_{m=-\infty}^{\infty} ~
\delta (y - 2 \pi m) ~,
\eeq
we obtain
\bea
\{ \psi_R (x), \psi_R (x^\prime) \} ~&=&~ 0 ~, \nonumber \\
{\rm and} \quad \{ \psi_R (x) , \psi_R^\dagger (x^\prime) \} ~&=&~ \delta 
(x - x^\prime) \quad {\rm for} \quad 0 \le x, x^\prime \le L ~. 
\eea

We define the vacuum or Fermi sea of the system to be the state $\vert 0 
\rangle$ satisfying
\bea
c_{R,k} ~\vert 0 \rangle ~&=&~ 0 \quad {\rm for} \quad k > 0 ~, \nonumber \\
c_{R,k}^\dagger ~\vert 0 \rangle ~&=&~ 0 \quad {\rm for} \quad k \le 0 ~,
\label{vac}
\eea
as shown in Fig. 3. (Following this definition of the vacuum state, some
people prefer to write the particle annihilation operator $c_{R,k}$ as a 
hole creation operator $d_{R,-k}^\dagger$ for $k \le 0$).
Given any operator $A$ which can be written as a product 
of a string of $c$'s and $c^\dagger$'s, we denote its normal ordered form 
by the symbol $: A :$. This new operator is defined by moving all the 
$c_k$ with $k>0$ and $c_k^\dagger$ with $k \le 0$ to the right of all
the $c_k$ with $k \le 0$ and $c_k^\dagger$ with $k>0$. This is achieved
by transposing as many pairs of creation and annihilation operators as 
necessary, remembering to multiply by a factor of $-1$ for each 
transposition. (It is sometimes claimed that 
$:A: = A - \langle 0 \vert A \vert 0 \rangle$. This is true if $A$ is
quadratic in the $c$'s and $c^\dagger$'s, but it is not true in general).

Next we define the fermion number operator
\beq
{\hat N}_R ~=~ \sum_{k=-\infty}^{\infty} ~: c_{R,k}^\dagger c_{R,k} : ~=~ 
\sum_{k>0} ~ c_{R,k}^\dagger c_{R,k} ~-~ \sum_{k \le 0} ~c_{R,k} 
c_{R,k}^\dagger ~.
\eeq
Thus ${\hat N}_R \vert 0 \rangle =0$. Now consider all possible states
$\vert \Psi \rangle$ satisfying ${\hat N}_R \vert \Psi \rangle = 0$. Clearly,
any such state can only differ from $\vert 0 \rangle$ by a certain number
of particle-hole excitations, $i.e.$, it must be of the form
\beq
\vert \Psi \rangle ~=~ c_{R,k_1}^\dagger c_{R,k_2} c_{R,k_3}^\dagger 
c_{R,k_4} c_{R,k_5}^\dagger c_{R,k_6} ... \vert 0 \rangle ~,
\label{plehole}
\eeq
where the $k_i$ are all different from each other, $k_1 , k_3 , ... > 0$,
and $k_2 , k_4, ... \le 0$. Two such excitations are shown in Fig. 4.
We will now see that all such states can be written in terms of certain
bosonic creation operators acting on the vacuum. Let us define the operators
\bea
b_{R,q}^\dagger ~&=&~ \frac{1}{\sqrt n_q} ~\sum_{k=-\infty}^{\infty} ~
c_{R,k+q}^\dagger c_{R,k} ~, \nonumber \\
b_{R,q} ~&=&~ \frac{1}{\sqrt n_q} ~\sum_{k=-\infty}^{\infty} ~
c_{R,k-q}^\dagger c_{R,k} ~, \nonumber \\
q ~&=&~ \frac{2\pi}{L} ~n_q ~,
\label{bos1}
\eea
where $n_q =1,2,3,...$. Note that we have defined the boson momentum label 
$q$ to be positive. Also, the fermion boundary condition parameter $\sigma$ 
does not appear in the definitions in Eq. (\ref{bos1}). We can check that
\bea
[ {\hat N}_R , b_{R,q} ] ~&=&~ [ {\hat N}_R , b_{R,q}^\dagger ] ~=~ 0 ~, 
\nonumber \\
{[} b_{R,q} , b_{R,q^\prime} {]} ~&=&~ 0 ~, \nonumber \\
{[} b_{R,q} , b_{R,q^\prime}^\dagger {]} ~&=&~ \delta_{q q^\prime} ~. 
\label{comm1}
\eea
Checking the last identity for $q=q^\prime$ is slightly tricky due to the
presence of an infinite number of fermion momenta $k$. One way to derive the
commutators is to multiply each $c_k$ and $c_k^\dagger$ by a factor of $\exp 
[ - \alpha \vert k \vert /2]$ in Eq. (\ref{bos1}), and to let 
$\alpha \rightarrow 0$ at the end of the 
calculation. We should emphasize that the length scale $\alpha$ is not
to be thought of as a short-distance cut-off like a lattice spacing; if we
had introduced a lattice, the number of fermion modes would have been finite,
and the bosonization formulas in Eq. (\ref{bos1}) would not have given the
correct commutation relations.

We see that the vacuum defined above satisfies $b_{R,q} \vert 0 \rangle =0$ 
for all $q$. If we consider any operator $A$ consisting of a string of $b$'s
and $b^\dagger$'s, we can define its bosonic normal ordered form $:A:$ by 
taking all the $b_q$'s to the right of all the $b_q^\dagger$'s by suitable 
transpositions. Given an operator $A$ which can be written in terms of either
fermionic or bosonic operators, normal ordering it in the fermionic and 
bosonic ways do not always give the same result. However, it will always be
clear from the context which normal ordering we mean.

We can now begin to understand why bosonization works.
First of all, note that there is a one-to-one correspondence between the
particle-hole excitations described in Eq. (\ref{plehole}) and 
the bosonic excitations created by the $b^\dagger$'s \cite{scho}. 
For instance, consider a bosonic excitation in which states with the
momenta labeled by the integers $n_1 \ge n_2 \ge ... \ge n_j >0$ (following 
the convention in Eq. (\ref{bos1})) are excited. Some of these integers
may be equal to each other; that would mean that particular momenta has an
occupation number greater than $1$. Now we can map this excitation to a 
fermionic excitation in which $j$ fermions occupying the states labeled by 
the momenta integers $0, -1, -2, ..., -j+1$ (following the convention in Eq. 
(\ref{wave})) are excited to momenta labeled by $n_1, n_2 -1, n_3 -2, ..., 
n_j - j+1$ respectively. This is clearly a one-to-one
map, and we can reverse it to uniquely obtain a bosonic excitation from a 
given fermionic excitation. This mapping allows us to show, once an
appropriate Hamiltonian is defined, that thermodynamic quantities like
the specific heat are identical in the fermionic and bosonic models.

The above mapping makes it plausible, although it requires more effort to 
prove, that {\it all} 
particle-hole excitations can be produced by combinations of
$b^\dagger$'s acting on the vacuum. For instance, the state in Fig. 4 (a) is 
given by $b_{R,1}^\dagger \vert 0 \rangle$. However the state in Fig. 4 (b)
has a more lengthy expression in terms of bosonic operators, although it
is also a single particle-hole excitation just like (a); to be explicit,
it is given by the linear combination $(1/6)[ 2 b_{R,3}^\dagger + 3 
b_{R,2}^\dagger b_{R,1}^\dagger + (b_{R,1}^\dagger )^3 ] \vert 0 
\rangle$.

Next, we define bosonic field operators and show that some
bilinears in fermionic fields, such as the density $\rho_R (x)$, have
simple expressions in terms of bosonic fields. Define the fields
\bea
\chi_R (x) ~&=&~ \frac{i}{2 {\sqrt \pi}} ~\sum_{q>0} ~\frac{1}{\sqrt n_q} ~
b_{R,q} ~e^{iqx - \alpha q/2} ~, \nonumber \\
\chi_R^\dagger (x) ~&=&~ - ~\frac{i}{2 {\sqrt \pi}} ~\sum_{q>0} ~
\frac{1}{\sqrt n_q} ~ b_{R,q}^\dagger ~e^{-iqx - \alpha q/2} ~, \nonumber \\
\phi_R (x) ~&=&~ \chi_R (x) ~+~ \chi_R^\dagger (x) ~-~ \frac{\sqrt \pi 
x}{L} ~{\hat N}_R ~.
\label{phi}
\eea
The last term in the definition of $\phi_R (x)$ has been put in for later
convenience; it simplifies the expressions for the 
Hamiltonian and the fermion 
density in terms of $\phi_R$. (Some authors prefer not to include that 
term in the definition of $\phi_R$ but add it separately in the Hamiltonian 
and density). Note that ${\hat N}_R$ commutes with both $\chi_R$ and 
$\chi_R^\dagger$. From the commutation relations in Eq. (\ref{comm1}), we 
see that
\bea
{[} \chi_R (x), \chi_R (x^\prime) {]} ~&=& 0 ~, \nonumber \\
{[} \chi_R (x), \chi_R^\dagger (x^\prime) {]} ~&=& - ~\frac{1}{4\pi} ~
\ln ~{[} ~1 ~-~ \exp ~(~ -~ \frac{2\pi}{L} (~\alpha + i (x- x^\prime ) ~
) ~{]} ~, \nonumber \\
&=& - ~\frac{1}{4\pi} ~\ln ~{[} ~\frac{2\pi}{L} (~\alpha + i ( x- x^\prime 
)~) ~{]} \quad {\rm in ~the ~limit} \quad L \rightarrow \infty ~.
\eea
Henceforth, the limit $L \rightarrow \infty$ will be assumed wherever
convenient. We find that
\bea
{[} \phi_R (x), \phi_R (x^\prime) {]} ~&=&~ \frac{1}{4\pi} ~\ln \Bigl[ ~
\frac{\alpha -i(x- x^\prime)}{\alpha + i(x-x^\prime)} ~\Bigr] ~, \nonumber \\
&=&~ - ~\frac{i}{4} ~{\rm sign} ~(x- x^\prime) \quad {\rm in ~the ~limit} 
\quad \alpha \rightarrow 0 ~.
\label{comm2}
\eea
Thus the commutator of two $\phi$'s looks like a step function which is 
smeared over a region of length $\alpha$. 

Now we use the operator identity 
\beq
\exp A ~\exp B ~=~ \exp ~(A+B+ \frac{1}{2} [A,B]) ~,
\label{bch}
\eeq
if $[A,B]$ commutes with both $A$ and $B$. It follows that 
\beq
\exp ~{[} i 2 {\sqrt \pi} \chi_R^\dagger (x) {]} ~\exp ~{[} i 2 {\sqrt \pi} 
\chi_R (x) {]} ~\exp ~{[} i \frac{2\pi x}{L} {\hat N}_R {]} = \Big( 
\frac{L}{2\pi \alpha} \Bigr)^{1/2} ~\exp ~{[} i2{\sqrt \pi} \phi_R (x) {]} ~.
\label{exphi}
\eeq
We observe that the left hand side of this equation is normal ordered while 
the right hand side is not; that is why the two sides are 
related through a divergent factor involving $L/ \alpha$.

We can show that the fermion density operator is linear in the bosonic
field, namely,
\bea
\rho_R (x) ~&=&~ : \psi_R^\dagger (x) {\psi_R} (x) : \nonumber \\
&=&~ \frac{1}{L} ~\sum_{q>0} ~{\sqrt n_q} ~(~ b_{R,q} e^{iqx} ~+~ 
b_{R,q}^\dagger e^{-iqx} ~)~ +~ \frac{1}{L} ~\sum_k ~: c_{R,k}^\dagger 
c_{R,k} : \nonumber \\
&=&~ - ~\frac{1}{\sqrt \pi} ~\frac{\partial \phi_R}{\partial x} ~.
\label{rhor}
\eea

We now go in the opposite direction and
construct fermionic field operators from bosonic ones. To do this, we 
first define the Klein factors $\eta_R$ and $\eta_R^\dagger$ which are 
unitary operators satisfying
\bea
[ {\hat N}_R , \eta_R^\dagger ] ~&=&~ \eta_R^\dagger ~, \quad [ {\hat 
N}_R , \eta_R ] ~=~ - ~\eta_R ~, \nonumber \\
{[} \eta_R , b_{R,q} {]} ~&=&~ {[} \eta_R , b_{R,q}^\dagger {]} ~=~ 0~.
\eea
Pictorially, in terms of Figs. 3 and 4, the action of $\eta_R^\dagger$ is 
to raise all the occupied fermion states by one unit of momentum, while the 
action of $\eta_R$ is to lower all the fermion occupied states by one unit of 
momentum. Although these actions are easy to describe in words, the explicit 
expressions for $\eta_R$ and $\eta_R^\dagger = {\eta_R}^{-1}$ in terms of 
the $c$'s and $c^\dagger$'s are rather complicated\cite{vond}. The Klein 
factors will be needed to ensure the correct anticommutation relations 
between the fermionic operators constructed below.

We observe that
\bea
{[} b_{R,q} , \psi_R (x) {]} ~&=&~ - ~\frac{e^{-iqx}}{\sqrt n_q} ~\psi_R 
(x) ~,
\nonumber \\
{[} b_{R,q}^\dagger ,\psi_R (x) {]} ~&=&~ -~\frac{e^{iqx}}{\sqrt n_q} ~
\psi_R (x)~.
\eea
Since $b_{R,q}$ annihilates the vacuum, we have
\beq
b_{R,q} ~\psi_R (x) ~\vert 0 \rangle ~=~ - ~\frac{e^{-iqx}}{\sqrt n_q} ~
\psi_R (x) ~ \vert 0 \rangle ~.
\eeq
Thus $\psi_R (x) \vert 0 \rangle$ is an eigenstate of $b_{R,q}$ for every 
value of $q$, namely, it is a coherent state. We therefore make the ansatz
\bea
\psi_R (x) ~\vert 0 \rangle ~&=&~ Q(x) ~\exp ~[~ - ~\sum_{q>0} ~
\frac{e^{-iqx}}{\sqrt n_q} ~b_{R,q}^\dagger ~\vert 0 \rangle ~] ~\vert 0 
\rangle ~, \nonumber \\
&=&~ Q(x) ~\exp ~[~ - i 2 {\sqrt \pi} ~\chi_R^\dagger (x) ~] ~\vert 0 
\rangle ~,
\eea
where $Q(x)$ is some operator which commutes with all the $b$'s and 
$b^\dagger$'s. Since $\psi_R$ reduces the fermion number by $1$, $Q$ must
contain a factor of $\eta_R$. Let us try the form $Q(x) = F(x) \eta_R$, 
where $F(x)$ is a $c$-number function of $x$. The form of $F$ is 
determined by computing
\bea
F (x) ~&=&~ \langle 0 \vert ~\eta_R^\dagger \eta_R F (x) ~\vert 0 \rangle 
\nonumber \\
&=&~ \langle 0 \vert ~\eta_R^\dagger \psi_R (x) ~\vert 0 \rangle \nonumber \\
&=&~ \frac{e^{-i\pi \sigma x/L}}{\sqrt L} ~.
\label{deff}
\eea
(The last line in Eq. (\ref{deff}) has been derived by using the actions 
of $\eta_R^\dagger$ above and of $\psi_R$ in Eq. (\ref{psi1}).
To see this explicitly, note that $\langle 0 \vert 
\eta_R^\dagger$ is the conjugate of the state in which the top most 
fermion has been removed from the vacuum. Hence, in $\psi_R \vert 0 \rangle$,
we only have to consider the state in which the top most fermion has been
removed; so we require the wave function of the state with $n_k =0$
in Eq. (\ref{wave})). We now obtain 
\bea
\psi_R (x) ~\vert 0 \rangle ~&=&~ \frac{e^{-i\pi \sigma x/L}}{\sqrt {2\pi 
\alpha}} ~\eta_R ~ e^{-i 2 {\sqrt \pi} \phi_R (x)} ~\vert 0 \rangle ~,
\eea
where we have used Eq. (\ref{exphi}) and the fact that $\chi_R (x)$ and 
${\hat N}_R$ annihilate the vacuum. We are thus led to the plausible 
conjecture 
\beq
\psi_R (x) ~=~ \frac{e^{-i\pi \sigma x/L}}{\sqrt {2\pi \alpha}} ~ \eta_R ~ 
e^{-i 2 {\sqrt \pi} \phi_R (x)} ~.
\label{bos2}
\eeq
To prove this, we need to show that the two sides of this equation have the
same action on {\it all} states, not just the vacuum. Such a proof is given
in Ref. 7. Eq. (\ref{bos2}) is one of the most important identities in 
bosonization.

We next introduce a non-interacting Hamiltonian by defining the energy of 
the fermion mode with momentum $k$ to be
\beq
\epsilon_k ~=~ v_F k 
\eeq
for all values of $k$. The Hamiltonian is
\bea
H_0 ~&=&~ v_F~\sum_{k=-\infty}^{\infty} ~k~: c_{R,k}^\dagger c_{R,k} : ~+~ 
\frac{\pi v_F}{L} ~{\hat N}_R^2 \nonumber \\
&=&~ - v_F ~\int_0^L ~dx ~: \psi_R^\dagger i \partial_x \psi_R : ~+~
\frac{\pi v_F}{L} ~{\hat N}_R^2 ~.
\label{ham2}
\eea
This defines the chiral Luttinger model. (The term proportional to 
${\hat N}_R^2$ has been introduced in Eq. (\ref{ham2}) so as to reproduce 
similar terms in Eq. (\ref{ham1}) after we introduce left-moving fields in 
the next section). We can check that $H_0 \vert 0 \rangle = 0$ and
\bea
{[} H_0 , b_{R,q} {]} ~&=&~ - v_F q ~b_{R,q} ~, \nonumber \\
{[} H_0 , b_{R,q}^\dagger {]} ~&=&~ v_F q ~b_{R,q}^\dagger ~,
\eea
To reproduce these relations in the bosonic language, we must have
\bea 
H_0 ~&=&~ v_F ~\sum_{q>0} ~q b_{R,q}^\dagger b_{R,q} ~+~ \frac{\pi 
v_F}{L} ~{\hat N}_R^2 \nonumber \\
&=&~ v_F ~\int_0^L ~dx ~: (\partial_x \phi )^2 : ~.
\label{ham3}
\eea

We can introduce an interaction in this model which is quadratic in the
fermion density. Let us consider the interaction
\beq
V ~=~ \frac{1}{2} ~\int_0^L ~g_4 ~{\rho_R}^2 (x) ~=~ \frac{g_4}{2\pi} ~
\sum_{q>0} ~q ~b_{R,q}^\dagger b_{R,q} ~.
\label{int1}
\eeq
Physically, such a term could arise if there is a short-range ($i.e.$,
screened) Coulomb repulsion or a phonon mediated attraction
between two fermions. We will therefore not make any assumptions about the 
sign of the interaction parameter $g_4$. If we add Eq. (\ref{int1})
to Eq. (\ref{ham3}), we see that the only effect of the interaction in this
model is to renormalize the velocity from $v_F$ to $v_F + (g_4 /2\pi)$.

In the next section, we will consider a model containing fermions with
opposite chiralities; we will then see that a density-density interaction
can have more interesting effects than just renormalizing the velocity.

\subsection{Bosonization of a fermion with two chiralities}

Let us consider a fermion with both right- and left-moving components
as depicted in Figs. 3 and 5 respectively. For the left-moving fermions
in Fig. 5, we define the momentum label $k$ as increasing towards the left;
the advantage of this choice is that the vacuum has the negative $k$ states 
occupied and the positive $k$ states unoccupied for both chiralities.
We introduce a chirality label $\nu$, such that $\nu = R$ and
$L$ refer to right- and left-moving particles respectively. Sometimes
we will use the numerical values $\nu = 1$ and $-1$ for $R$ and $L$;
this will be clear from the context. Let us choose periodic boundary 
conditions on the circle so that $\sigma =0$. Then the Fermi fields are 
given by
\bea
\psi_\nu (x) ~&=&~ \frac{1}{\sqrt L} ~\sum_{k=-\infty}^{\infty} ~
c_{\nu,k} ~e^{i\nu kx} ~, \nonumber \\
k ~&=&~ \frac{2\pi}{L} ~n_k ~,
\label{psi2}
\eea
where $n_k = 0, \pm 1, \pm 2, ...$, and
\bea
\{ c_{\nu ,k} , c_{\nu^\prime , k^\prime} \} ~&=&~ 0 ~, \nonumber \\
\{ c_{\nu ,k} , c_{\nu^\prime , k^\prime}^\dagger \} ~&=&~ 
\delta_{\nu \nu^\prime} ~\delta_{k k^\prime} ~. 
\eea
The vacuum is defined as the state satisfying
\bea
c_{\nu ,k} ~\vert 0 \rangle ~=~ 0 \quad {\rm for} \quad k > 0 ~, 
\nonumber \\
c_{\nu ,k}^\dagger ~\vert 0 \rangle ~=~0 \quad {\rm for} \quad k \le 0~.
\eea
We can then define normal ordered fermion number operators ${\hat N}_\nu$ 
in the usual way.

Next we define bosonic operators
\bea
b_{\nu ,q}^\dagger ~&=&~ \frac{1}{\sqrt n_q} ~\sum_{k=-\infty}^{\infty} ~
c_{\nu ,k+q}^\dagger c_{\nu ,k} ~, \nonumber \\
b_{\nu ,q} ~&=&~ \frac{1}{\sqrt n_q} ~\sum_{k=-\infty}^{\infty} ~
c_{\nu ,k-q}^\dagger c_{\nu ,k} ~.
\label{bos3}
\eea
Note that $b_{R,q}^\dagger$ and $b_{L,q}^\dagger$ create excitations with
momenta $q$ and $-q$ respectively, where the label $q$ is
always taken to be positive. We can show as before that
\beq
{[} b_{\nu ,q} , b_{\nu^\prime , q^\prime} {]} ~=~ 0 ~, \quad {\rm 
and} \quad {[} b_{\nu ,q} , b_{\nu^\prime , q^\prime}^\dagger {]} ~=~ 
\delta_{\nu \nu^\prime} ~\delta_{q q^\prime} ~.
\eeq

The unitary Klein operators $\eta_\nu$ ($\eta_\nu^\dagger$) are 
defined to be operators which raise (lower) the momentum label $k$ of 
all the occupied states for fermions of type $\nu$. We then have
\bea
\{ \eta_R ,\eta_L \} ~&=&~ \{ \eta_R ,\eta_L^\dagger \} ~=~0~, \nonumber \\
{[} {\hat N}_\nu , \eta_{\nu^\prime}^\dagger {]} ~&=&~ \delta_{\nu 
\nu^\prime} ~ \eta_{\nu^\prime}^\dagger ~, \quad {[} {\hat N}_\nu , 
\eta_{\nu^\prime} {]} ~=~ - ~\delta_{\nu \nu^\prime} ~
\eta_{\nu^\prime} ~, \nonumber \\
{[} \eta_\nu , b_{\nu^\prime ,q} {]} ~&=&~ {[} \eta_\nu , 
b_{\nu^\prime , q}^\dagger {]} ~=~ 0~.
\eea

We now define the chiral creation and annihilation fields
\bea
\chi_\nu (x) ~&=&~ \frac{i\nu}{2 {\sqrt \pi}} ~\sum_{q>0} ~\frac{1}{
\sqrt n_q} ~b_{\nu ,q} ~e^{i\nu qx - \alpha q/2} ~, \nonumber \\
\chi_\nu^\dagger (x) ~&=&~ - ~\frac{i\nu}{2 {\sqrt \pi}} ~\sum_{q>0} ~
\frac{1}{\sqrt n_q} ~b_{\nu ,q}^\dagger ~e^{-i\nu qx - \alpha q/2} ~. 
\eea
Then
\bea
{[} \chi_\nu (x), \chi_\nu^\prime (x^\prime ) {]} &=& 0 ~, \nonumber \\
{[} \chi_\nu (x), \chi_{\nu^\prime}^\dagger (x^\prime ) {]} &=& - ~
\frac{1}{4\pi} ~\delta_{\nu \nu^\prime} ~\ln ~{[} ~\frac{2\pi}{L} 
(~\alpha - i \nu ( x- x^\prime )~) ~{]} \quad {\rm in ~the ~limit} \quad L 
\rightarrow \infty ~.
\label{comm3}
\eea
The chiral fields
\beq
\phi_\nu (x) ~=~ \chi_\nu (x) ~+~ \chi_\nu^\dagger (x) ~-~ \frac{
\sqrt \pi x}{L} ~{\hat N}_\nu ~,
\eeq
satisfy
\beq
{[} \phi_\nu (x) , \phi_{\nu^\prime} (x) {]} ~=~ - \frac{i\nu}{4} ~
\delta_{\nu \nu^\prime} ~{\rm sign} ~(x-x^\prime ) 
\eeq
in the limit $\alpha \rightarrow 0$. Finally, we can define two fields 
dual to each other
\bea
\phi (x) ~&=&~ \phi_R (x) ~+~ \phi_L (x) ~, \nonumber \\
\theta (x) ~&=&~ -~ \phi_R (x) ~+~ \phi_L (x) ~,
\eea
such that $[ \phi (x), \phi (x^\prime)] = [ \theta (x), \theta (x^\prime)] 
= 0$, while
\beq
{[} \phi (x) , \theta (x^\prime) {]} ~=~ \frac{i}{2} ~{\rm sign} ~(x-
x^\prime) ~.
\label{comm4}
\eeq

The fermion density operators $\rho_\nu (x) = :\psi_\nu^\dagger (x)
\psi_\nu (x) :$ satisfy $\rho_\nu = \partial_x \phi_\nu /{\sqrt 
\pi}$. Hence the total density and current operators are given by
\bea
\rho (x) ~&=&~ \rho_R + \rho_L ~=~ - ~\frac{1}{\sqrt \pi} ~\partial_x \phi ~,
\nonumber \\
j (x) ~&=&~ v_F (\rho_R - \rho_L ) ~=~ \frac{v_F}{\sqrt \pi} ~\partial_x 
\theta ~,
\eea
where $v_F$ is a velocity to be introduced shortly.

We can again show that the fermionic fields are given in terms of the bosonic
ones as
\bea
\psi_R (x) ~&=&~ \frac{1}{\sqrt {2\pi \alpha}} ~\eta_R ~e^{-i 2 {\sqrt \pi} 
\phi_R} ~, \nonumber \\
\psi_L (x) ~&=&~ \frac{1}{\sqrt {2\pi \alpha}} ~\eta_L ~e^{i 2 {\sqrt \pi} 
\phi_L} ~.
\label{bos4}
\eea

As before, we introduce a linear dispersion relation $\epsilon_{\nu ,k} = 
v_F k$ for the fermions. The non-interacting Hamiltonian then takes the form
\bea
H_0 &=& v_F \sum_{k=-\infty}^\infty ~k ~{[} ~: c_{R,k}^\dagger c_{R,k} ~+~ 
c_{L,k}^\dagger c_{L,k} ~: ~{]} ~+~ \frac{\pi v_F}{L} ({\hat N}_R^2 + {\hat 
N}_L^2 ) \nonumber \\
&=& -v_F \int_0^L dx ~[: \psi_R^\dagger (x) i \partial_x \psi_R (x) ~-~
\psi_L^\dagger (x) i \partial_x \psi_L (x) : ] ~+~ \frac{\pi v_F}{L} 
({\hat N}_R^2 + {\hat N}_L^2 )
\eea
in the fermionic language, and
\bea
H_0 ~&=&~ v_F ~\sum_{q>0} ~q ~(~ b_{R,q}^\dagger b_{R,q} ~+~ 
b_{L,q}^\dagger b_{L,q} ~)~ +~ \frac{\pi v_F}{L} ({\hat N}_R^2 + 
{\hat N}_L^2 ) \nonumber \\
&=&~ v_F ~\int_0^L ~dx ~[ ~: (\partial_x \phi_R )^2 ~+~ (\partial_x \phi_L 
)^2 : ~] \nonumber \\
&=&~ \frac{v_F}{2} ~\int_0^L ~dx ~[ ~: (\partial_x \phi )^2 ~+~ (\partial_x 
\theta )^2 : ~]
\eea
in the bosonic language. If we use this Hamiltonian to transform all the
fields to time-dependent Heisenberg fields, we find that $\psi_R ,\phi_R$ 
become functions of $x_R = x -v_F t$ while $\psi_L ,\phi_L$ become 
functions of $x_L = x +v_F t$.

{} From Eq. (\ref{comm4}), we see that the field canonically conjugate to 
$\phi$ is given by
\beq
\Pi ~=~ \partial_x \theta ~.
\eeq
Thus
\beq
{[} \phi (x) , \Pi (x^\prime) {]} ~=~ i \delta (x-x^\prime) ~,
\eeq
and 
\beq
H_0 ~=~\frac{v_F}{2} ~\int_0^L ~dx ~[ ~\Pi^2 ~+~ (\partial_x \phi )^2 ~] ~.
\eeq

We now study the effects of four-fermi interactions. In the beginning it is 
simpler to work in the Schr\"odinger representation in which the 
fields are time-independent; we will transform to the Heisenberg 
representation later when we compute the correlation functions. Let us 
consider an interaction of the form
\beq
V ~=~ \frac{1}{2} ~\int_0^L ~dx ~[~ 2g_2 ~\rho_R (x) \rho_L (x) ~+~g_4 ~
(~ \rho_R^2 (x) ~+~ \rho_L^2 (x) ~) ~] ~.
\label{int2}
\eeq
Physically, we may expect an interaction such as $g :\rho^2 (x):$, so that
$g_2 = g_4 =g$. However, it is instructive to allow $g_2$ to differ from 
$g_4$ to see what happens. For reasons explained before, we will again
not assume anything about the signs of $g_2$ and $g_4$. In the fermionic
language, the interaction takes the form
\bea
V= \frac{1}{2L} \sum_{k_1 , k_2 ,k_3 =-\infty}^{\infty} [ && 2 g_2 
c_{R,k_1 + k_3}^\dagger c_{R,k_1} c_{L,k_2 +k_3}^\dagger c_{L,k_2} \nonumber \\
&& + g_4 ( c_{R,k_1 + k_3}^\dagger c_{R,k_1} c_{R,k_2 -k_3}^\dagger c_{R,
k_2} ~+~ c_{L,k_1 + k_3}^\dagger c_{L,k_1} c_{L,k_2 -k_3}^\dagger c_{L,k_2} )].
\label{int3}
\eea
{} From this expression we see that $g_2$ corresponds to a two-particle 
scattering involving both chiralities; in this model, we can call it 
either forward scattering or backward scattering since there is no 
way to distinguish between the two processes
in the absence of some other quantum number such as spin. 
The $g_4$ term corresponds to a scattering between two fermions with the
same chirality, and therefore describes a forward scattering process. 

The quartic interaction in Eq. (\ref{int3}) seems very difficult to analyze. 
However we will now see that it is easily solvable in the bosonic 
language; indeed this is one of the main motivations behind bosonization. 
The bosonic expression for the total Hamiltonian $H = H_0 +V$ is found to be 
\bea
H = && \sum_{q>0} ~q~ {[} v_F~ (b_{R,q}^\dagger b_{R,q} + 
b_{L,q}^\dagger b_{L,q} )~+~ \frac{g_2}{2\pi} (b_{R,q}^\dagger 
b_{L,q}^\dagger + b_{R,q} b_{L,q} ) ~+~ \frac{g_4}{2\pi} (b_{R,q}^\dagger 
b_{R,q} + b_{L,q}^\dagger b_{L,q} ) {]} \nonumber \\
&& +~ \frac{\pi v_F}{L} ~ (~ {\hat N}_R^2 ~ +~ {\hat N}_L^2 ~) ~
+~ \frac{g_2}{L} ~{\hat N}_R {\hat N}_L ~+~ \frac{g_4}{2L} ~ (~
{\hat N}_R^2 ~ +~ {\hat N}_L^2 ~) ~.
\label{ham4}
\eea
The $g_4$ term again renormalizes the velocity. The $g_2$ term can then be
rediagonalized by a Bogoliubov transformation. We first define two
parameters
\bea
v ~&=&~ \Bigl[ ~(~ v_F + \frac{g_4}{2\pi} - \frac{g_2}{2\pi} ~)~
(~ v_F + \frac{g_4}{2\pi} + \frac{g_2}{2\pi} ~) ~\Bigr]^{1/2} ~, \nonumber \\
K ~&=&~ \Bigl[ ~(~ v_F + \frac{g_4}{2\pi} - \frac{g_2}{2\pi}~) ~/~ (~v_F + 
\frac{g_4}{2\pi} + \frac{g_2}{2\pi} ~) ~\Bigr]^{1/2} ~.
\eea
Note that $K < 1$ if $g_2$ is positive (repulsive interaction), and $> 1$ 
if $g_2$ is negative (attractive interaction). (If $g_2$ is so large that
$v_F + g_4/(2\pi) - g_2/(2\pi) < 0$, then our analysis breaks down. The 
system does not remain a Luttinger liquid in that case, and is likely to go 
into a different phase such as a state with charge density order). 
The Bogoliubov transformation now takes the form
\bea
{\tilde b}_{R,q} ~&=&~ \frac{b_{R,q} ~+~ \gamma ~
b_{L,q}^{\dagger}}{\sqrt {1 - \gamma^2}} ~, \nonumber \\
{\tilde b}_{L,q} ~&=&~ \frac{b_{L,q} ~+~ \gamma ~
b_{R,q}^{\dagger}}{\sqrt {1 - \gamma^2}} ~, \nonumber \\
\gamma ~&=&~ \frac{1-K}{1+K} ~,
\label{bogo}
\eea
for each value of the momentum $q$. The Hamiltonian is then given by
the quadratic expression
\bea
H ~=~ & & \sum_{q>0} ~v q ~{[}~ {\tilde b}_{R,q}^{\dagger} {\tilde 
b}_{R,q} ~+~ {\tilde b}_{L,q}^{\dagger} {\tilde b}_{L,q} ~{]} \nonumber \\
& & +~ \frac{\pi v}{2L} ~[~ \frac{1}{K} ~({\hat N}_R + {\hat N}_L )^2 ~+~ K~
({\hat N}_R - {\hat N}_L )^2 ~]~ .
\label{ham5}
\eea
Equivalently,
\beq
H ~=~ \frac{1}{2} ~\int_0^L ~dx ~[~ vK \Pi^2 ~+~ \frac{v}{K} ~(\partial_x
\phi)^2 ~] ~.
\label{ham6}
\eeq

The old and new fields are related as
\bea
\phi_R ~&=&~ \frac{(1+K) ~\tphi_R ~-~ (1-K) ~\tphi_L}{2 {\sqrt K}} ~, 
\nonumber \\
\phi_L ~&=&~ \frac{(1+K) ~\tphi_L ~-~ (1-K) ~\tphi_R}{2 {\sqrt K}} ~, 
\nonumber \\
\phi ~&=&~ {\sqrt K} ~\tphi \quad {\rm and} \quad \theta ~=~ 
\frac{1}{\sqrt K} ~{\tilde \theta} ~.
\label{oldnew}
\eea

Note the important fact that the vacuum changes as a result of the 
interaction; the new vacuum $\vert {\tilde 0} \rangle$ is the state 
annihilated by the 
operators ${\tilde b}_{\nu ,q}$. Since the various correlation functions
must be calculated in this new vacuum, they will depend on the interaction
through the parameters $v$ and $K$. In particular, we will see in the next
section that the power-laws of the correlation functions are governed by $K$.

Given the various Hamiltonians, it is easy to guess the forms of the 
corresponding Lagrangians. For the non-interacting theory ($g_2 = g_4 =0$), 
the Lagrangian density describes a massless Dirac fermion,
\beq
{\cal L} ~=~ i \psi_R^\dagger (\partial_t + v_F \partial_x ) \psi_R ~ +~ i 
\psi_L^\dagger (\partial_t - v_F \partial_x ) \psi_L 
\label{lag1}
\eeq
in the fermionic language, and a massless real scalar field,
\beq
{\cal L} ~=~ \frac{1}{2v_F} ~(\partial_t \phi )^2 ~ -~ \frac{v_F}{2} ~( 
\partial_x \phi )^2 
\eeq
in the bosonic language. For the interacting theory in Eq. (\ref{ham6}), we 
find from Eq. (\ref{oldnew}) that
\beq
{\cal L} ~=~ \frac{1}{2vK} ~(\partial_t \phi )^2 ~-~ \frac{v}{2K} ~( 
\partial_x \phi )^2 ~=~ \frac{1}{2v} ~(\partial_t \tphi )^2 ~-~ 
\frac{v}{2} ~( \partial_x \tphi )^2 ~.
\label{lag2}
\eeq

The momentum operator in Eq. (\ref{ham1}) has the same expression in terms 
of the old and new fields, namely, 
\beq
P ~=~ k_F ({\hat N}_R - {\hat N}_L ) ~+~ \int_0^L ~dx ~\partial_x \phi 
\partial_x \theta ~.
\eeq
We can check that $[P, \phi ] = -i \partial_x \phi$ and $[P, \theta ] = -i 
\partial_x \theta$.

Let us now write down the fields $\tphi$ and $\tilde \theta$ in 
the Heisenberg representation. This is simple to do once we realize that the
right- and left-moving fields must be functions of $x_R =x-vt$ and $x_L =
x+vt$ respectively. We find that
\bea
\tphi (x,t) = \frac{i}{2{\sqrt \pi}} ~\sum_{q>0} \frac{1}{\sqrt n_q} [ 
&& {\tilde b}_{R,q} ~e^{iq(x_R +i\alpha /2)} ~- ~{\tilde b}_{R,q}^{
\dagger}~ e^{-iq(x_R -i\alpha /2)} ~-~ {\tilde b}_{L,q} ~e^{-iq(x_L -i
\alpha /2)} \nonumber \\
&& + ~{\tilde b}_{L,q}^{\dagger} ~e^{iq(x_L +i\alpha /2)} ~] \nonumber \\
- ~\frac{\sqrt \pi}{L} [ && \frac{x}{\sqrt K} ~({\hat N}_R ~+~ 
{\hat N}_L) ~-~ {\sqrt K} vt ({\hat N}_R ~-~ {\hat N}_L ) ~]~, \nonumber \\
{\tilde \theta} (x,t) =~ \frac{i}{2{\sqrt \pi}} ~\sum_{q>0} 
\frac{1}{\sqrt n_q} [ && ~- ~{\tilde b}_{R,q} ~e^{iq(x_R +i\alpha /2)} ~+~ 
{\tilde b}_{R,q}^{\dagger} ~e^{-iq(x_R -i\alpha /2)} ~-~ {\tilde b}_{L,q} ~
e^{-iq(x_L -i\alpha /2)} \nonumber \\
&& + ~{\tilde b}_{L,q}^{\dagger} ~e^{iq(x_L +i\alpha /2)} ~] \nonumber \\
+ ~\frac{\sqrt \pi}{L} [ && {\sqrt K} x ~({\hat N}_R ~-~ {\hat N}_L) ~
-~ \frac{vt}{\sqrt K} ~({\hat N}_R ~+~ {\hat N}_L ) ~]~.
\label{field}
\eea
We observe that the coefficients of ${\hat N}_R$ and ${\hat N}_L$ have terms 
which are linear in $t$. This is necessary because we want the conjugate 
momentum field to satisfy
\beq
\tPi ~=~ \frac{1}{v} ~\partial_t \tphi ~=~ \partial_x {\tilde \theta} ~.
\eeq
We note that a dual equation holds, namely,
\beq
\frac{1}{v} ~\partial_t {\tilde \theta} ~=~ \partial_x \tphi ~.
\eeq
(One can check from Eq. (\ref{field}) that $\tphi_R$ and $\tphi_L$ 
are functions of $x_R$ and $x_L$ alone). In terms of $\theta$, the 
Lagrangian density is 
\beq
{\cal L} ~=~ \frac{K}{2v} ~(\partial_t \theta )^2 ~-~ \frac{Kv}{2} ~( 
\partial_x \theta )^2 ~=~ \frac{1}{2v} ~(\partial_t {\tilde \theta} )^2 ~-~ 
\frac{v}{2} ~( \partial_x {\tilde \theta} )^2 ~.
\label{lag3}
\eeq
Although the Lagrangians in Eq. (\ref{lag2}) and Eq. (\ref{lag3}) have 
opposite signs, the Hamiltonians derived from the two are identical.

Before ending this section, let us comment on a global symmetry of all
these models. It is known that fermionic systems with a conserved charge
are invariant under a global phase rotation
\beq
\psi_R ~\rightarrow ~e^{i\lambda} ~\psi_R ~, \quad {\rm and} \quad \psi_L ~
\rightarrow ~e^{i\lambda} ~\psi_L ~, 
\label{glo1}
\eeq
where $\lambda$ is independent of $(x,t)$. Eq. (\ref{bos4}) then implies 
that the corresponding bosonic theories must remain invariant under
\beq 
\phi ~\rightarrow ~\phi ~, \quad {\rm and} \quad \theta ~\rightarrow 
\theta + ~\frac{\lambda}{\sqrt \pi} ~.
\label{glo2}
\eeq
This provides a constraint on the kinds of terms which can appear in the
Lagrangians of such theories.

\subsection{Field theory of modes near the Fermi momenta}

In the last section, we discussed bosonization for a model of fermions
which has the following properties.
 
\noindent (i) There are an infinite number of right- and left-moving modes
with the momenta going from $- \infty$ to $\infty$, and

\noindent (ii) the relation between energy and momentum is linear for all 
values of the momentum.

\noindent Neither of these properties is true in condensed matter systems
which typically are non-relativistic and have a finite (though possibly
very large) number of states. The question is the following: can bosonization 
give useful results even if these two 
properties do not hold? We will see that the answer is yes, provided that 
we are only interested in the long-wavelength, low-frequency and
low-temperature properties of such systems.

In an experimental system, the fermions may be able to move either on a 
discrete lattice of points such as in a crystal, or in a continuum such as 
the conduction electrons in a metal. For instance, non-interacting fermions
moving in a continuum have a dispersion $\epsilon_k = k^2 /2m$, while 
fermions hopping on a lattice have a dispersion such as $\epsilon_k = - 
t \cos (ka)$ if $a$ is the lattice spacing and $t$ is the nearest neighbor 
hopping amplitude. In either case, a non-interacting system in one-dimension 
will, at zero temperature, have a Fermi surface consisting of two points in 
momentum space given by $k= \pm k_F$ (see Fig. 2). As stated before, we define 
the one-particle energy to be zero at the Fermi points. At low temperatures 
$T$ or low frequencies $\omega$, the only modes which can contribute are the 
ones lying close to those points, $i.e.$, with excitation energies of the 
order of or smaller than $k_B T$ or $\omega$. Near the Fermi points, we can 
approximate the dispersion relation by a linear one, with the velocity being 
defined to be $v_F = (d \epsilon_k / dk)_{k= k_F}$. 
We thus restrict our attention to the right-moving modes with momenta 
lying between $k_F - \Lambda$ and $k_F + \Lambda$, and the left-moving modes
with momenta lying between $-k_F - \Lambda$ and $-k_F + \Lambda$. Here 
$\Lambda$ is taken to be much smaller than the full range of the momentum 
(which is $2\pi /a$ on a lattice if the lattice spacing is $a$), 
but $v_F \Lambda$ is much larger than the temperatures or frequencies of 
interest. If we include only these regions of momenta, 
then the second quantized Fermi field can be written in the approximate form 
\beq
\psi (x,t) ~=~ \psi_R (x,t) ~e^{ik_F x} ~+~ \psi_L (x,t) ~e^{-ik_F x} ~,
\label{fer1}
\eeq
where $\psi_R$ and $\psi_L$ vary slowly over spatial regions which are
large compared to the distance scale $1/\Lambda$. The momentum components
of these slowly varying fields are related to those of $\psi$ as
\bea
\psi_{R,k} (t) ~&=&~ \psi_{k+k_F} (t) ~, \nonumber \\
\psi_{L,k} (t) ~&=&~ \psi_{-k-k_F} (t) ~,
\label{fer2}
\eea
where $-\Lambda \le k \le \Lambda$. These long-wavelength fields are the
ones to which the technique of bosonization can be applied.

The definitions in Eqs. (\ref{fer1}-\ref{fer2}) tell us the forms of the 
various terms in a microscopic model and also tell us which of them survive 
in the long-wavelength limit. For instance, the density is given by
\bea
\rho ~&=&~ : \psi^\dagger \psi : ~=~ : \psi_R^\dagger \psi_R ~+~ 
\psi_L^\dagger \psi_L ~+~ e^{-i2k_F x} ~\psi_R^\dagger \psi_L ~+~ 
e^{i2k_F x} ~ \psi_L^\dagger \psi_R : \nonumber \\
&=& ~- ~\frac{1}{\sqrt \pi} ~\frac{\partial \phi}{\partial x} ~+~\frac{1}{2\pi
\alpha} ~{[} \eta_R^\dagger \eta_L e^{i(2 {\sqrt \pi} \phi - 2k_F x)} ~+~
\eta_L^\dagger \eta_R e^{-i(2 {\sqrt \pi} \phi - 2k_F x)} ~{]} ~.
\label{lat1}
\eea
The terms containing $\exp (\pm i2k_F x)$ in Eq. (\ref{lat1}) vary on a 
distance scale $k_F^{-1}$ which is typically of the same order as the inverse 
particle density $\rho^{-1}$. These terms can therefore be ignored if we 
are only interested in the asymptotic behavior of 
correlation functions at distances much larger than $k_F^{-1}$. In a lattice
model, we have to be more careful about this argument since the lattice 
momentum only needs to be conserved modulo $2\pi /a$ in any process. However,
since $0 < k_F < \pi /a$ in general,
and $x/a$ is an integer, we see that the last two terms in 
Eq. (\ref{lat1}) vary on the scale of the lattice unit $a$; we can 
therefore ignore those terms if we are only interested in phenomena at 
distance scales which are much larger than $a$.

On the other hand, there are situations when a density term like $\rho
\cos (2k_F x)$ is generated in the model; for instance, this happens below a 
Peierls transition if the fermions are coupled to lattice phonons. We then 
find that the slowly varying terms in the continuum field theory are given by 
\bea
\cos (2k_F x) ~\rho ~&=&~ \frac{1}{2} ~[~\psi_R^\dagger \psi_L ~+~ 
\psi_L^\dagger \psi_R ~] \quad {\rm in ~general} ~, \nonumber \\
&=&~ \psi_R^\dagger \psi_L ~+~ \psi_L^\dagger \psi_R \quad {\rm if} \quad
e^{i4k_F x} ~=~ 1 ~.
\label{lat2}
\eea
The second possibility can arise in a lattice model if $4 k_F a= 2 \pi$, 
$i.e.$, at half-filling; we then call it a dimerized system. We will call the 
term on the right hand sides of Eq. (\ref{lat2}) the mass operator. We will 
see below that for any value of $K < 2$, this term produces a gap in the 
low-energy spectrum. This is called the dimerization gap if it occurs in a 
lattice system.

We should emphasize an important difference between models defined in 
the continuum and those defined on a lattice. In the continuum,
$\psi_R^2 (x) = \psi_L^2 (x) =0$ due to the anticommutation relations.
Therefore a term like $\psi_R^{\dagger 2} (x) \psi_L^2 (x)$ is equal to
zero in the continuum. However such a term need not vanish on a lattice,
if we take the two factors of $\psi_R^\dagger$ (or $\psi_L$) as coming from
two neighboring sites separated by a distance $a$. In fact, this term
is allowed by momentum conservation on a lattice if $4k_F a =2\pi$, and it
leads to umklapp scattering.

\section{Correlation functions and dimensions of operators}

We will now use bosonization to compute the correlation functions of 
some fermionic operators in the interacting theory discussed above. The
power-law fall-offs of the correlation functions will tell us the 
dimensions of those operators.

The bosonic correlation function can be found from the commutation relations
in Eq. (\ref{comm3}), remembering that all normal-orderings have to be done 
with respect to the new vacuum $\vert {\tilde 0} \rangle$. (Henceforth we 
will omit the tilde denoting the new vacuum, but we will continue to use the 
tilde for the new $\phi$ fields). For instance,
\beq
\langle 0 \vert ~T \tphi (x,t) \tphi^{\dagger} (0,0) ~\vert 0 
\rangle ~=~\frac{1}{4\pi} ~\ln ~\Bigl[ ~\Bigl( ~\frac{2\pi}{L} ~
\Bigr)^2 ~\Bigl( ~x^2 ~-~ (vt -i \alpha ~{\rm sign} (t) )^2 ~\Bigr) ~\Bigr] ~.
\label{corr1}
\eeq

We can use the expressions in Eq. (\ref{bos4}) and identities like Eq. 
(\ref{exphi}) to obtain the correlation functions of various operators. For 
instance, 
\bea
\langle 0 \vert ~T e^{i2 {\sqrt \pi}\beta \tphi_R (x,t)} e^{-i2 {\sqrt \pi}
\beta \tphi_R (0,0)} ~\vert 0 \rangle ~& \sim & ~\Bigl( ~\frac{\alpha}{vt-x-i
\alpha ~{\rm sign} (t)} \Bigr)^{\beta^2} ~, \nonumber \\ 
\langle 0 \vert ~T e^{i2 {\sqrt \pi}\beta \tphi_L (x,t)} e^{-i2 {\sqrt \pi}
\beta \tphi_L (0,0)} ~\vert 0 \rangle ~& \sim & ~\Bigl( ~\frac{\alpha}{vt+x -i
\alpha ~{\rm sign} (t)} \Bigr)^{\beta^2} ~,
\eea
in the limit $L \rightarrow \infty$; we will assume henceforth that this limit 
is taken in the calculation of all correlation functions. Consider now the 
positive-chirality fermion field; according to Eq. (\ref{bos4}), 
\beq
\psi_R ~=~ \frac{1}{\sqrt {2\pi \alpha}} ~\eta_R ~e^{-i2 {\sqrt \pi} 
\phi_R} ~,
\eeq
where $\phi_R$ is given in Eq. (\ref{oldnew}) in terms of $\tphi_R$ and 
$\tphi_L$. Hence its time-ordered correlation function takes the form
\beq
\langle 0 \vert T \psi_R (x,t) \psi_R^\dagger (0,0) \vert 0 
\rangle \sim \frac{\alpha^{(1-K)^2 /2K}}{2\pi (vt-x-i\alpha {\rm sign} 
(t))^{(1+K)^2 /4K} (vt+x-i\alpha {\rm sign} (t))^{(1-K)^2 /4K}}.
\label{corr2}
\eeq
We see that the correlation function falls off at large space-time distances 
($i.e.$, large compared to $\alpha$) with the power $(1+K^2 )/(2K)$. This 
means that the scaling dimension of the operator $\psi_R$ or $\psi_R^\dagger$ 
is $(1+K^2 )/4K$; this agrees with the familiar value of $1/2$ for 
non-interacting fermions.

If we set $x=0$ in Eq. (\ref{corr2}), and Fourier transform over time, we 
find that the one-particle density of states (DOS) has a power-law form 
near zero frequency,
\beq
{\tilde n} (\omega ) ~\sim ~ \vert \omega \vert^\beta ~,
\eeq
where 
\beq
\beta ~=~ \frac{(1-K)^2}{2K} ~.
\eeq
The same result holds for the DOS of the
negative-chirality fermions. We therefore see that for any non-zero
interaction, either repulsive or attractive, the one-particle DOS vanishes 
as a power. (This result is not to be confused with the {\it bosonic} DOS 
which, from Eq. (\ref{ham5}), is a constant near zero energy since the energy
is linearly related to the momentum which has a constant density. That leads 
to a specific heat which is linear in the temperature at low temperatures). 
Alternatively, we may set $t=0$ in Eq. (\ref{corr2})
and Fourier transform over space, with a factor of $\exp (ik_F x)$ since
the momentum of the right-chirality fermions is measured with respect to
the Fermi momentum $k_F$. We then see that the momentum distribution function
is continuous at $k_F$ with a power-law form,
\beq
n (k) ~= ~ n (k_F ) ~+~~ {\rm constant} ~\cdot ~{\rm sign} (k-k_F ) ~\vert 
k - k_F \vert^\beta ~,
\eeq
as we have sketched in Fig. 1 (b). These expressions for ${\tilde n} 
(\omega )$ and $n(k)$ are characteristic features of a Luttinger liquid. 

Next let us compute the correlation function of an operator which
is bilinear in the fermion fields, namely, the mass operator 
\beq
M ~=~ \psi_R^\dagger \psi_L ~+~ \psi_L^\dagger \psi_R ~=~\frac{1}{2\pi 
\alpha} ~\Bigl[ ~\eta_R^\dagger \eta_L ~e^{i2 {\sqrt \pi} \phi} ~+~ 
\eta_L^\dagger \eta_R ~e^{-i2 {\sqrt \pi} \phi} ~\Bigr] ~.
\eeq
Using the same technique as before, we find that
\beq
\langle 0 \vert ~T M (x,t) M (0,0) ~\vert 0 \rangle ~\sim ~
\frac{\alpha^{2(K-1)}}{4\pi^2 ~((vt-i\alpha ~{\rm sign} (t))^2 ~-~ x^2)^K} ~. 
\label{corr3}
\eeq
This shows that the scaling dimension of the mass operator is $K$. For the
non-interacting case $K=1$, we see that the addition of such a term to the
Lagrangian density in Eq. (\ref{lag1}) makes the Dirac fermion massive; this 
is why we have called it the mass operator. (For convenience, we 
will sometimes omit the Klein factors when writing fermionic operators in
the bosonic language. We will of course need to restore those factors when 
calculating the correlation functions; clearly, correlation functions
will vanish if the numbers of $\eta_R$ and $\eta_R^\dagger$ (or $\eta_L$ and
$\eta_L^\dagger$) are not equal).

An important operator to consider is the density $\rho$. From Eqs. 
(\ref{lat1}), (\ref{corr1}) and (\ref{corr3}), we see that the 
density-density equal-time correlation function is asymptotically 
given by
\beq
\langle 0 \vert ~\rho (x,0) \rho (0,0) ~\vert 0 \rangle ~= ~-~ \frac{K}{2
\pi^2 x^2} ~+~~ {\rm const} ~\cdot ~\frac{\cos (4k_F x)}{x^{2K}} ~.
\label{corr4}
\eeq
We should emphasize that this is only the asymptotic expression;
the complete expression generally contains oscillatory terms like $\cos 
(4nk_F x) /x^{2n^2 K}$ for all positive integers $n$. However the form of the
denominator shows that these terms decay rapidly with $x$ as $n$ increases.

In general, we can consider an operator of the form
\beq
O_{m,n} ~=~ e^{i2 {\sqrt \pi} (m \phi + n \theta)} ~.
\label{omn}
\eeq
(Such an operator can arise from a product of several $\psi$'s and 
$\psi^\dagger$'s if we ignore the Klein factors; then Eq. (\ref{bos4}) 
implies that $m \pm n$ must take integer values). We then find the following 
result for the two-point correlation function
\bea
&& \langle 0 \vert ~T O_{m,n} (x,t) O_{m^\prime , 
n^\prime }^\dagger (0,0) ~\vert 0 \rangle \nonumber \\
&& \sim \delta_{mm^\prime } \delta_{nn^\prime } \frac{\alpha^{2(m^2 K + 
n^2 /K)}}{(vt-x-i\alpha {\rm sign} ( t))^{(m {\sqrt K} - n/ {\sqrt K})^2 } 
(vt+x-i\alpha {\rm sign} (t))^{(m {\sqrt K} + n/ {\sqrt K})^2} } ~, 
\label{corr5}
\eea
where we have taken the limit $L \rightarrow \infty$ as usual. (If $L$ had 
been kept finite, the correlation function in Eq.
(\ref{corr5}) would have been 
non-zero even if $m \ne m^\prime$ or $n \ne n^\prime$. This may seem 
surprising since the global phase invariance in Eqs. (\ref{glo1} - \ref{glo2})
should lead to the Kronecker $\delta$'s in Eq.
(\ref{corr5}) even for finite 
values of $L$. The resolution of this puzzle is that we need to include the
appropriate Klein factors in the definition in (\ref{omn}) to show that the
correlation function of a product of fermionic operators is zero if it is 
not phase invariant). We conclude that the scaling dimension of $O_{m,n}$ 
is given by
\beq
d_O ~=~ m^2 K ~+~ \frac{n^2}{K} ~.
\label{dim}
\eeq

The appearance of the cut-off $\alpha$ in the expressions for the various
correlation functions may seem bothersome. This may be eliminated by 
redefining the operators $O_{m,n}$ in Eq. (\ref{omn}) by multiplying them 
with appropriate 
$K$-dependent powers of $\alpha$; then the two-point correlation function 
has a well-defined limit as $\alpha \rightarrow 0$. The important point 
to note is that all the 
correlation functions fall off as power-laws asymptotically, and that the 
exponents give the scaling dimensions of those operators. The significance of
the scaling dimension will be discussed in the next section.

For certain applications of bosonization, it is useful to know the forms 
of the correlation functions in imaginary time. From the various expressions 
above, it is clear that if $x$ is held fixed at some non-zero value, then
the poles in the complex $t$ plane are either in the first or in the third
quadrant. We may therefore rotate $t$ by $\pi /2$ without crossing any 
poles. After doing this, we write $t=i\tau$ where $\tau$ is a real
variable. Eq. (\ref{corr5}) then takes the form
\beq
\langle ~O_{m,n} (x,t) O_{m^\prime , n^\prime }^\dagger (0,0) \rangle ~
\sim ~\delta_{mm^\prime } \delta_{nn^\prime } ~e^{i4mn \zeta} ~\Bigl( ~
\frac{\alpha^2}{x^2 + v^2 \tau^2} ~\Bigl)^{m^2 K + n^2 /K} ~,
\label{corr6}
\eeq
where $\zeta = \tan^{-1} (v\tau /x)$, and we have dropped the 
$\alpha$-dependent terms in the denominator since there are no longer any
poles for non-zero values of $x$.

\section{Renormalization group analysis of perturbed models}

We will now study the effects of some perturbations on the low-energy
properties of Luttinger liquids. A standard way to do this is to use the 
renormalization group (RG) idea. Suppose that we are given an action at a 
microscopic length scale which may be a lattice spacing $a$; the action 
contains some small perturbations proportional to certain dimensionless 
parameters 
$\lambda_i$, such that for $\lambda_i =0$, we have a gapless system with
an infinite correlation length $\xi$, $i.e.$, all correlations fall off as 
power laws. Then the RG procedure typically consists of the following steps. 

\noindent (i) First, a small range of high momentum modes of the various
fields are integrated out. Specifically, we will assume that the momenta
lie in the range $[-\Lambda ,\Lambda ]$ while the frequencies go all the way
from $-\infty$ to $\infty$. Then we will 
integrate out the modes with momenta lying in 
the two intervals $[-\Lambda , - \Lambda /s]$ and $[\Lambda /s , \Lambda ]$ 
and with all frequencies from $-\infty$ to $\infty$. Here $s = e^{dl}$ 
where $dl$ is a small positive number. The asymmetry between
the momentum and frequency integrations is necessary to ensure that the
action remains local in time at all stages. (Note that we are using
sharp momentum cut-offs in this section, whereas we used a smooth momentum
cut-off with the parameter $\alpha$ in the previous sections).

\noindent (ii) Secondly, the space-time coordinates, the fields and the 
various parameters are rescaled by appropriate powers of $s$ so that the 
new action looks exactly like the old action. This new action is 
effectively at a larger length scale equal to $ae^{dl}$. Clearly, the 
changes in the parameters $\lambda_i$ must be proportional to the small 
number $dl$. Since we are going to repeat the process of integrating out
high momentum modes, we introduce the idea of an effective length 
scale $a(l) = ae^l$; we also define length scale dependent parameters 
$\lambda_i (l)$, where $\lambda_i (0)$ denote the values of
$\lambda_i$ in the original action. We then define the $\beta$-functions 
\beq
\beta (\lambda_i ) ~=~ \frac{d \lambda_i (l)}{dl} ~.
\eeq
These are functions of all the $\lambda_i (l)$'s so that we get a set of
coupled non-linear equations in general. In principle, the $\beta$-functions
are given by infinite power series in the $\lambda_i$, but in practice, we
can only obtain the first few terms depending on the number of loops of
the various Feynman diagrams that we can compute. The RG analysis is 
therefore usually limited to small values of $\lambda_i$.

\noindent (iii) Finally, we integrate the RG equations, $i.e.$,
the differential equations described by the $\beta$-functions, in 
order to obtain the functions $\lambda_i (l)$. For simplicity, let us 
consider the case of a single perturbation with a coefficient $\lambda$. 
Then one of three things can happen as $l$ increases from $0$. Either 
$\lambda (l)$ goes to zero in which case we recover the unperturbed theory 
at long distances; or $\lambda$ does not change with $l$; or $\lambda (l)$ 
grows with $l$ till its value becomes of order $1$. In the last case, the RG 
equation cannot be trusted beyond that length scale since the 
$\beta$-functions are generally only known up to some low order in the 
$\lambda$'s. All that we can say is that beyond the length scale $ae^l$ where 
$\lambda (l)$ becomes of order $1$, a completely new kind of action is likely 
to be required to describe the system. Large perturbations of a gapless 
system often (but not always) correspond to a gapped system whose correlation 
length $\xi$ (which governs the exponential decay of various correlation 
functions) is of the same order as that length scale $ae^l$. Thus, although 
the blowing up of a parameter $\lambda$ at some scale does not tell us 
what the new action must be beyond that scale, it can give us an idea
of the correlation length of that new theory. This is the main use that is
made of RG equations. To complete the picture and find the new theory
beyond the scale $\xi$, one usually has to do some other kind of analysis.

Let us now examine in a little more detail the various kinds of RG equations 
which can arise at low orders. Suppose that to first
order, the RG equation for a single perturbing term is given by
\beq
\frac{d \lambda}{dl} ~=~ b_1 \lambda ~,
\label{bet1}
\eeq
where $b_1$ is some constant.
If $b_1 < 0$, any non-zero value of $\lambda$ at $l=0$ flows to $0$ as 
$l$ increases. Such a perturbation is called irrelevant. If $b_1 >0$,
it is called a relevant perturbation. A small perturbation then grows 
exponentially with $l$ and reaches a number of order $1$ at a distance 
scale given by $e^{b_1 l} \sim 1/ \lambda (0)$. In many models, this gives an 
estimate of the correlation length $\xi$ and of the energy gap $\Delta E$ 
of the system, namely,
\bea
\xi ~&=&~ ae^l ~=~ \frac{a}{\lambda (0)^{1/b_1}} ~, \nonumber \\
{\rm so} \quad \Delta E ~&=&~ \frac{v}{\xi} ~=~ \frac{v\lambda 
(0)^{1/b_1}}{a} ~.
\eea
Finally, if $b_1 =0$, the perturbation is called marginal. One then has 
to go to second order in $\lambda$. If the RG equation takes the form
\beq
\frac{d \lambda}{dl} ~=~ b_2 \lambda^2 ~,
\label{bet2}
\eeq
then a small perturbation of one particular sign flows to zero and is called 
marginally irrelevant, while a small perturbation of the opposite sign grows 
and is called marginally relevant. For instance, suppose that $b_2 > 0$.
Then the above equation gives
\beq
\lambda (l) ~=~ \frac{\lambda (0)}{1 ~-~ b_2 \lambda (0) l} ~.
\eeq
If we start with a negative value of $\lambda (0)$, $\lambda (l)$ flows to
$0$. For large $l$, $\lambda (l)$ goes to zero logarithmically in the 
distance scale
as $-1/(b_2 l)$ independently of the starting value. (It turns out that
this produces logarithmic corrections to the power-law fall-offs of the 
correlation functions at large distances and the various excitation energies 
\cite{aff2}). On the other hand, if we start with a small positive value of 
$\lambda (0)$, then $\lambda (l)$ grows and becomes of order $1$ at a 
distance scale which we identify with a correlation length
\beq
\xi = a e^{1/(b_2 \lambda (0))} ~.
\eeq
The corresponding energy gap $\Delta E = v/\xi$ is extremely small for small
values of $\lambda (0)$; it may be very hard to distinguish this kind of
a system from a gapless system by numerical studies. This is in 
sharp contrast to the situation with a relevant perturbation where the gap 
scales as a power of $\lambda (0)$.

There is a simple relation between the scaling dimension of an operator 
$O$ (assumed to be hermitian for simplicity), and the first-order 
coefficient $b_1$ in its $\beta$-function. We recall that the scaling 
dimension $d_O$ is defined as half of the exponent appearing
in the two-point correlation function at large distances, namely,
\beq
\langle ~O (x,0) O (0,0) ~\rangle ~=~ \vert x \vert^{- 2d_{\cal O}} ~.
\label{corr7}
\eeq
It is convenient to define the normalization of $O$ in such a way that the 
right hand side of Eq. (\ref{corr7}) has a prefactor equal to $1$. 
Consequently, $O$ has the engineering dimensions of $a^{-d_{\cal O}}$. Let us 
now add a perturbation to the Hamiltonian (or to the Lagrangian with a 
negative sign) of the form 
\beq
\delta H ~=~ \lambda a^{d_{\cal O} -2}v ~\int ~dx ~O ~,
\eeq
where the factors of $a$ and $v$ (the velocity of the unperturbed
Luttinger liquid) are put in to make $\lambda$ dimensionless; note that
$v/a$ has the dimensions of energy. Then the first-order RG equation for 
$\lambda$ must take the form given in Eq. (\ref{bet1}) with
\beq
b_1 ~=~ 2 ~-~ d_O ~.
\eeq
This important statement will be proved below for the class of operators
$O_{m,n}$ introduced in Eq. (\ref{omn}). If $d_O = 2$, the perturbation is
marginal and we have to proceed to Eq. (\ref{bet2}). It turns out that 
$b_2$ can be obtained from a {\it three-point} correlation function, but 
we will not pursue that here \cite{aff2}.

It will not come as a surprise that the RG equations for 
interacting quantum systems in one dimension can often be derived in 
two different ways, namely, using the fermionic theory or the bosonic one. 
Although both the derivations are limited in practice to small values of the 
perturbations $\lambda_i$, we will see that the bosonic derivation is 
superior because it can handle the interactions in Eq. (\ref{int2}) 
exactly. In the fermionic derivation, we have to assume that not only the 
$\lambda$'s but also the interaction parameters $g_2$ and $g_4$ are small. 
We will now discuss some simple examples of $\beta$-function calculations to
first order in the two kinds of theories.

As a particularly simple exercise, consider a non-interacting massive Dirac 
theory, where the mass term is to be treated as a perturbation. We define 
the Fourier components of $\psi_\nu$ as
\bea
\psi_R (x,t) ~&=&~ \int_{-\Lambda}^{\Lambda} ~\frac{dk}{2\pi} ~
\int_{\infty}^{\infty} ~\frac{d\omega}{2\pi} ~e^{i(kx-\omega t)} ~\psi_R (k, 
\omega ) ~, \nonumber \\
\psi_L (x,t) ~&=&~ \int_{-\Lambda}^{\Lambda} ~\frac{dk}{2\pi} ~
\int_{\infty}^{\infty} ~\frac{d\omega}{2\pi} ~e^{-i(kx+\omega t)} ~\psi_L (k, 
\omega ) ~. 
\eea
Then the action takes the form
\bea
S [\psi_\nu , \psi_\nu^\dagger ] = \int_{-\Lambda}^{\Lambda} 
\frac{dk}{2\pi} \int_{\infty}^{\infty} \frac{d\omega}{2\pi} \Bigl[ && 
\psi_R^\dagger (k,\omega ) (\omega - vk) \psi_R (k, \omega ) + 
\psi_L^\dagger (k,\omega ) (\omega - vk) \psi_L (k, \omega) \nonumber \\
&& - \mu \Bigl( \psi_R^\dagger (-k,\omega ) \psi_L (k,\omega ) +
\psi_L^\dagger (-k,\omega ) \psi_R (k,\omega ) \Bigr) ~\Bigr] ~.
\label{act1}
\eea
Since $\mu$ has the dimensions of energy, the dimensionless parameter 
must be taken to be 
\beq
\lambda ~=~ \frac{a\mu}{v} ~.
\eeq
(The value of $a$ is completely arbitrary here and it will not appear in 
any physical quantity as we will see). We consider the partition function 
in the functional integral representation,
\beq
Z ~=~ \int {\cal D} \psi_\nu {\cal D} \psi_\nu^\dagger ~e^{iS} ~.
\eeq
We integrate out the modes in the momentum and frequency ranges
specified in step (i) of the RG procedure outlined above. Since Eq.
(\ref{act1})
describes non-interacting fermions, the mode integration produces an action
which looks exactly the same, except that the momentum integrations go
from $-\Lambda /s$ to $\Lambda /s$. To restore this to the original range
of $[-\Lambda ,\Lambda ]$, we define the new (primed) quantities
\bea
k^\prime ~&=&~ s k ~, \nonumber \\
\omega^\prime ~&=&~ s \omega ~, \nonumber \\
\psi_\nu^\prime (k^\prime ,\omega^\prime ) ~&=&~ s^{-3/2} ~\psi_\nu (k, 
\omega ) ~, \nonumber \\
\lambda^\prime ~&=&~ s \lambda ~. 
\label{ren1}
\eea
The resultant action in terms of the new variables and fields looks exactly
the same as the original action in terms of the old variables. Note that
we had to rescale the mass parameter also in order to achieve this. Since
$s=e^{dl}$, we obtain the RG equation
\beq
\frac{d\lambda (l)}{dl} ~=~ \lambda (l) ~.
\eeq
Clearly, this describes a relevant perturbation, and $\lambda (l)$ grows to
$1$ at a length scale
\beq
\xi ~=~ ae^l ~=~ \frac{v}{\mu} ~.
\eeq
The energy gap is $\Delta E = v/\xi = \mu$ as expected.

Now let us add density-density interactions as in Eq. (\ref{int2}) to the 
above massive theory. The question is the following: do $\xi$ and $\Delta E$ 
scale in the same way with $\mu$ as they do in the non-interacting theory? 
Clearly, it is not easy to answer this in the fermionic language since the 
interactions themselves are not easy to handle in that language, and the mass 
perturbation is an additional complication. But bosonization comes to our 
rescue here since the bosonic theory remains quadratic even after including 
the four-fermi interactions; hence the mass perturbation is the only thing 
that needs to be studied. 

Let us consider a more general perturbing operator of the form
\beq
O ~=~ O_{m,0} ~+~ O_{m,0}^\dagger ~,
\label{om}
\eeq
where $O_{m,n}$ is defined in Eq. (\ref{omn}); the reason for setting $n=0$ 
will be explained later. From Eq. (\ref{dim}), the scaling dimension of $O$ is 
given by $d_O = m^2 K$; note that this contains the effects of the 
four-fermion interaction in a non-trivial way through the parameter $K$. In 
the bosonic language, the perturbed action has the sine-Gordon form,
\beq
S [\tphi ] ~=~ \int ~dx dt ~\Bigl[ ~\frac{1}{2v} ~(\partial_t 
\tphi )^2 ~-~ \frac{v}{2} ~(\partial_x \tphi )^2 ~-~ 
\frac{v \lambda}{a^2} ~\cos (2 m {\sqrt {\pi K}} \tphi ) ~\Bigr] ~,
\label{act2}
\eeq
where we have changed variables from $\phi$ to $\tphi$ using Eq.
(\ref{oldnew}). We now have to apply the RG procedure to this action. We 
introduce the Fourier components of $\tphi$ as
\beq
\tphi (x,t) ~=~ \int_{-\Lambda}^{\Lambda} ~\frac{dk}{2\pi} ~
\int_{\infty}^{\infty} ~\frac{d\omega}{2\pi} ~e^{i(kx-\omega t)} ~
\tphi (k, \omega ) ~.
\eeq
(In principle, the momentum cut-offs for fermion and boson fields need not
be equal, but we will use the same symbol $\Lambda$ for convenience). Next 
we consider the partition function
\beq
Z ~=~ \int ~{\cal D} \tphi ~e^{iS} ~,
\eeq
and expand $e^{iS}$ in powers of $\lambda$ to obtain an infinite series. Let 
us write the field $\tphi$ as the sum
\beq
\tphi ~=~ \tphi_< ~+~ \tphi_> ~,
\eeq
where both $\tphi_<$ and $\tphi_>$ contain all frequencies, but $\tphi_<$
only contains momenta lying in the range $[-\Lambda /s, \Lambda /s]$, 
whereas $\tphi_>$ only contains momenta lying in the ranges 
$[-\Lambda , -\Lambda /s]$ and $[\Lambda /s, \Lambda]$.
Following step (i) of the RG procedure, we have to perform the functional 
integration over $\tphi_>$, and then re-exponentiate the infinite 
series to obtain the new action in terms of $\tphi_<$. We will do this
calculation only to first order in $\lambda$. This is not difficult since
$e^{\pm i2m{\sqrt {\pi K}} \tphi}$ can be written as the product of 
exponentials of $\tphi_<$ and $\tphi_>$, while the
quadratic part of the action decouples as $S_0 [\tphi ] =S_0 
[\tphi_< ] + S_0 [\tphi_> ]$. Let us denote the expectation 
value of a functional $F [\tphi_> ]$ as
\beq
\langle ~F [ \tphi_> ] ~\rangle ~=~ \int ~{\cal D} \tphi_> ~
e^{iS_0 [\tphi_> ]} ~F [\tphi_> ] ~.
\eeq
Now we have to compute
\beq
\langle ~e^{\pm i2m {\sqrt {\pi K}} \tphi_> (x,t) } ~\rangle ~.
\label{expec1}
\eeq
By translation invariance, the value of this is independent of the 
coordinates $(x,t)$, so we can evaluate it at the point $(0,0)$. We then 
use the fact that $\langle \tphi_>^n (0,0) \rangle =0$ if $n$ is odd, while 
\beq
\langle ~\tphi_>^n (0,0) ~\rangle ~=~ (n-1)(n-3) \cdot \cdot \cdot 1 ~
\langle ~\tphi_>^2 (0,0) ~\rangle^{n/2} 
\eeq
if $n$ is even. Thus the expectation value in (\ref{expec1}) is given by
\bea
\langle ~e^{\pm i2m {\sqrt {\pi K}} \tphi_> (0,0)} ~\rangle ~&=&~ 
\sum_{n=0}^\infty ~\frac{1}{n!} ~(\pm i 2m{\sqrt \pi K})^n ~\langle ~
\tphi_>^n (0,0) ~\rangle ~\nonumber \\
&=&~ e^{-2 m^2 \pi K \langle \tphi_>^2 (0,0) \rangle } ~.
\label{expec2}
\eea
Now we use the fact that 
\beq
\langle ~\tphi_>^2 (0,0) ~\rangle ~=~ 2 \int_{\Lambda /s}^\Lambda ~
\frac{dk}{2\pi} ~\int_{-\infty}^\infty ~\frac{d\omega}{2\pi} ~\frac{i}{
\omega^2 /v ~-~ vk^2 + i \epsilon} ~=~ \frac{\ln s}{2\pi} 
\eeq
to show that the left hand side of 
Eq. (\ref{expec2}) is equal to $s^{-m^2 K}$.
Putting everything together, we find the new action to be
\beq
S [\tphi_< ] ~=~ \int ~dx dt ~\Bigl[ ~\frac{1}{2v} ~(\partial_t 
\tphi_< )^2 ~-~ \frac{v}{2} ~(\partial_x \tphi_< )^2 ~-~ 
\frac{v \lambda s^{- m^2 K}}{a^2} ~\cos (2m {\sqrt {\pi K}} \tphi_< ) ~
\Bigr] ~,
\eeq
where the momentum integrals only go from $-\Lambda /s$ to $\Lambda /s$.
To restore the range of the momentum to $[-\Lambda ,\Lambda ]$ and to recover 
the form of the action in Eq. (\ref{act2}), we have to define
\bea
k^\prime ~&=&~ s k ~, \quad {\rm and} \quad x^\prime ~=~ s^{-1} ~x ~, 
\nonumber \\
\omega^\prime ~&=&~ s \omega ~, \quad {\rm and} \quad t^\prime ~=~ s^{-1} ~
t ~, \nonumber \\
\tphi^\prime (k^\prime ,\omega^\prime ) ~&=&~ \tphi_< (k, \omega ) ~, 
\nonumber \\
\lambda^\prime ~&=&~ s^{2- m^2 K}~ \lambda ~, 
\label{ren2}
\eea
and write the action in terms of primed variables. Since $s=e^{dl}$, we see 
that $d\lambda = \lambda^\prime - \lambda$ satisfies the RG equation 
\beq
\frac{d \lambda}{dl} ~=~ (2 ~-~ m^2 K) \lambda ~.
\eeq
This proves the relation between the first-order $\beta$-function
coefficient $b_1$ and the scaling dimension $d_O$. 
Note that the $\beta$-functions of the parameters $v$ and $K$ remain zero 
up to this order in the perturbation. However they do get a 
contribution to second order in $\lambda$ as shown in Ref. 3.

The mass perturbation corresponds to the special case of Eq. (\ref{om}) with
$m=1$. We now see that it is marginal for $K=2$ and is relevant if $K <2$. In 
the latter case, $\lambda (l)$ grows till we reach a length scale $\xi = a / 
\lambda (0)^{1/(2-K)}$ where the length scale of the coefficient of the
cosine term in the Lagrangian becomes of the same order as $a$; that is the
appropriate point to stop the RG flow of $\lambda$. The expression for $\xi$ 
implies that the energy gap of the system is given by 
\beq
\Delta E ~=~ \frac{v}{a} ~\lambda (0)^{1/(2-K)} ~. 
\label{de}
\eeq
Thus the effect of the renormalization is to produce a sine-Gordon theory 
with the Lagrangian density
\beq
{\cal L} ~=~ \frac{1}{2v} ~(\partial_t \tphi )^2 ~-~ \frac{v}{2} ~
(\partial_x \tphi )^2 ~-~~ {\rm const} ~\cdot ~ 
\frac{(\Delta E)^2}{v} ~\cos (2 {\sqrt {\pi K}} \tphi ) ~,
\label{lag4}
\eeq
where $x$ and $t$ in this expression denote the {\it original} coordinates,
and it is understood that this Lagrangian is {\it not} to be renormalized any
further. This theory is exactly solvable and its spectrum is known in detail 
\cite{raja}. It has both bosonic and fermionic (soliton) excitations, and both 
of them have energy gaps of the order of $\Delta E$ given in Eq. (\ref{de}).

Finally, let us briefly consider some other relevant and
marginal perturbations that can appear in a system which, at 
the microscopic model, involves fermions on a lattice. If the model has the 
global phase invariance discussed in Eqs. (\ref{glo1} - \ref{glo2}), then 
the operators $O_{m,n}$ appearing in the bosonized theory must necessarily 
have $n=0$. The scaling dimension is then $d_O = m^2 K$. Since $m \ge 1$, 
there is only a finite number of relevant operators possible depending on the 
value of $K$. For $K > 2$, there are no relevant operators at all. For $1/2 < 
K < 2$, the mass operator is the only relevant term, and so on. 

Turning to the possible marginal operators, we see that the umklapp operator 
$O_{2,0} =\psi_R^{\dagger 2} \psi_L^2$ is marginal for $K =1/2$. 
This is a particularly important case to consider because a Luttinger 
liquid at $K=1/2$ is known to have a global $SU(2)$ symmetry; it therefore 
describes a large number of gapless systems involving spins. From conformal 
field theory, the value of $b_2$ in the RG equation Eq.
(\ref{bet2}) for the
umklapp operator $O$ is exactly known to be $4\pi /{\sqrt 3}$ for the
normalization given in Eq. (\ref{corr7}). The coefficient of $O$ in the 
Hamiltonian, namely $\lambda$, depends on the microscopic parameters of 
the model. In general, a system will have a non-zero value of $\lambda$. As 
discussed above, for one sign of $\lambda$, the system remains gapless but 
with logarithmic corrections to various physical quantities; for instance, 
a $1/\ln T$ term appears in the magnetic susceptibility of a spin system at 
low temperatures. For the other sign of $\lambda$, the system spontaneously 
dimerizes producing a finite correlation length and an energy gap; this 
leads to an exponentially vanishing susceptibility at low temperatures.

\vskip .5 true cm
\section{Applications of Bosonization}

We will now study various applications of the
method of bosonization. The method, as you have learned, can only be
applied in one dimension, so we restrict ourselves to one-dimensional
models. As you have also seen, the main
advantage of the method of bosonization is that many interacting fermion
theories can often be recast (within some approximations) 
as non-interacting boson theories. This
enables the explicit calculation of correlation functions. This is
an advantage, even in Bethe ansatz solvable one-dimensional
models, because it is often not possible to compute correlation
functions using the Bethe ansatz.

We will concentrate on the applications of the bosonization technique
in the following problems - (i) the quantum antiferromagnetic spin 1/2
chain, (ii) the Hubbard model in one dimension, (iii) transport in
clean quantum wires and (iv) transport through isolated impurities. 
Since the physics of each of these applications is a huge subject by
itself, here we will only concentrate on explaining the model and the
quantities that we can obtain through the use of bosonization, rather
than go into details of its phenomenology.

\section{Quantum antiferromagnetic spin 1/2 chain}
\vskip 0.3cm 

\noindent{\bf The model}

The first problem that we shall study is the model of a spin 1/2
antiferromagnetic chain. We are picking this model, since you have
already learned a lot about the model from the course on quantum spin
chains and spin ladders. Here, we will restrict ourselves to just the study 
of the spin 1/2 anisotropic Heisenberg model with the Hamiltonian given by
\beq
H = \sumi ~[~\frac{J}{2} ~(S_i^+ S_{i+1}^- + S_i^- S_{i+1}^+) ~+~ J_z ~S_i^z 
S_{i+1}^z ~]~,
\label{model}
\eeq
where the interactions are only between nearest neighbor spins, and
$J > 0$. $S_i^+= 
S_i^x+iS_i^y$ and $S_i^-=S_i^x-iS_i^y$ are the spin raising and lowering
operators. Although this model can be exactly solved using the Bethe
ansatz and one has the explicit result that the model is gapless for 
$-J \le J_z \le J$ and gapped for $J_z>J$, (there is a phase transition 
exactly at the isotropic point $J_z=J$), it is not easy to compute explicit
correlation functions in that approach. Hence, it is more profitable
to use field theory methods. 

\vskip 0.3cm 
\noindent {\bf Symmetries of the model}

Note that this spin model has a global $U(1)$ invariance, which is rotations 
about the $S^z$ axis. Precisely when $J_z=J$, the $U(1)$ invariance is 
enhanced to an $SU(2)$ invariance, because at this point the model can simply 
be written as $H=J\sum_i{\bf S_i}\cdot{\bf S_{i+1}}$. The model also has
discrete symmetries under $S^x \rightarrow -S^x$, $S^y \rightarrow S^y$,
and under $S^z \rightarrow -S^z$. 
Note also that one can change the sign of the $XY$ part of the
Hamiltonian by making a rotation by $\pi$ about the $S^z$ axis on
alternate sites, without affecting the $J_z$ term, although this is
not an extra symmetry of the model. 

\vskip 0.3cm 
\noindent {\bf Aside on non-linear sigma models}

Even using field theory methods, there are two distinct approaches to
the problem. In the large-$S$ limit, there exists a semiclassical
field theory approach to this model, which leads to an $O(3)$
non-linear sigma model ($NL\sigma M$), with integer and half-integer
spins being distinguished by the absence or presence of a Hopf term in
the action. In this approach, it is easy to see that integer spin
 models have a gap in the spectrum. However, it is less easy to study
the effect of the Hopf term and show that 1/2-integer spin models are
gapless. In fact, in this case, it was the spin model which gave
information about field theories with the Hopf term! 

\vskip 0.3cm 
\noindent {\bf Jordan-Wigner transformation}

For spin 1/2 models, it is possible to fermionize
and then bosonize the spin model and study its spectrum. That is the
approach we will follow in the rest of this lecture.
First, we will try to convince you that it is possible to 
rewrite the spin model in terms of spinless fermions.
The spin 1/2 model has two states possible at every site - spin
$\uparrow$ or spin $\downarrow$. Hence,it can be mapped to another two
state model which we can construct in terms of fermions. 
We shall assume that an $\uparrow$ spin or $\downarrow$ can be denoted by the
presence or absence of a fermion at that site. Since no more than a
single spinless fermion can sit at a site, the degrees of freedom in
both the models are the same. 
This mapping is implemented by introducing a fermion annihilation 
operator $\psi_i$ at each site and writing the spin at the site as
\bea
S_i^z &=& \psi_i^{\dagger} \psi_i -1/2 = n_i-1/2 \nonumber\\
S_i^- &=& (-1)^i ~\psi_i e^{i\pi \sum_j n_j} ~,
\eea 
where the sum runs from one boundary of the chain up to the
$(i-1)^{\rm th}$ site and $S_i^+$ is the hermitian conjugate of $S_i^-$. 
So an $\uparrow$-spin is denoted by $n_i=1$ and
the $\downarrow$-spin by $n_i=0$ at the site $i$. One might have 
naively guessed that the spin-lowering operator should be expressed 
by $\psi_i$ which denotes annihilation of a fermion (with the spin
raising operator being given by the hermitian conjugate). One can
explicitly check that this gives the correct commutation relations of
the spin operators at a site because $[S_i^+,S_i^-] = 2S_i^z$ just
reproduces the correct anticommutation relations for the fermions 
$\{\psi_i,\psi_i^{\dagger}\} = 1$. The extra string factor has 
to be added in order to correct for different site statistics - the
fermions at different sites anticommute, whereas the spin operators
commute. In fact, it is instructive to check explicitly that the 
string operator changes the commutation relation on different sites. \\ 
(H.W. Exercise 1. Check the above explicitly). 

\vskip 0.3cm
\noindent{\bf The Hamiltonian}

Now, we rewrite the spin model in terms of the fermions. We find that 
\bea
H &=& -~ \frac{J}{2} ~\sum_i ~[\psi_i^{\dagger}e^{i\pi n_i} \psi_{i+1} +
h.c.] ~+~ J_z ~\sum_i ~[(n_i-1/2)(n_{i+1}-1/2)] \nonumber \\
&=& -~ \frac{J}{2} ~\sum_i ~[\psi_i^{\dagger}\psi_{i+1} +h.c.] ~+~
J_z ~\sum_i ~[(n_i-1/2)(n_{i+1}-1/2)] ~.
\label{Hamiltonian} 
\eea
The point to notice is that the string operator has cancelled out in the
nearest neighbor interaction, except for a phase term, which also can be 
explicitly shown to be just $1$ because $e^{i\pi n_i}$ precedes a creation 
operator $\psi_i^{\dagger}$ which can only act if $n_i = 0$. The spin-flip 
terms are like the hopping terms in the fermion Hamiltonian and give rise to 
motion of fermions whereas the $S^z$-$S^z$ interaction term leads to a four
fermion interaction between fermions on adjacent sites (the analog
of the on-site Hubbard interaction for spinless fermions). So for non-zero
$J_z$, the fermionic model is non-trivial. There exists a competition
between the hopping term or kinetic energy term, which gains in energy
when the electrons are free to hop from site to site, and the potential
energy which costs $J_z$ if there are electrons present on adjacent sites. 
So naively, for large $J_z$, one expects the potential energy to win and
electrons to be localized on non-adjacent sites, and for small $J_z$,
one expects the kinetic energy to win and to have delocalized fermions. 
Let us see whether this expectation is true and how it comes about.

\vskip 0.3cm
\noindent {\bf Set $J_z=0$}

To make the problem simpler, we first consider the case where 
$J_z=0$ or where there are no interaction terms. Then this is just 
the model of free spinless fermions. By Fourier
transforming the fermions, - $\psi_j = \sum_k \psi_k
e^{ikja}/\sqrt{N}$, ( a is the lattice
spacing) where the $k$ sum is over momentum values in the first
Brillouin zone, - we find that the Hamiltonian is given by 
\beq
H ~=~ -~ J\sum_k ~\cos (ka) ~\psi_k^{\dagger}\psi_k ~.
\label{Hamk}
\eeq 
(H.W. Exercise 2. Obtain the above Hamiltonian explicitly). \\
The discrete symmetry of the model under $S^-_i 
\leftrightarrow -S_i^+$ and $S_i^z\rightarrow -S_i^z$ implies a particle-hole
symmetry $\psi_i \leftrightarrow \psi_i^{\dagger}$ 
in the fermion language. Thus, the ground state has to have total spin
$M\equiv\sum_i S_i^z = 0$ or equivalently in the fermionic language, the
ground state is precisely half-filled. This symmetry can be broken by
the addition of a magnetic field term that couples linearly to $S^z$. 
In the fermionic language, this is equivalent to adding a chemical
potential term (which couples to $n_i$ which is the $S^z$ term) in which
case, the ground state no longer has $M=0$ and the fermion model is no
longer half-filled. Thus, for $M=0$, the band is precisely half-filled 
and the Fermi surface ($E=0$) occurs exactly at $ka = \pm \pi/2 \equiv k_F a$
(because the density of states is symmetric about $E=0$. )
Low energy excitations are particle-hole excitations about the Fermi 
surface, which can occur either at a single Fermi point $i.e.$, $k\sim 0$ 
modes, or across Fermi points, which are the $k\sim 2k_F=\pi/a$ modes. 

\vskip 0.3cm
\noindent{\bf Effective field theory}

The next step is to write down an effective field theory for the low energy 
modes. Now comes the approximation. Let us make the assumption that it is
only the modes near the Fermi surface (or here, the Fermi points),
which are relevant at low energies. Hence, we are only interested in
$ka$ values near $ka=\pm\pi/2$ and we may approximate the dispersion
relation around the Fermi points to be linear - $i.e.$, $\cos (ka) =
\cos (\pm k_F a + k'a) = \cos (\pm \pi/2 + k'a) = \mp \sin (k'a) = \mp
(k'a)$. We introduce the labels left and
right to denote fermion modes near $ka = -\pi/2$ and $ka=\pi/2$
respectively and henceforth drop the primes on the momenta and assume
that they are always measured from the Fermi points; as before, we take $k$
as increasing towards the right near the right Fermi point, and increasing
towards the left near the left Fermi point. 
If we want to solve the problem without any approximations, we have
to allow for excitations about the Fermi points with arbitrary
$k$. The approximation that is made is that we only allow small values of $k$
compared to $k_F$. This is why the excitations around the left and
right Fermi points can be thought of as independent excitations. 
In this approximation, the Hamiltonian breaks up into 
\beq
H = Ja\sum_k ~k ~(\psi_{R,k}^{\dagger}\psi_{R,k} + 
\psi_{L,k}^{\dagger}\psi_{L,k}) ~.
\eeq
(Note that we have incorporated the change in sign mentioned below
Eq. (\ref{Hamk})). The fermions around the Fermi points are 
Dirac fermions since we have linearized the dispersion. These fields
do not contain any high momentum modes. In real space, 
the original non-relativistic fermion field, which has high energy modes 
(rapidly oscillating factors), 
has been split up as exponential prefactors times smoothly varying fields - 
\bea
\psi_j &\sim& e^{-ik_F ja} \int_{-k_Fa-\Lambda}^{-k_Fa+\Lambda} ~{d(ka)\over 
2\pi} e^{ikja} \psi_k ~ +e^{ik_F ja} \int_{k_Fa-\Lambda}^{k_Fa+\Lambda} ~
{d(ka)\over 2\pi} e^{ikja}\psi_k \nonumber \\
&\equiv& e^{-ik_F ja}\psi_{Lj} +e^{ik_F ja}\psi_{Rj} ~.
\label{lowenft}
\eea
We assume that the $\Lambda <<k_Fa$ and that it
is sufficient to keep just these modes, if we are interested in
physics at length scales much greater than $1/\Lambda$, which is
of course much greater than the lattice spacing. (The real physical
cutoff is the lattice length or in momentum space, the Fermi momentum.
As a low energy approximation, we are introducing the larger length
cutoff $1/\Lambda$ or the smaller momentum cutoff $\Lambda$). For both 
$R$ and $L$ fermions, states with $k>0$ are empty and correspond to electron
operators ($c_k$), while states with 
$k<0$ are filled and correspond to hole operators ($d_k^{\dagger}$). 
(See Fig. 2). In terms of these operators, the Hamiltonian can be rewritten as 
\beq
H=Ja\sum_{k>0} ~k ~(c_{L,k}^{\dagger}c_{L,k} +d_{L,k}^{\dagger}d_{L,k}
+c_{R,k}^{\dagger}c_{R,k} +d_{R,k}^{\dagger}d_{R,k}) ~.
\label{lowenop}
\eeq
We now introduce continuum fermion fields made up of particle
(electron) and anti-particle (hole) operators at the left and right
Fermi points as 
\bea
\psi_R(x,t) &=& {1\over\sqrt{Na}} \sum_{k>0} ~
[c_{R,k}e^{-ik(vt-x)} +d_{R,k}^{\dagger} e^{ik(vt-x)}]\nonumber \\
\psi_L(x,t) &=& {1\over\sqrt{Na}} \sum_{k>0} ~
[c_{L,k}e^{-ik(vt+x)} +d_{L,k}^{\dagger} e^{ik(vt+x)}] ~,
\label{conti} 
\eea
where $v=Ja$ is defined to be the velocity. Note that the factor of 
$1/\sqrt{a}$ is needed to relate continuum fermions to 
lattice fermions. (The factor of
$\sqrt{a}$ is needed to get the dimensions right. The lattice fermions
satisfy $\{\psi_i, \psi_j^{\dagger}\} = \delta_{ij}$, whereas continuum 
fermions satisfy $\{\psi(x), \psi^{\dagger}(y)\} = \delta(x-y)$ where the
Dirac $\delta$-function has the dimension of 1/length. Also $Na=L$ gives the 
conventional box normalization of the continuum fermions). Note also that 
the standard inclusion of the $e^{-ikvt}$ for the particle fields and
the $e^{ikvt}$ for the anti-particle fields, show that the right-movers
are a function only of $x_R = x-vt$ and 
the left-movers are a function only of $x_L=x+vt$. This 
observation will come in useful when we compute
correlation functions. In general, we only need to compute equal time
correlation functions. The time-dependent correlations are then obtained 
by replacing $x$ by $x_R$ for right-movers and by $x_L$ for left-movers. 

In terms of the continuum fields, the Hamiltonian is obtained as 
\beq
H=iv\int dx ~[-\psi_R^{\dagger} {d\over dx} \psi_R + \psi_L^{\dagger} 
{d\over dx} \psi_L] ~.
\label{nine}
\eeq
(H.W. Exercise 3. Check that this Hamiltonian reduces to the one in
Eq. (\ref{lowenop}) using Eqs. (\ref{conti})). \\
We see that the corresponding Lagrangian density is just the standard one 
for free fermions given by
\beq
L=i\psi_R^{\dagger}(\partial_t+ v\partial_x) \psi_R +
i\psi_L^{\dagger} (\partial_t -v\partial_x) \psi_L ~. 
\eeq
Using the standard rules of bosonization, 
this Lagrangian can also be rewritten as 
\beq
L= ~{1\over 2v}(\partial_t\phi)^2 -{v \over 2}(\partial_x\phi)^2 ={1\over
2}~ \partial_\mu\phi\partial^\mu\phi ~,
\eeq
where the last equality requires setting $v=1$. {\bf Note}:
It is also worth checking to see that the same Hamiltonian in Eq. (\ref{nine}) 
is obtained by directly starting with the real space lattice model given in
Eq. (\ref{Hamiltonian}),
rewriting the lattice fermions in terms of the continuum fermions
remembering the $\sqrt{a}$ conversion factor, using $\sum_i a =\int
dx$ and using $\psi_{i+1}= \psi_i + a\partial_x \psi_i$. 

\vskip 0.3cm
\noindent {\bf Correlation functions}

Thus, we have a Lorentz-invariant massless Dirac fermion field
theory in the low energy approximation. All low energy properties can
be obtained from the field theory, which in fact are trivially
computed, since this is a free massless field theory. As far as fermionic
correlation functions are concerned, one does not even require bosonization. 
However, for the spin correlations, it depends on how the spins 
can be expressed in terms of fermions. For instance, we can
explicitly obtain the following spin-spin correlation function 
\beq
G^{zz} (x,t) \equiv <S^z(x,t) S^z(0,0)>
\eeq
simply using Wick theorem.
We start by writing $S_j^z$ in term of the fermions as $S_j^z=n_j-1/2
= \psi_j^{\dagger}\psi_j-1/2 = :\psi_j^{\dagger}\psi_j:$ , since the
expectation value of $n_j$ is half. Since the lattice fermion can be
written in terms of the continuum fermions as
\beq
\psi_j = \sqrt{a} ~[e^{ik_Fj}\psi_R (x=ja)+e^{-ik_Fj}\psi_L(x=ja)] ~,
\label{contLR}
\eeq
and since $e^{i2k_Fj} = e^{(i\pi)x/a} = (-1)^{x/a}$, 
we find that the spin operator can be written as
\beq
S_j^z/a = S^z(x=ja,t) = :\psi_L^{\dagger}\psi_L: +:\psi_R^{\dagger}
\psi_R: +(-1)^{x/a}[\psi_R^{\dagger}\psi_L + \psi_L^{\dagger}\psi_R] ~.
\label{scont}
\eeq
Directly using the Wick theorem and the fermion correlators
\bea
< T\psi_L(x,t)\psi_L^{\dagger}(0,0)> &=& {-i\over 2\pi(x_L -i\alpha ~
{\rm sign} (t))} \nonumber \\
{\rm and}~~~ <T \psi_R(x,t)\psi_R^{\dagger}(0,0)> &=& {i\over 2\pi(x_R + 
i\alpha ~{\rm sign} (t))} ~,
\eea
we see that 
\beq
G^{zz} (x,t) = -{a^2\over 4\pi^2} ~[({1\over x_R^2} +{1\over x_L^2})-
(-1)^{x/a}{1\over x_R x_L}] ~,
\eeq
where $x_R = x-t$ and $x_L = x+t$. \\
(H.W. Exercise 4. Obtain this explicitly). \\
This can also be computed using bosonization. Note that even without 
doing the calculation, one could have guessed that the
four-point correlation of the fermions must go as $1/l^2$, where $l$
is a distance, because in 1+1 dimensions, the fermion field has
a mass dimension of 1/2 or distance dimension of $-1/2$. So, in the
absence of any other scale in the problem (the fermion field is
massless and there are no interactions to cause divergences or 
introduce any anomalous mass
scale), as long as the spin correlations can be expressed purely in
terms of local fermion fields, no calculations are needed to see that
correlations go as $1/l^2$. But we do need to calculate to get the
explicit coefficients of $1/x_R^2$, etc, because they could be
multiplied by dimensionless quantities like $f(x_R /x_L )$, $etc$. 

However, to obtain the correlation function $<S^+(x,t)S^-(0,0)>$ in
the fermionic language is more difficult because of the non-local
string operator. Here, simple dimensional analysis is not sufficient
to give the answer and one actually needs bosonization. The
correlation function can be written as
\bea
G^{+-}(x,t) &=& <S^+(x,t)S^-(0,0)> \nonumber \\ 
&=& (-1)^{x/a} ~[e^{-ik_Fx/a}\psi_R^\dagger (x,t)+e^{ik_Fx/a}\psi_L^\dagger 
(x,t)] \times \nonumber\\
&~~& [e^{i\pi\int_0^{x} (:\psi^{\dagger}(x',t)\psi(x',t): +1/2a)dx'} + h.c.]
\times \nonumber\\
&~~& [\psi_R(0,0) + \psi_L(0,0)] ~,
\label{offd}
\eea
where the string operator stretches between the two positions of the
spin operator. (The other terms cancel out between $S^-$ and $S^+$).
Also, we have explicitly made the string operator hermitian, since it
is hermitian in the lattice model. The reason bosonization comes in
handy here is because the non-local operator when written in terms of
bosons, turns out to be perfectly simple. We just use the bosonization
identity
\bea
\int_0^x dx':\psi^{\dagger}(x',t)\psi(x',t): &=& -~{1\over \sqrt{\pi}}\int_0^x
dx' \partial_{x'} \phi ~=~ -~ {1\over \sqrt{\pi}} ~[\phi(x,t)-\phi(0,t)]
\nonumber\\
&=& -~{1\over \sqrt{\pi}} ~[\phi_R(x,t)+\phi_L(x,t)-\phi_R(0,t)-\phi_L(0,t)]~.
\eea
Substituting this in Eq. (\ref{offd}), and substituting for the other
fermion operators in terms of bosons, we get
\bea
G^{+-}(x,t) &=& (-1)^{x/a} ~{a\over 2\pi\alpha}[\eta_R^{\dagger}e^{-ik_Fx/a}
e^{i2\sqpi \phi_R(x,t)} +\eta_L^{\dagger}
e^{ik_Fx/a}e^{-i2\sqpi\phi_L(x,t)}~]\times\nonumber\\
&&[e^{ik_Fx/a -i{\sqrt \pi} (\phi_R(x)+\phi_L(x)-\phi_R(0) -
\phi_L(0))} +e^{-ik_Fx/a +i {\sqrt \pi} (\phi_R(x)+\phi_L(x)-
\phi_R(0)- \phi_L(0))} ~]\times\nonumber\\
&&[\eta_R e^{-i2\sqpi\phi_R(0,0)} +\eta_L e^{i2\sqpi\phi_L(0,0)}~]
\eea
fully in terms of bosons. Now we use the operator identity $e^{A+B} = e^A e^B 
e^{-[A,B]/ 2}$ to write each of the 8 terms that appear in
the above equation in terms of products of exponential factors. Just for 
illustration, we explicitly write the first term which appears by multiplying 
the first term in each of the square brackets in the above equation. 
\beq
G^{+-}(x,t) = {a\over 2\pi\alpha} ~[\eta_R^{\dagger}\eta_R
e^{i\sqpi\phi_R(x,t)}e^{-i\sqpi\phi_L(x,t)}
e^{-i\sqpi\phi_R(0,0)}e^{i\sqpi\phi_L(0,0)} + 7 ~~{\rm other ~terms}] ~.
\eeq 
Now, we use the standard commutators \\ 
$[\phi_{R/L}(x),\phi_{R/L}(y)] = (-/+) ~i~ {\rm sign} (x-y)/4$, and
$[\phi_{R/L}(x), \phi_{L/R}(y)] = 0$ (since we are using Klein factors),
and the standard algorithm for computing the correlation function
\bea 
<e^{i2\sqpi m_1\phi_{L}(x)}e^{-i2\sqpi m_2\phi_{L}(0)}> &\sim &
Lim_{\alpha\rightarrow 0} ~\Bigl( {\alpha\over x_L - i \alpha ~{\rm sign} 
(t)} \Bigr)^{m_1 m_2}\nonumber \\
{\rm and} \quad <e^{i2\sqpi m_1\phi_{R}(x)}e^{-i2\sqpi m_2\phi_{R}(0)}> 
&\sim & Lim_{\alpha\rightarrow 0} ~\Bigl( {\alpha\over x_R + i \alpha ~
{\rm sign} (t)} \Bigr)^{m_1 m_2} ~,
\eea 
when $m_1$ and $m_2$ have the same sign and vanish when they have
opposite signs \cite{shan1}. This implies that of the 8 terms
above, four of them give zero contribution. Adding up the
contributions of the remaining four, we obtain
\beq
G^{+-}(x,t) \sim {1\over (x_R x_L)^{1/4}} ~[ (-1)^{x/a} ~+~ {\rm const} ~
({1\over x_R^2} + {1\over x_L^2})] ~.
\eeq 
Note that the Klein factors always come as $\eta_i^{\dagger}\eta_i =1$ 
in this correlation function. Also note that one cannot fix the
arbitrary constant that can appear between the uniform and the
alternating parts of the correlation function because of the normal
ordering ambiguities. It is only the exponents which can be found. \\
(H.W. Exercise 5. Obtain the above explicitly). \\
Thus even for the non-interacting theory or purely the $XY$ model,
bosonization comes in handy to compute the correlation functions.
As we have already said, the reason the correlation functions are not
obtainable just by naive scaling arguments is because the expression
for the `off-diagonal' spin correlations in terms of the fermion
operators is non-trivial, because of the presence of the string term. 
These are the only non-zero correlators in
the theory. The other correlators such as $G^{z+}$ or $G^{++}$ are
zero by symmetry - $i.e$ because of $U(1)$ invariance in the spin
model or because of charge conservation in the fermion model.

\newpage
\noindent {\bf Case when $J^z\ne 0$}

We now consider the Hamiltonian in Eq. (\ref{Hamiltonian}). In the
fermionic language, the last term is given by
\beq
\delta H = J_z \sum_j :\psi_j^{\dagger}\psi_j: 
:\psi_{j+1}^{\dagger}\psi_{j+1}: ~.
\eeq
At very large $J_z$, we would expect electrons to be localized on 
every alternate site so that adjacent sites are not occupied. However, 
this will not be true for small $J_z$, so the point of the exercise is 
to see when this happens and what the ground state 
looks like, for both small $J_z$ and large $J_z$. 
In the low energy limit, we can rewrite this term in terms of the
continuum Dirac fermions at the Fermi points (use Eq. (\ref{contLR})) as
\bea
\delta H = aJ_z\int dx &[& :\psi_R^{\dagger}(x)\psi_R(x)+ \psi_L^{\dagger}
(x) \psi_L(x) +(-1)^{x/a} M (x):] \times \nonumber \\
&[& :\psi_R^{\dagger}(x+a)\psi_R(x+a)+\psi_L^{\dagger}(x+a)\psi_L(x+a)
+(-1)^{x/a+1} M (x+a) : ], \nonumber \\
{\rm where} \quad M(x) &=& \psi_R^{\dagger}(x)\psi_L(x)+ \psi_L^{\dagger}(x) 
\psi_R(x) ~.
\label{delh}
\eea
Using the notation $\rhl (x)=\psi_L^{\dagger}(x)\psi_L(x)$ and $\rhr (x)=
\psi_R^{\dagger}(x)\psi_R(x)$, ($\rhr +\rhl $ is the charge density, and $\rhr 
-\rhl $ is the current density; $\rhl$ and $\rhr$ are also called the left and 
right moving currents respectively), we can rewrite Eq. (\ref{delh}) as 
\bea
\delta H = a J_z\int dx &[& \rhr (x) \rhr (x+a) + \rhl (x) \rhl (x+a) + 
\rhr (x) \rhl (x+a) + \rhl (x) \rhr (x+a) \nonumber \\
&& - M(x) M(x+a) ] ~. 
\eea
Here we have used the fact that oscillatory factors integrate to zero. (More 
precisely, they give rise to higher dimension operators, which, however, are
irrelevant and ignored in this analysis). In the current-current
terms, we can use the expansions $\rhl (x+a) = \rhl (x) + a \partial_x
\rhl (x), \psi_L(x+a) = \psi_L(x) + a \partial_x\psi_L(x)$, 
$etc.$, and the fact that square of a Fermi field vanishes, $e.g.$, 
$\psi_L^2(x) =0$, to deduce that terms of the form $\rhl (x) \rhl (x+a)$ are 
higher dimension operators ( they have four fermion operators and at 
least one derivative term) and renormalize to zero in the $a\rightarrow 0$
limit. So among the current-current terms, we are only left with 
$\rhl $-$\rhr $ cross terms of the form $\rhl (x)\rhr (x)$ as the lowest
dimension operators. For the four fermion terms in the second line
also, we apply the same expansion. Dropping higher derivative terms,
we see that the only term which
survives in the product of the curly brackets is of the form $-\rhr (x) 
\rhl (x) -\rhl (x) \rhr (x)$. The extra negative sign is because we need to
anticommute one of the fields. This adds to the $\rhr \rhl$ term coming 
from the first line and we are finally left with 
\beq
\delta H = 4 J_z a\int dx ~ \rhl (x)\rhr (x) ~.
\eeq
This is a four fermion term which in continuum quantum field theory is
called the Thirring term. In the fermionic language, this is an interacting 
quantum field theory. However, it is easy to solve by bosonization. 

By the standard rules of bosonization for non-interacting fermions, we 
can write 
\beq 
\rhl ={1\over 2\sqpi}~ ({1\over v}\partial_t+\partial_x) ~\phi ~, \quad {\rm 
and} \quad \rhr ={1\over 2\sqpi} ~ (-{1\over v}\partial_t+\partial_x) ~\phi ~.
\eeq
In that case, using the units that $Ja=v=1$, we get 
\beq
\delta L = -\delta H ={J_z\over J\pi}~\partial_{\mu}\phi\partial^{\mu}\phi ~,
\eeq
where $\delta L$ denotes the change in the Lagrangian.
This is precisely of the same form as the bosonization of the free
fermion Hamiltonian. So the new Lagrangian is given by
\beq
L= {1\over 2K}~\partial_{\mu}\phi\partial^{\mu}\phi ~,
\label{lagint}
\eeq
where 
\beq
{1\over K}=1+{2J_z\over J\pi} ~.
\label{rsquared}
\eeq
This can be made to look like the free
term by redefining the field $\phi$ - $i.e.$, we define a new field
$\tphi = \phi/\sqk$, so that in terms of $\tphi$, the Lagrangian is just 
${1\over 2}(\partial_{\mu}\tphi\partial^{\mu}\tphi)$. However, the canonical 
momentum obtained from the rescaled Lagrangian is just $\tPi = \partial_0 
\tphi$ whereas the momentum obtained from the Lagrangian in 
Eq. (\ref{lagint}) $\Pi = \partial_0
\phi/K$. Hence, the momentum gets rescaled compared to the original
momentum as $\tPi = \sqk\Pi$. Clearly, the new coordinate and momenta
satisfy the canonical commutation relations, since they are rescaled
in opposite ways. (Remember that $\phi$ takes values on a compact circle, 
since the original spin operators are defined in terms of exponential of the 
boson fields and are invariant under periodic changes of $\phi$). 
But since the right and left mover fields are defined by
taking both the field and the canonical momentum, and they scale in
different ways, one can no longer write the right and left moving
fields in the tilde representation as just scaled versions of the
right and left moving fields of the original theory - in fact, they
mix up the left and right moving fields. Explicitly,
\bea
\tphi_R(t,x) &=& {1\over 2}~[\tphi(t,x) -\int_{-\infty}^x dx' ~\tPi(t,x')] = 
{1\over 2} ~[\phi(t,x)/\sqk -\int_{-\infty}^x dx' ~\sqk\Pi(t,x')] \nonumber \\
&=& {(\phi_R+\phi_L)\over 2\sqk} - {\sqk(\phi_L-\phi_R)\over 2} \nonumber\\
&=& {1\over 2}({1\over\sqk}+\sqk)~ \phi_R+ {1\over 2}({1\over\sqk}-\sqk) ~
\phi_L = {\rm cosh}\beta ~\phi_R +{\rm sinh}\beta ~\phi_L ~,
\eea
where $e^{-\beta} =\sqk$. Similarly 
\beq
\tphi_L(t,x)= {\rm cosh}\beta ~\phi_L + {\rm sinh}\beta ~\phi_R ~.
\eeq 
One can now express the spin fields in terms of the $\tphi$
fields. They are given by
\bea
S^z (x,t) &\simeq& \sqrt{K \over \pi} \partial_x \tphi +(-1)^{x/a} ~{\rm
const} ~e^{i2 \sqrt{\pi K} \tphi} \nonumber\\
S^- (x,t) &\simeq& (-1)^{x/a} e^{i{\sqrt{\pi /K}} (\tphi_R-\tphi_L)} + 
{\rm const} ~\times \nonumber \\
\quad &&[e^{i(2\sqrt{\pi K}(\tphi_R+\tphi_L)+{\sqrt{\pi /K}} 
(\tphi_R-\tphi_L))} +e^{i(-2\sqrt{\pi K}(\tphi_R+\tphi_L)+
{\sqrt{\pi /K}}(\tphi_R-\tphi_L))} ]~.
\label{spinbosi} 
\eea
With these substitutions, it is trivial (albeit algebraically more
tedious!) to recalculate the spin-spin correlators $G^{zz} (x,t)$ and
$G^{+-}(x,t)$. Since the method is exactly the same as for the free
case, we just quote the answers here.
\bea
G^{zz} (x,t) &\simeq& -{K\over 4\pi}({1\over x_L^2} +
{1\over x_R^2}) +(-1)^{x/a} ~{\rm const} ~(x_R x_L)^{-K}\nonumber \\
G^{+-}(x,t) &\simeq& (-1)^{x/a}(x_R x_L)^{-1/4K} + ~{\rm const} ~
(x_R x_L)^{-({1\over 2\sqk} - \sqk)^2} ({1\over x_L^2} + {1\over x_R^2}) ~.
\eea 
(H.W. Exercise 6: Obtain the above expressions). \\
Note that at $K=1/2$, the two correlations above are the same.

\vskip 0.3cm
\noindent{\bf Limitations of this calculation}

So the end result is that we have now obtained spin-spin correlation
functions even including $J_z$. But since we have made a low energy
continuum approximation and included only a few
low-lying modes around the Fermi point, this derivation of the
correlation functions is not true for arbitrary $J_z$. For instance, we 
left out terms that were irrelevant by naive
power counting, which only works in the non-interacting case. Once
we have interactions, some of those operators could acquire anomalous
dimensions and hence become relevant. In other words, we have seen
that interactions change the dimensions of operators. However, we have
only studied operators of the form $\rhl (x)\rhr (x)$, which were marginal
to start with and seen how they evolved. But we did not keep all the
irrelevant operators and see how they evolved. Sometimes, they
will also become relevant with sufficiently strong interactions.

\vskip 0.3cm
\noindent{\bf A more general effective action approach}

However, one can try to understand what can possibly
change if we include other corrections that we left out in our
approximation. One way of doing this is to look at all possible
relevant terms that can appear consistent with the symmetries of the
problem. The idea is not to try and derive these terms but to write
them down in the effective Lagrangian assuming that if they are not
explicitly prohibited by a symmetry, then they will appear. This is the
philosophy behind what are called effective field theories. 

\vskip 0.3cm
\noindent {\bf Aside on how to `read off' dimensions of operators}

We know that to see whether an operator is relevant or irrelevant, we
have to compute its correlation function and find out its scaling
dimension. Then, we have to check whether the scaling dimension is
such that the coefficient of the operator grows or becomes smaller as
the energy scale is reduced. 
So given any operator $O_i$ in terms of bosons, we first compute the
correlation function $<O_i(x,t)O_i(x',0)>$ which goes as
$1/(x-x')^{2d_i}$. For a free fermion theory with no interactions,
(equivalently a boson theory with the interaction parameter $K=1$) 
we know that $<O_i(x,t)O_i(x',0)>=1/(x-x')^{2\td_i}$ where $\td_i$ is
just the naive scaling dimension or the engineering dimension of the
operator $O_i$. The difference between $\td_i$ and the $d_i$ that
appears when we actually compute the correlation function is because
of the interactions and is called the anomalous dimension of the
operator. As was explained in the other courses in this
school \cite{SB}, the extra dimensional parameter comes from the
cutoff scale. We shall use the term scaling dimension to mean $d_i$ itself. 
For an operator of the form $O_i\sim e^{i\sqfp \beta(\phi_L+\phi_R)}$, 
the scaling dimension is given by $d_i=\beta^2$, for the standard
(non-interacting) form of the Hamiltonian. Since the space-time
dimension is two, it is clear that $d_i>2$ implies that the coefficient 
$\lambda_i$ of the operator
has to have dimension $2-d_i<0$. So each time the cutoff is scaled
down by a factor $\Lambda$, $\lambda_i \rightarrow \lambda_i
\Lambda^{2-d_i}$. Hence, after successive rescalings, this term in the
action is irrelevant and scales to zero. On the other hand, $d_i<2$ denotes 
relevant operators, whose coefficients grow under scaling downs of the
cutoff. $d_i=2$ is a marginal operator, whose coefficient remains
unchanged under rescalings. (We will come back to this when we study 
impurity scattering and scaling dimensions of `boundary operators'). 

\vskip 0.3cm
\noindent {\bf Back to the effective action} 

The only possible
Lorentz-invariant relevant terms that can be added to the Lagrangian is
either $\cos\sqfp\beta(\tphi_L+\tphi_R)$ or $\cos\sqfp
\beta(\tphi_L-\tphi_R)$; both of these have dimension
$\beta^2$ and are thus relevant for $\beta<\sqrt{2}$. 
(The real problem on a lattice, of course, does not have
Lorentz invariance. However, in the long distance or low energy limit,
all such Lorentz non-invariant interactions will probably be irrelevant).
Of these, the $U(1)$ symmetry under rotations about the
$z$-axis in fermion language implies that $\psi_L$ and $\psi_R$ have
to be multiplied by the same phase (because $S_z$ which has terms of
the form $\psi_L^{\dagger}\psi_R+h.c.$ should not change). 
This in turn means that $\tphi_L\rightarrow \tphi_L+c$ and
$\tphi_R\rightarrow \tphi_R-c$ so that $\tphi_L+\tphi_R
\rightarrow\tphi_L+\tphi_R$ and
$\tphi_L-\tphi_R\rightarrow \tphi_L-\tphi_R$+constant. 
Thus, to be consistent with this
symmetry, we can only allow $\cos\sqfp\beta(\tphi_L+\tphi_R)$. 

Furthermore, since the
spin operators are all expressed in terms of exponentials of the boson
fields, (see Eq. (\ref{spinbosi})) 
the boson fields need to be 'compactified on a circle'. This only
means that the boson fields are periodic -
\beq
\tphi \leftrightarrow \tphi +\sqrt{\pi\over K}
\eeq
since the spin fields cannot distinguish between $\tphi$ and $\tphi
+\sqrt{\pi/K}$. This restricts $\beta$ in
$\cos\sqfp\beta(\tphi_L+\tphi_R)$ 
to be of the form $n\sqk$ where $n$ is an integer.

Finally, we use an unusual feature which occurs in the continuum field
theories of many lattice spin models. The translational symmetry of
the lattice spin model by one site ( or more sites for more general
models) maps to a discrete symmetry 
in the continuum model, which is distinct from translational
symmetry. This can be seen from the continuum definition of the spin
in terms of the Dirac fermions - Eq. (\ref{scont}). When we change $j$ to
$j+1$, the oscillatory factor $(-1)^j \rightarrow -(-1)^j$. This is a
drastic change from site to site. So if in the continuum version, we
want to define smooth fields without having this rapid oscillations,
we need to define one field for every pair of sites. Thus invariance under
translation by $2a$ on the lattice becomes translational invariance in the
continuum model. But from Eq. (\ref{scont}), we see that 
translational symmetry through a single site
corresponds to the discrete symmetry
\beq
\psi_L\rightarrow i\psi_L, \quad \psi_R\rightarrow-i\psi_R ~.
\eeq
In the bosonic language, this corresponds to 
\beq
\tphi_L\rightarrow \tphi_L+{1\over 2}{\sqrt{\pi\over K}}, \quad {\rm and} 
\quad \tphi_R\rightarrow \tphi_R+ {1\over 2}{\sqrt{\pi\over K}} ~.
\eeq
This symmetry implies that the only terms that can be added to the
Lagrangian are of the form $\cos\sqfp 2n\sqk(\tphi_L+\tphi_R)$, so that 
$\beta = 2n\sqk$. This is relevant
when $K<1/2$ when $n=1$. So the system is in a massless phase
till $K$ reaches $1/2$ below which it develops a relevant
interaction, and a mass gap. 

However, we cannot use our low energy approximate result to estimate 
the point at which the spin model develops a relevant interaction. 
Besides adding the possible relevant term mentioned above, the most
general thing the other terms that we have neglected can do is to
change the relation between $K$ and $J_z$ in an unpredictable way.
In fact, the low energy result relating $K$ to the perturbation $J_z$ 
is only true to lowest order in $J_z/J$. This particular spin-chain model 
is, in fact, solvable by Bethe ansatz and the exact answer is
\beq
{1\over K}=1 + {2\over \pi}\sin^{-1}({J_z\over J})
\eeq
which, to lowest order in $J_z/J$, gives us the relation in
Eq. (\ref{rsquared}). From this, we see that $1/K\rightarrow 2$ at $J_z=J$. 
This is precisely the $K$ value for which the cosine interaction term becomes
relevant. To prove that a relevant interaction necessarily leads to a
mass gap is non-trivial, but it is certainly plausible. Once,
there is a relevant interaction, its coefficient grows under
renormalization. It becomes divergent as we make the energy scale
lower and lower, so we have to cut it off at some scale, which is the
mass scale associated with the theory. However, 
it could also lead us to a new fixed point, which may not have a mass gap. 

To get higher orders in $J_z/J$ in this effective field theory
approach is not easy, because one needs to go beyond the region of
linear dispersion. Also, once one starts including modes with $k\simeq
k_F$, we need to be careful to put in the restriction that $-k_F<k<k_F$.
Also, since the Bethe ansatz already gives the exact answer to all orders,
there may not be much point in trying to do this for this problem. 

So what does all this formalism gain us?
How does the ground state evolve as $J_z$ changes? For small values
of $J_z$ all that happens is that the spin-spin correlations have a
slightly different power law fall-off with anomalous non-integer
exponents. Does this continue for all values of $J_z$? No. Once, 
$J_z$ reaches $J_z=J$, the isotropic point, there
exists a phase transition to a massive phase where spin-spin correlations 
fall off exponentially fast at large separations. In this particular problem,
of course, one knew this answer from the Bethe ansatz, but the point is
that the bosonization method can be used even for other models, which are
not exactly solvable by the Bethe ansatz. But without the
Bethe ansatz, one cannot analytically find the value of the parameter 
where the phase transition into a massive phase occurs. 
The other important gain that we have in this method is that it allows
the computation of correlation functions, which is not possible using
the Bethe ansatz. Finally, since it is a symmetry analysis, it tells us
that for any Hamiltonian of this type, the model is likely to be
massless only when the theory has $U(1)$ symmetry and the $Z_2$ symmetry of
translation by a single site and even then, only for some restricted
values of the parameter space.
 
The best reference for this application is Affleck's
lectures \cite{aff1} on field theories and critical phenomena, which
we have followed fairly faithfully.

\vskip 0.5cm
\section{Hubbard model}

The Hubbard model is one of the simplest realistic models that one can study
which has a competition between the kinetic energy and the potential energy. 
The kinetic energy or the hopping term gains, or rather the energy gets 
lowered, if the fermions are delocalized - free to move throughout the
sample. In this model, the potential energy represents screened Coulomb
interactions between electrons and the model is constructed so that it
costs energy to put two electrons at the same place. So the potential
energy prefers each electron to sit at its own site. The model is
given by
\beq
H = -{t\over 2}\sum_{j\alpha} ~(\psi_{j\alpha}^{\dagger} \psi_{j+1\alpha} 
+h.c.) + U\sum_j ~n_{j\uparrow}n_{j\downarrow} +\mu\sum_{j,\alpha} ~
\psi_{j\alpha}^{\dagger} \psi_{j\alpha} ~,
\label{hubbard} 
\eeq
where $t$ is the hopping parameter,
$U$ is a positive constant denoting the repulsion between two
electrons at a site, $\mu$ is the chemical potential and 
$\alpha$ is the spin index which can be
$\uparrow$ or $\downarrow$. This model is very similar to the fermion
model we studied for the spin chain except that the electrons have
spin and the chemical potential term allows for arbitrary fillings. 
The $U$ term or Hubbard term is analogous to the nearest neighbor $J_z$ 
interaction term for spinless electrons.

At half-filling (one electron/site, since a filled band
implies two electrons/site), for large $U$, the model is expected to 
describe an insulator. One can easily understand
this, because at infinite $U$, the ground state will have one electron
at every site. Any excitation will cost an energy of $U$. So there is
a gap to excitations and the model behaves as an insulator. It is
called a Mott-Hubbard insulator (as opposed to other band insulators)
because here the insulating gap is created by interactions. 

The question that one would like to ask is, at what value of $U$ does
the Mott-Hubbard gap open, because naively one may think that at very
small values of $U$, the model allows free propagation of electrons
and describes a metal. Using bosonization, we will show that 
in one dimension, this expectation is wrong. The Mott-Hubbard gap
opens up for any finite $U$ if the filling is half and not otherwise. 
For any other filling, the model at low
energies is an example of a Luttinger liquid with separate spin and
charge excitations. The spin modes are always gapless
whereas the charge modes are gapless at any filling other than half-filling;
precisely at half-filling a charge gap opens up. The model for arbitrary 
filling (other than half-filling) and positive $U$ is said to be in the
Luttinger liquid phase. Spin and charge correlations fall-off as power
laws and we expect power law transport. At half-filling, the model is
in a charge-gapped phase called the charge-density wave phase. \\
\noindent{\bf Aside}: For
negative $U$, it is found that the spin excitations are always
gapped. Here, the model is said to be in the Luther-Emery phase or
spin gapped phase.

\vskip 0.3cm
\noindent{\bf Bosonization of the model without interactions}

How do we go about seeing all that we have described above? In higher
dimensions, we would do a mean field theory, but in one space
dimension, we know that a mean field analysis is not very useful
because of the infrared divergences of the low energy fluctuations. 
(In other words, if we write down a mean field theory and then try to
do systematic corrections about the mean field theory, then order by
order in perturbation theory, we find that the integrals which appear
in the corrections are divergent).
So it seems like a good idea to try and use bosonization. In fact, 
the way this model is analyzed is very similar to the way we analyzed
the spinless fermion model in the previous section. We first switch
off the interactions and start with the Fourier decomposition 
\beq
\psi_{j\alpha} = {1\over\sqrt{N}}\sum_k ~\psi_{k\alpha} e^{ikja} ~.
\eeq
We rewrite the Hamiltonian as 
\beq
H_0=\sum_{k\alpha}~(\mu-t \cos ka )~\psi_{k\alpha}^{\dagger}\psi_{k\alpha} ~,
\eeq
where the $k$ values go from $-\pi/a$ to $\pi/a$. In the ground state, all 
states with $|k|<k_F$ are filled, where $k_F$ is determined by the chemical 
potential from the equation $\mu=\cos k_F a$. Just as in the spinless 
case, we will look only at the low energy modes near the Fermi surface, so
that each fermion is written as
\bea
\psi_{j\alpha} &\sim& e^{-ik_F ja} \int_{-k_Fa-\Lambda}^{-k_Fa+\Lambda} ~
{d(ka)\over 2\pi} ~e^{ikja}\psi_{k\alpha} +e^{ik_F ja} \int_{k_Fa-\Lambda}^{k_F
a+\Lambda} {d(ka)\over 2\pi} ~e^{ikja} \psi_{k\alpha} \nonumber \\
&\equiv& e^{-ik_F ja}\psi_{Lj\alpha} +e^{ik_F ja}\psi_{Rj\alpha} ~,
\label{lowenftspin}
\eea
so that the $\psi_{Lj\alpha}$ and $\psi_{Rj\alpha}$ do not contain high 
energy modes. Substituting this expression in Eq. (\ref{hubbard}), we get
\bea 
H_0=-{t\over 2} \sum_{j\alpha} &[& (e^{-ik_Fa}\psi_{Lj\alpha}^{\dagger}
\psi_{Lj+1\alpha}+ e^{ik_Fa}\psi_{Rj\alpha}^{\dagger}\psi_{Rj+1\alpha} + 
\nonumber \\
&& \quad e^{-i2k_Fja-ik_Fa}\psi_{Lj\alpha}^{\dagger}\psi_{Rj+1\alpha} + 
e^{i2k_Fja+ ik_Fa}\psi_{Rj\alpha}^{\dagger}\psi_{Lj+1\alpha})+h.c.] \nonumber\\
+\mu \sum_{j\alpha}&[& \psi_{Lj\alpha}^{\dagger}\psi_{Lj\alpha}+ \psi_{Rj
\alpha}^{\dagger}\psi_{Rj\alpha} + e^{-i2k_Fja}\psi_{Lj\alpha}^{\dagger}
\psi_{Rj\alpha} + e^{i2k_Fja}\psi_{Rj\alpha}^{\dagger}\psi_{Lj\alpha}] ~.
\eea
The oscillatory terms do not contribute because they have the form 
\beq
\sum_j e^{i2k_Fja}\psi_{Rj\alpha}^{\dagger}\psi_{Lj\alpha}\sim \sum_j 
e^{i2k_Fja} \sum_{k,k'} e^{-ikja}e^{ik'ja}\psi_{k\alpha}^{\dagger}
\psi_{k'\alpha} \sim \sum_{k,k'} \psi_{k\alpha}^{\dagger}\psi_{k'\alpha} 
\sum_j e^{i2k_F ja-ikja+ik'ja} ~, 
\eeq
and the sum over $j$ in the last expression produces $\delta_{2k_F,k-k'}$
which cannot be satisfied for small values of $k,k'$. (Note that even for
the half-filled case where $k_Fa=\pi/2$, this cannot be satisfied). 
Hence, we may drop these terms and we are left with only
\bea
H_0 =&-&{t\over 2}\sum_{j\alpha} ~[(e^{-ik_Fa}\psi_{Lj\alpha}^{\dagger}
\psi_{Lj+1\alpha}+ 
e^{ik_Fa}\psi_{Rj\alpha}^{\dagger}\psi_{Rj+1\alpha}) +h.c.] \nonumber \\ 
&+& \mu \sum_{j\alpha} ~(\psi_{Lj\alpha}^{\dagger}\psi_{Lj\alpha}+
\psi_{Rj\alpha}^{\dagger}\psi_{Rj\alpha}) ~.
\eea
Now, we expand $\psi_{j+1\alpha}\equiv \psi_\alpha (j+1)=
\psi_\alpha(j)+a \partial_x \psi_\alpha(j)$ + higher order irrelevant
terms and use the fact that $\cos k_Fa=\mu$ (which means that part of
the hopping term cancels with the $\mu$ term) to get 
\bea
H&=&-{at\over 2}\sum_{j\alpha} ~(e^{-ik_Fa}\psi_{Lj\alpha}^{\dagger}
\partial_x \psi_{Lj\alpha}+ e^{ik_Fa}\psi_{Rj\alpha}^{\dagger}\partial_x
\psi_{Rj\alpha} +h.c.) \nonumber\\
&=& iat \sin (k_Fa) \sum_{j\alpha}~ 
(\psi_{Lj\alpha}^{\dagger} \partial_x \psi_{Lj\alpha}- 
\psi_{Rj\alpha}^{\dagger} \partial_x \psi_{Rj\alpha}) ~, 
\eea
where we have also integrated the hermitian conjugate terms by part to
get it in the form of the second equation above. Finally, 
we can rewrite this Hamiltonian as a continuum Hamiltonian in terms of
continuum fields (defined with the usual factor of $\sqrt{a}$ as
$\psi_{\alpha}(j)/\sqrt{a} = \psi_{\alpha}(x)$ ) and using $\sum_j a =
\int dx$ 
\beq
H_0=it \sin (k_Fa)\sum_{\alpha} ~
\int dx ~[\psi_{L\alpha}^{\dagger}(x) \partial_x \psi_{L\alpha}(x)- 
\psi_{R\alpha}^{\dagger}(x) \partial_x \psi_{R\alpha}(x) ~]~,
\label{real} 
\eeq
where we call $t \sin (k_Fa)=v_Fa$, the Fermi velocity times $a$.
The derivation here is very similar to the one for spinless fermions,
except that here we have carried it out in real space instead of
momentum space. This Hamiltonian can be bosonized using the usual
rules of bosonization and we get
\beq
H_0 = \frac{v_F a}{2} \sum_{\alpha}~\int dx ~[\Pi_{\alpha}^2 +
(\partial_x\phi_\alpha)^2] ~.
\label{ho}
\eeq
(H.W. Exercise 7. Derive the Hamiltonian in Eq. (\ref{real}) through a
momentum space derivation).

\vskip 0.3cm
\noindent{\bf Bosonization of the interaction term}

The next step is to figure out the low energy part of the on-site
Hubbard interaction. Here, again, the principle is the same. We
rewrite the four-fermion term written in terms of the original
fermions in terms of the low energy Dirac fermion modes. Just as in
the spin model, the $S^z-S^z$ term or the four fermion term 
corresponded to a product of normal ordered bilinears, here also the 
four fermion term in Eq. (\ref{hubbard}) can be written in terms of the
product of normal ordered bilinears if we subtract the average charge
densities of the $\ua$ and $\da$ fields. So we may write
\beq
H_{\rm int} = U\sum_j ~n_{j\uparrow}n_{j\downarrow} 
= U\sum_i ~:n_{j\uparrow}::n_{j\downarrow}: ~.
\eeq
In terms of the Dirac fields, this becomes
\beq
H_{\rm int} = U\sum_j [(:\psi_{jL\ua}^{\dagger}\psi_{jL\ua}: +
:\psi_{jR\ua}^{\dagger}\psi_{jR\ua}:
+\psi_{jR\ua}^{\dagger}\psi_{jL\ua}e^{-i2k_Fja} + \psi_{jL\ua}^{\dagger}
\psi_{jR\ua}e^{i2k_Fja}) \times (\ua \rightarrow \da ) ].
\label{osc}
\eeq
We now expand the products and keep only the terms with no oscillatory
factor, to get
\beq
H_{\rm int} = U\sum_j ~(J_{jR\ua}+J_{jL\ua})(J_{jR\da}+J_{jL\da}) + U\sum_j ~
(\psi_{jR\ua}^{\dagger}\psi_{jL\ua}\psi_{jL\da}^{\dagger}\psi_{jR\da} +h.c.)~.
\eeq
The remaining terms have the oscillatory factors of either
$e^{i2k_Fja}$ or $e^{i4k_Fja}$ and can be set to zero for arbitrary
filling. Notice however, that $e^{i4k_Fja} = 1$ and is not oscillatory
at half-filling since $k_Fa=\pi/2$. We will come back to this point
later. Now we first express these fields in terms of the continuum
fields and just use the standard bosonization formulae to get
\beq
H_{\rm int} = Ua\int dx ~[{1\over\pi}\partial_x \phi_\ua \partial_x 
\phi_\da + \eta_{R\uparrow}^{\dagger}\eta_{L\downarrow} \eta_{L
\downarrow}^{\dagger}\eta_{R\uparrow} {Ua\over 2\pi^2 \epsilon^2} \cos
\sqrt{4\pi} (\phi_{R\ua}+\phi_{L\ua} -\phi_{R\da}-\phi_{L\da})] ~.
\label{Hint}
\eeq
(H.W. Exercise 8. Derive the above). \\
The interesting point to note here is that the cosine term only
depends on $\phi_{\ua}-\phi_\da$. (We use the earlier defined notation that
$\phi=\phi_L+\phi_R$ and $\theta =-\phi_R+\phi_L$). 
So if we define the charge and spin
fields 
\beq
\phi_{c} = {\phi_\ua+\phi_\da\over \sqrt{2}} ~, \quad {\rm and} \quad 
\phi_{s} = {\phi_\ua-\phi_\da\over \sqrt{2}} ~,
\eeq
the Hamiltonian is completely separable in terms of these two fields
and we may write $H =H_0+H_{\rm int} = H_c+H_s$ with
\bea
H_c &=& {v_F\over 2} \int dx ~[\Pi_c^2 + (1+{U \over \pi v_F})(\partial_x
\phi_c)^2] \nonumber\\
H_s &=&{v_F\over 2} \int dx ~[\Pi_s^2 + (1-{U \over \pi v_F})(\partial_x
\phi_s)^2 + {Ua\over 2\pi\epsilon^2} \cos\sqrt{8\pi}\phi_s ]~,
\eea
where the bosonized form of the kinetic energy term given by $H_0$ in 
Eq. (\ref{ho}) along with the first term in Eq. (\ref{Hint}) ($
U\partial_x\phi_\ua \partial_x\phi_\da /\pi$) 
can also be written in terms of the charge
and spin fields as above. The charge sector is massless, but for the
spin sector, one has a cosine term in the Hamiltonian. From our
earlier experience of spinless models, we know that a cosine term can
lead to a mass gap, when it becomes relevant. So we need to compute
the dimension of the operator and see when it becomes relevant. Note
that we have chosen the product of the Klein factors to be
unity\footnote{For single chain problems, the Klein factors usually cause
no problems and can be set to be unity, in most cases. 
The only care that we need 
to take is to remember the negative sign that one gets when two of them
are exchanged. But for multi-chain models, when more than four explicit Klein
factors exist, one needs to be more careful.}. 
But we only know how to compute correlation functions when 
the quadratic Hamiltonian is in the standard form. To get that, we need 
to rescale the $\phi$ fields and their conjugate momenta (in the
opposite way so that the commutation relations are preserved) as 
\beq
\bphi_c =(1+{U \over \pi v_F})^{1/4} \phi_c ~, \quad {\rm and} \quad 
\bPi_c =(1+{U \over \pi v_F})^{-1/4}\Pi_c ~,
\eeq
and similarly for the spin fields to get the
Hamiltonian in the standard form, from which we can directly read out
the dimensions of the operators. In terms of the bar fields, we see that
\bea
H_c &=& (1+{U \over \pi v_F})^{1/2}{v_F a\over 2}\int dx [\bPi_c^2 +
(\partial_x\bphi_c)^2] \nonumber\\
H_s &=& (1-{U \over \pi v_F})^{1/2}{v_F a\over 2}\int dx [\bPi_s^2 +
(\partial_x\bphi_s)^2 + {Ua\over 2\pi\epsilon^2} \frac{1}{(1-{U \over \pi 
v_F})^{1/2}} \cos\sqrt{8\pi\over (1-{U \over \pi v_F})^{1/4}}\bphi_s ].
\nonumber \\
&&
\eea
The charge sector is purely quadratic (both before and after rescaling!) and 
remains massless, whereas for the spin sector, the rescaling was necessary 
to `read off' the dimension of the cosine operator. Since its scaling 
dimension is given by $d=2/(1-{U \over \pi v_F})^{1/4}$, 
it is irrelevant ($d<2$) for any weak positive $U$ and the spin sector is 
also massless. On the other hand, for any negative $U$, this term has 
dimension $d>2$ and is relevant. As we explained in the spinless case, this 
means that the spin sector acquires a mass gap for all negative $U$.

Also note that the velocities of the charge and the spin modes have got 
renormalized in different ways. $v_c=(1+{U \over \pi v_F})^{1/2}v_F$ is the
velocity of the charge mode and $v_s=(1-{U \over \pi v_F})^{1/2}v_F$ 
is the velocity of the spin mode. Thus, spin and charge move
independently. This is one of the hallmarks of Luttinger liquid
behavior in one-dimensional fermion models. It is only for $U=0$,
that the spin and the charge modes move together.

How does one look for such spin-charge separation in one-dimensional
models? Experimentally, one has to look at different susceptibilities
and measure the Wilson ratio, which is the ratio of the spin
susceptibility to the specific heat coefficient. The specific heat
coefficient depends both on spin and charge modes and is given by 
\beq
{\gamma\over\gamma_0} = {1\over 2}~({v_F\over v_c} + {v_F\over v_s}) ~,
\eeq
where $\gamma_0$ is the specific heat of non-interacting electrons with
velocity $v_F$. However, spin susceptibility only depends on the spin
mode and is given by 
\beq
{\chi\over\chi_0} = {v_F\over v_s} ~.
\eeq
Thus, the Wilson ratio is given by
\beq
R_W = {\chi/\chi_0\over\gamma/\gamma_0} = {2v_c\over v_c+v_s}.
\eeq
Clearly, when there is no spin-charge separation, this is given by
one. So deviations of the Wilson ratio 
from unity are a sign of spin-charge separation in real systems.

Finally, let us consider the case exactly at half-filling, $k_Fa = \pi/2$. In 
this case, the $e^{i4k_Fja}$ term we neglected in Eq. (\ref{osc}) as 
oscillatory, is no longer oscillatory, since $e^{i4k_Fa}=1$. 
In this case, there exists a term in the Hamiltonian of the form
\beq
H_{\rm umklapp} = U\sum_j ~(\psi_{jR\ua}^{\dagger}\psi_{jL\ua}
\psi_{jR\da}^{\dagger}\psi_{jL\da} +h.c.) ~.
\eeq
Note that this term destroys
two right movers and creates two left movers or vice-versa. So there
is an overall change in momentum by $4k_Fa = 2\pi$, which has to be
absorbed by the lattice. It is an umklapp process unlike the earlier
interaction term for arbitrary filling which created and destroyed a
particle at the left Fermi point and also created and destroyed a
particle at the right Fermi point and did not change any momentum. It
is easy to see that this term also gives rise to a cosine term by bosonizing, 
which, after rescaling gives ${Ua\over 2\pi\epsilon^2 (1+{U / \pi 
v_F})^{1/2}} \cos \sqrt{8\pi\over (1+{U / \pi v_F})^{1/4}}\bphi_c$ 
neglecting Klein factors. Thus it appears in the Hamiltonian of the charge 
sector. This term is irrelevant for any negative $U$, but relevant for any 
positive $U$. Thus, precisely at half-filling, the charge sector has a
gap. This is similar to the case for spinless fermions where the spin
model actually corresponded to a half-filled spinless fermion model. But 
unlike the case for the spinless fermions where the gap only opens up at 
$J=J_z$, here the gap opens up for any positive $U$, however small.

Is there any way one can understand these results in a physical way?
For negative $U$, we found that the spin sector has a gap. This can be
understood by saying that since there is an attractive interaction
between the spin $\ua$ and the spin $\da$ densities of fermions, they
will like to form singlets and sit on a single site. So to make a spin
excitation, one needs to break a pair and this costs energy. But 
charge excitations can move around as bound spin singlet pairs with 
no cost in energy. On the other hand, for
positive $U$, there exists a repulsion between two electrons at a
site. So each electron will tend to sit on a different site. At
half-filling, hence, there is no way for an electron to move, without
trying to sit at a site, at which an electron is already present. And
this costs a repulsive energy $U$. Hence, there is a gap to charge
excitations. But one can flip spins at a site and hence have spin
excitations with no cost in energy. 

So what results has bosonization given us here?
We started with electrons with spin
and charge moving together via a hopping term, but with a strong on-site
Coulomb repulsion. This term could not be treated
perturbatively. However, when we rewrote the theory in terms of
bosons, with one boson for the $\uparrow$ spin and one for the
$\downarrow$ spin, we found that the theory decoupled in terms of new
spin and charge bosons. For a generic filling, the charge boson was
just a massless free boson excitation, whereas the spin boson
Hamiltonian had a cosine term, which was relevant when $U$ was
negative, but irrelevant for positive $U$. But at half-filling, the
charge excitations develop a gap, for any positive $U$. The most
important thing to note here is that the charge and spin degrees of
freedom have completely decoupled. Since the two fields are scaled
differently, they move with different velocities in the system. This is
a result that one could never have obtained perturbatively.
Thus, at any filling other than half-filling, the low energy limit of
the Hubbard model is a Luttinger liquid with massless spin and charge
excitations moving with different velocities. A good reference for
this part is Shankar's article \cite{shan1} which also explains in
great detail how to compute correlation functions.

\vskip 0.5cm
\section{Transport in a Luttinger liquid - Clean Wire}

The last two applications involved the study of correlation functions,
with the aim of finding out the different phases possible in a
one-dimensional system of interacting fermions. 
In this part of the course, we will study another application of
bosonization, which is to study transport, in particular the DC (or zero 
frequency) conductivity in one-dimensional wires of interacting fermions. 

Firstly, are one-dimensional wires experimentally feasible? The general idea 
to make narrow wires is to 'gate' 2D electron gases. In recent times, 
technology has developed enough to make these wires so narrow, that they 
contain only one transverse channel. So these are good enough approximations
to one-dimensional wires. Another good approximation to
coupled chains of one-dimensional models are carbon
nanotubes, though those are not the kind of models we will study here.

The next point to note is that even at a qualitative level, transport
in low dimensional systems is extremely different from transport in
higher dimensions. To understand this point, we will first make
qualitative statements about transport and conductivity before we
explicitly start computing it using bosonization. The usual aim is to
compute the conductance as a
function of the voltage, temperature, presence of impurities or
disorder and so on. Normally, when currents are measured in wires, one
does not worry about quantum effects, because wires are still
macroscopic objects, but that is clearly not the case here, since we
are interested in one-dimensional wires. 
In fact, whenever the physical dimensions of the conductor becomes
small, (it need not be really one-dimensional), the usual Ohmic
picture of conductance where the conductance is given by
\beq
G=\sigma ~~{W\over L} =\sigma ~~{{\rm width ~~of ~~conductor}\over {\rm length
~~of ~~conductor}} ~,
\eeq
where $\sigma$ is a material dependent quantity called conductivity, 
breaks down. A whole new field called `mesoscopic physics' 
has now been created to deal with electronic transport in such systems. 
The term `mesoscopic' in between microscopic and
macroscopic is used for systems, where the sizes of the devices are
such that it is comparable with a) the de Broglie wavelength ( or
kinetic energy) of the electron, b) the mean free path of the electron
and c) the phase relaxation length ( the length over which the
particle loses memory of its phase) of the electron. Ohmic behavior
is guaranteed only when all these length scales are small compared to
the size which happens for any macroscopic object. These
lengths actually vary greatly depending on the material and also on
the temperature. Typically, at low temperatures, they vary between a
nanometer for metals to a micrometer for quantum Hall systems. 

For mesoscopic wires, in general, quantum effects need to be taken into 
account. One way of computing these conductances is by using the quantum 
mechanical formulation of transmission and reflection through impurities and 
barriers. This formulation is called the Landauer-Buttiker formulation and 
works for Fermi liquids. However, it does not include interactions. But for 
one dimensional wires, interactions change the picture dramatically, since the 
quasi-particles are no longer fermion-like. Hence the Landauer-Buttiker 
formalism cannot be directly applied and one needs to compute
conductances in Luttinger wires taking interactions into account right
from the beginning. One way of doing this is by using bosonization 
and this is the method that we will follow here.

The aim is to compute the conductance of a one-dimensional
wire. First, we will compute the conductance through a clean wire ( no
impurities or barriers) and argue why the conductance is not
renormalized by the interaction. Then we will study the conductance
again after introducing a single impurity. Here, we will see that the
interactions change the picture dramatically. For a non-interacting
one-dimensional wire, from just solving usual one-dimensional quantum 
mechanics problems, we know that we can get both transmission and reflection
depending on the strength of the scattering potential. But for an
interacting wire, we shall find that for any scattering potential,
however small, for repulsive interactions between the electrons, there
is zero transmission and full reflection (implies conductance is zero,
or that the wire is `cut')
and for attractive interactions between electrons (which is of course
possible only for some renormalized `effective' electrons), there is
full transmission and zero reflection (implying perfect conductance
or `healing' of the wire). 

\vskip 0.3cm 
\noindent {\bf Ballistic conductor}

Let us first define the conductance of a mesoscopic ballistic conductor 
($i.e.$, a conductor with no scattering) without taking interactions 
into account. We said earlier
that the usual definition of conductance as $G=\sigma {W\over L}$ 
breaks down for mesoscopic systems. For instance, it is seen that 
instead of the conductance smoothly going down as a function of the area
or width of the wire $W$, it starts going down discretely in steps,
each of height $2e^2/h$. Also as $L$ decreases, instead of increasing
indefinitely, $G$ saturates at some limiting value $G_c$. The general
understanding now, is that as the wire becomes thinner and thinner,
the current is carried in a very few channels, each of them carrying a
current of $2e^2/h$ (two for spin degeneracy) 
until we reach the lowest value which is just a
single channel (which we interpret as the lowest eigenstate of the
transverse Hamiltonian) carrying this current. Moreover, as the length
decreases, the resistance does not decrease indefinitely but instead
reaches a limiting value. One way of understanding this is to simply
consider this to be a contact
resistance, independent of the length of the wire, which arises simply
because the conductor and the contacts are different. One cannot make
the contacts the same as the conductor, because then our assumption
that the voltage drop is across the conductor alone does not make
sense. That makes sense only if we assume that the contacts are
infinitely more conducting than the conductor. So we are finally left
with a non-zero resistance and the wire does not become infinitely conducting. 
In fact, in this limit, the conductance or resistance of the wire is
purely a 'boundary' property and the `conductivity' of the wire has no
real significance. In fact, whether we get a finite conductivity or
infinite conductivity depends on how one defines it. 

However, for a single channel wire, clearly, the wire is
one-dimensional and we know that interactions can change the picture
drastically. The question that we want to answer here is precisely
that. What is the conductance of a clean one-dimensional interacting 
wire or Luttinger wire? 

\vskip 0.3cm
\noindent{\bf Computing conductance of a clean one-dimensional
(mesoscopic) wire}

\vskip 0.3cm
\noindent{\bf (a)} Without leads

First, we shall perform a calculation to compute the conductance of a
Luttinger liquid without any consideration of contacts or leads.
(We shall restrict ourselves to spinless fermions since spin only
increases the degrees of freedom and gives an overall multiplicative
factor of two in the conductance). 
The conductance of a wire is calculated by applying an electric field
to a finite region $L$ of an infinitely long wire and the current $I$
is related to the field as
\beq
I(x)=\int_0^L dx' \int {d\omega\over 2\pi} ~e^{-i\omega
t}\sigma_{\omega}(x,x') E_{\omega}(x') ~,
\label{current}
\eeq
where $E_{\omega}(x')$ is the frequency $\omega$ component of the
time Fourier transform of the electric field. The conductivity
$\sigma_{\omega}(x,x')$, in turn, is related to the (imaginary time) 
current-current correlation function by the usual Kubo formula as
\beq
\sigma_{\omega}(x,x') = -{e^2\over {\bar\omega}} \int_0^{\beta} ~d\tau ~< 
T_\tau j(x,\tau)j(x',0)> e^{-i{\bar\omega}\tau} ~,
\label{sigma}
\eeq
where $\tau =it$, $\omega = i{\bar\omega} + \epsilon$, $T_\tau$ is the
(imaginary) time ordering operator and
$j(x,\tau)$ is the current operator. Both these
formulae are standard in many books \cite{MAHAN} on many body
techniques, so here we will confine ourselves to just describing what they
mean. The first equation describes the current as a
response to an electric field (externally applied plus induced) of frequency
$\omega$. The proportionality function is the conductivity. To get
the usual Ohmic formula, all we need to do is replace $\sigma =
\sigma_0\delta(x-x')$ or remember that the $\sigma(x,x')$ is generally
a function which is centered around $x \simeq x'$ and which falls off
sufficiently fast elsewhere. The point for mesoscopic systems is that
the length of the wire is roughly comparable with the range of
$\sigma(x,x')$. Hence, the current gets contributions from the
electric field all over the wire, which is different from what happens in 
the usual case, where the current at a point gets contributions only from the
electric field very near that point. The
second equation tells us that the conductivity is related to the
current-current correlation function. This is derived by computing the
current $I(x)$ in a Hamiltonian formulation to first order in the
perturbation which is the applied electric field. The Euclidean
formulation is used so that the generalization to 
finite temperature calculations is straightforward, but we shall only
work at zero temperature and hence take the $\beta\rightarrow \infty$ limit.

Our aim here will be to compute the current-current correlation
function and hence the conductance for a Luttinger wire using
bosonization. We shall denote the Euclidean time action of a generic
Luttinger liquid as 
\beq
S_E = {1\over 2K}\int d\tau \int dx ~[{1\over v}(\partial_\tau\phi)^2
+ v (\partial_x\phi)^2] ~.
\label{Eucaction}
\eeq
(Note that in the spin model and Hubbard model, 
$\tau$ was replaced by $it$).
The current can directly be expressed in terms of the boson operators as
\beq
j(x,\tau) \equiv v (\rhr -\rhl ) =- {i\over \sqpi}\partial_\tau \phi ~.
\eeq
(The extra factor of $i$ is because we are now using imaginary time $\tau$).
Our first step is to obtain the correlation function 
$<j(x,\tau)j(x',0)>$ which is similar to the correlation functions for
spinless fermions that we computed earlier when we were
studying spin models, except that we are now interested in the
Euclidean correlation functions. Since we can pull out the
$\partial_\tau$ outside the correlation function\footnote{See
R. Shankar in \cite{shan2} for subtleties in pulling out the derivative
outside the time ordering operator. One gets an extra term which
cancels another term that we have ignored here, a singular $c$-number
term.}, all we have to do is compute the propagator given by
\beq
G(\tau, x,x') = <T_\tau \phi(x,\tau)\phi(x',0)> ~,
\eeq
or equivalently, its Fourier transform 
\beq
G_\bo(x,x') =\int_0^\beta d\tau <T_\tau \phi(x,\tau)\phi(x',0)>
e^{-i\bo\tau} ~.
\eeq
The conductivity is then given by 
\bea
\sigma_{\bo}(x,x') &=& {e^2\over \bo \pi} \int_0^\beta d\tau 
<T_\tau \partial_\tau\phi(x,\tau)\partial_{\tau} \phi (x',0)> e^{-i\bo\tau}
\nonumber \\
&=& {e^2 \bo\over\pi}G_{\bo}(x,x') ~.
\eea
So now, to compute the conductance, all we have to do is compute the
propagator for the boson with a free Euclidean action. 
The propagator satisfies the equation
\beq
{1\over K}~(-v\partial_x^2 + {\bo^2\over v})G_\bo(x,x') = \delta(x-x') ~,
\eeq
from which upon integrating once, we get
\beq
{v\over K}\partial_x G(x,x')|_{x=x'-0}^{x=x'+0} = -1 ~.
\label{delt}
\eeq
The solution to the differential equation is given by
\bea
G(x,x') &=& Ae^{|\bo|(x-x')/v}, \quad x<x' \nonumber\\ 
&=& Ae^{-|\bo|(x-x')/v}, \quad x>x' ~.
\eea
Using this in Eq. (\ref{delt}), we see that 
\bea
({2\bo\over K}) A &=&1, \\
{\rm so ~that} \quad G_\bo(x,x') &=& {K\over 2\bo}e^{|\bo|(x-x')/v},
\quad x<x' \nonumber \\ &=& {K\over 2\bo}e^{-|\bo|(x-x')/v},
\quad x>x' \\
{\rm leading ~to}\quad \sigma_{\bo}(x,x') &=& {Ke^2\over 2\pi}
e^{|\bo|(x-x')/v},
\quad x<x' \nonumber \\ &=& {Ke^2\over 2\pi}e^{-|\bo|(x-x')/v},
\quad x>x' ~.
\eea
The point to note is that in the $\bo\rightarrow 0$ limit or static limit, 
the conductivity is finite and does not drop down to zero even for large
$|x-x'|$. This is the main
difference from macroscopic conductivities which always decay to zero
as $|x-x'|\rightarrow \infty$. Furthermore, for $x=x'$, even for
arbitrary $\bo$, the Green's function has a finite value, which is
responsible for the saturation value of the conductance. This only
happens for a one-dimensional Green's function. In any other
dimension, the Green's function and hence conductivity 
will be divergent at $x=x'$. Using this in the
equation for the current, Eq. (\ref{current}) for a static electric field
$\bar E_\omega (x)= 2\pi \delta(\omega) E(x)$, we finally get 
\beq
I(x) = {Ke^2\over 2\pi} \int_0^L dx' ~E(x') = {Ke^2\over 2\pi}(V_L-V_0) 
\eeq
which gives the final result for the conductance as
\beq
g= {Ke^2 \over 2\pi} ~.
\label{nolead}
\eeq
There are several subtle points to note in this calculation. One is that
we have taken the $\omega\rightarrow 0$ before $|x-x'|\rightarrow \infty$, 
which is opposite to the usual order of limits in the Kubo formula. The
physical justification for the usual order of limits in the Kubo
formula comes from the fact that if we first take $\omega$ to zero,
then we have a static electric field, which is periodic in space. This means 
that the charge will seek an equilibrium distribution after which there 
will be no flow of current. Setting $|x-x'| \rightarrow \infty$, 
on the other hand, means taking the
thermodynamic limit or infinite length limit first, which allows for
an unlimited supply of electrons and is probably equivalent to having
reservoirs even if we have do not really have infinite length wires. For the 
mesoscopic systems, however, it is not correct to take the thermodynamic limit
first. The physical situation here, is that one applies a static
electric field to a finite length of the wire $L$, which in fact, is
comparable to the range of the conductivity. If we take the
$|x-x'|\rightarrow \infty$ limit first, then it is as if we are looking
at a long length of the wire beyond the range of the conductivity. This
is the usual limit and we will get the usual Ohm's law, which however,
is wrong in this context. In fact, it is
instructive to try out the calculation with the other order of limits
-$i.e.$ by computing $\sigma_{\bo}(q)$ and
taking the $q\rightarrow 0$ limit first. \\
(H.W. Exercise 9: Try the above).
 
The second point is something we have mentioned earlier - $i.e.$, we have 
not taken contacts or leads into account. This was the initial computation by
Kane and Fisher \cite{KF} and they obtained the answer in
Eq. (\ref{nolead}) that the conductance of the clean wire depends on
the interaction parameter $K$.

\vskip 0.3cm
\noindent{\bf (b)} Including leads

When any experiment is done, however, one does have explicit contacts
or leads. In fact, when a measurement was actually done under conditions
where one expected to measure the Luttinger parameter $K$, it was
found to $\simeq 1$, instead of 0.7, which was expected from other
measurements of the $K$ value of the wire. (We will see how else $K$
can be measured after considering impurity scattering). So we need
to understand what happens when we actually try to measure the
conductance of a Luttinger wire.

How do we model the leads? The simplest model to consider is
that the Luttinger wire is connected to Fermi liquid leads on either
side. (See Fig. 6). So the regions A and C can be modelled by the same 
bosonic model with $K_L=1$ and the wire in region B can be modelled as before
as a Luttinger wire with $K=K$. But now, we have to put appropriate
boundary conditions at the points P and P' between A and B and 
between B and C respectively. Note that we are making the assumption
that one has the same $\phi$ field or same quasiparticle in all the 
three regions and it is only the LL parameters which are
changing. Although, it is interesting to compute the conductance in
this case, it is still not clear that this brings the calculation any
closer to real experiments, because real experiments will have three
dimensional reservoirs. 

We start with the action in all the three regions in Euclidean space as
\beq
S_E = {1\over 2}\int_0^\beta d\tau \int_0^L dx 
~[{(\partial_\tau\phi)^2\over
K(x)v(x)} + {v(x)\over K(x)}(\partial_x\phi)^2] ~,
\label{action}
\eeq
with $K(x)=K_L$, $v(x)=v_L$ in regions A and C and $K(x)=K$, $v(x)=v$
in region B. This is just the free action of a scalar field in all the three
regions. Fourier transforming the imaginary 
time variable with respect to $\bo$, we obtain 
\beq
S_E = {1\over 2}\int_0^\beta d\tau \int_0^L dx 
~[{\bo^2\phi^2\over K(x)v(x)} + 
{v(x)\over K(x)}(\partial_x\phi)^2] ~,
\eeq
from which we see that the propagator satisfies the equation 
\beq 
\{ -\partial_x({v(x)\over K(x)}\partial_x) + {\bo^2\over
K(x)v(x)} \} ~G_\bo(x,x') = \delta(x-x') ~.
\eeq
Now let us consider the four regions. We assume that the interaction parameter 
changes abruptly at P and P', but that the Green's function is 
continuous and the derivative of the Green's function has the correct 
discontinuity at all the boundaries. So now, we need to solve the Green's 
function equation subject to these boundary conditions. Let us choose $x'$ to 
lie between $0$ and $L$. It is then easy to see that the solution is of the 
form
\bea
G_\bo(x,x') = && A e^{|\bo|x/v} \quad\quad {\rm for} ~~ x\le 0 \nonumber\\
= &&Be^{|\bo|x/v} + Ce^{-|\bo|x/v} \quad {\rm for} ~~ 0< x\le x' \nonumber\\
= &&De^{|\bo|x/v} + Ee^{-|\bo|x/v} \quad {\rm for} ~~ x'< x\le L \nonumber\\
= &&Fe^{-|\bo|x/v}\quad \quad {\rm for} ~~ x> L 
\eea 
for semi-infinite leads, because we have assumed that the lengths of
the leads are sufficiently long compared to $L$ so that we do not need
to put any further boundary conditions on them. Note that here the
Green's functions will no longer be functions of $x-x'$ since we have
explicitly broken translational invariance. The constants $A, B, ..., F$ are 
found by matching the boundary conditions. Since we are interested in the DC
conductance, we only need the solutions for
$\bo\rightarrow 0$ which are easy to obtain and are given by
\bea
A = F = {K_L\over 2\bo},\quad B=E={K_L+K\over 4\bo}, \quad 
C=D={K_L-K\over 4\bo} ~.
\eea
From this, we see that $\sigma_\bo(x,x')$ is $x$ and $x'$ independent in
the $\bo\rightarrow 0$ limit and is equal to $K_Le^2/2\pi$ in all regions 
from which we find the conductance (using Eq. (\ref{current})) given by 
\beq
g\equiv {I\over V} = {K_Le^2\over 2\pi} ~.
\eeq
is the same in all the
regions. Thus, the conductance is determined by the $K_L$ of the leads, which
is just $K_L=1$ for Fermi liquid leads and does not depend on
interactions in the wire. This is a highly counter-intuitive
answer! It is telling us that whether we measure the conductance in
the leads or in the quantum wire, we get the same answer, so long as
we take into account the fact that we are attaching leads, which allow
for the fermions to enter and leave the quantum wire. At a very
naive level, one may understand this by saying that 
since the wire itself has no impurities, the only
source of resistance is the contact effect between the leads and the
wire, which has nothing to to do with the interactions in the wire. However,
remember that we have taken semi-infinite leads and abrupt contacts
and we are only looking for DC conductance. If any of these
assumptions are relaxed, certainly, there are differences in the three
regions and one could get more interesting answers. 

In fact, using a Landauer-Buttiker scattering approach \cite{SAFI}, 
there has been some attempt to understand these results more intuitively. 

Physically, the difference between this computation and the earlier one 
is that any real measurement requires Fermi liquid leads.
So the end result is that the measurable conductance of an interacting
one-dimensional wire is simply given by $g=e^2/h$ for spinless
fermions and $g=2e^2/h$ for fermions with spin \cite{MS}.

\section{Transport in the presence of isolated impurities}

\noindent {\bf Computing conductance with a single impurity}

Now let us consider the case when there is a single impurity at the
origin. At first, we will model the impurity as a weak barrier and add
a term to the action of the form
\beq 
S_{\rm int} = \int dx d\tau ~V(x)\psi^{\dagger}(x)\psi(x) ~.
\eeq 
We assume that $V(x)$ is weak and is centred around the origin. For
instance, we can choose $V(x) = \lambda\delta(x)$, where $\lambda$ is
much less than the Fermi energy. 

First, let us think of what happens when we introduce such a
perturbation in a non-interacting wire. In that case, all one has is a
one-dimensional quantum mechanics problem with a $\delta$-function
potential at the origin. We can find the reflection and transmission
probabilities for a single particle with momentum $k$ as 
\beq
R= {\lambda^2\over \lambda^2+k^2}\quad {\rm and} \quad T={k^2\over
\lambda^2+k^2} ~.
\eeq
So for any $\lambda$, one gets both reflection and transmission. To
get the total current, we just have to sum up the contributions of
all the electrons close to the Fermi surface. But it is clear
that there will be non-zero conductance for any potential, with the amount
of current being transmitted depending on the strength of the potential. 
However, for the Luttinger wire, since there exists
interactions between electrons in the wire, and no convenient quasiparticle
picture, one cannot solve the
problem this way. We have to use the bosonized field theory and
include the impurity potential as a perturbation.

Let us first rewrite the impurity potential in terms of the left- and
right- moving low energy Dirac modes. We find that
\bea
\psi^{\dagger}(x)\psi(x) &=& (\psi_R^{\dagger} e^{-ik_Fx} + \psi_L^{\dagger}
e^{ik_Fx}) (\psi_R e^{ik_Fx} + \psi_L e^{-ik_Fx}) \nonumber \\
&=& \psi_R^{\dagger}\psi_R +\psi_L^{\dagger}\psi_L + e^{-i2k_Fx}
\psi_R^{\dagger}\psi_L + e^{i2k_Fx}\psi_L^{\dagger}\psi_R \nonumber \\
&=& - {1\over\sqpi}\partial_x\phi + {1\over 2\pi\alpha} (\eta_R^\dagger\eta_L
e^{i\sqfp(\phi_R+\phi_L +2k_Fx)}+\eta_L^\dagger\eta_R
e^{-i\sqfp(\phi_R+\phi_L +2k_Fx)}) , \nonumber \\
& &
\label{imp} 
\eea
where the last line is obtained using standard bosonization.
So the full action is given by 
\beq
S=S_E+S_{\rm int} = S_E -{\lambda\over\sqpi} \partial_x \phi (0) + 
{\lambda\over 2\pi\alpha}\int d\tau \cos \sqfp [\phi_R(0)+\phi_L(0)] ~,
\eeq
where $S_E$ is given in Eq. (\ref{Eucaction}) and 
we have incorporated the fact that the potential only acts at
the origin. Moreover, we have simply set both $\eta_R^\dagger\eta_L$ 
and $\eta_L^\dagger\eta_R$ to be one, with the knowledge that in
correlation functions, we will compute $O(\tau)O^{\dagger}(0)$ so
that the Klein factors disappear using $\eta_{R/L}\eta_{R/L}^\dagger =
\eta_{R/L}^\dagger\eta_{R/L} =1$. The first term due to the interaction can 
be taken care of by a simple redefinition of $\partial_x\phi \rightarrow 
\partial_x\phi' = \partial_x\phi+\lambda/2\sqpi$, which makes no difference 
to the conductance. This could have been seen even from the fermion terms
from which it came. The $\psi_R^{\dagger}\psi_R
+\psi_L^{\dagger}\psi_L$ term only causes scattering at the same Fermi
point with momentum transfers $q<<2k_F$. This does not change the direction 
of propagation of the particles and hence does not affect conductance in any 
appreciable way. But the cosine term, on the other hand, occurs because of
backscattering of fermions from the origin. These represent
scattering with $q\sim |2k_F|$ - $i.e.$, from the left branch to the
right branch and vice-versa and change the direction of propagation of the 
particles. These scatterings will definitely affect the conductance. The 
action with this perturbation is no longer quadratic and cannot be exactly 
solved. However, since $\lambda$ is a weak perturbation, one can try to use
perturbation theory and the renormalization group approach to see the
relevance of this perturbation at low energies.

What is the question that we want to answer?
We want to compute the conductance through this barrier at low
energies. One way to do that is to see whether this barrier coupling 
strength grows or becomes smaller as we go to lower energy scales. To check
that, we need to perform the usual steps of a renormalization group analysis. 

Here since the perturbation term is fixed in space, it is more convenient
to first integrate out the variables away from the origin and write down the
action purely in terms of the $\phi(x=0,\tau)$ variables. Since integrating 
out quadratic degrees of freedom is equivalent to using equations of
motion for those degrees of freedom, we write down the equations of
motion for the action $S_0$ as
\beq
\partial_x^2\phi -{\bo^2\over v^2} \phi = 0 \Rightarrow 
\partial_x^2\phi -k^2 \phi = 0 ~.
\eeq
The solution to the above equations are given by
\bea
\phi &=& Ae^{|k|x}, \quad x>0 \nonumber \\
 &=& Ae^{-|k|x}, \quad x<0 ~,
\eea
where $A\equiv \phi(x=0,\tau)$. Using this solution in the action, we
get the effective action in terms of $\phi(\bo) =\int
\phi(x=0,\tau)e^{i\bo\tau}d\tau$ as
\bea
S_{\rm eff}&=& {1\over 2K}\int \frac{d\bo}{2\pi}
\int_{-\infty}^0 dx ~[v\phi^2k^2e^{2kx}+
{\bo^2\over v}e^{2kx}] + {1\over 2K}\int \frac{d\bo}{2\pi} \int_{0}^\infty dx~
[v\phi^2k^2e^{-2kx}+{\bo^2\over v}e^{-2kx}] \nonumber \\
&=& {1\over 2K}\int \frac{d\bo}{2\pi} ~{2\phi^2\bo^2\over v} ~[~{e^{2kx}\over
2k}|_{-\infty}^0 + {e^{-2kx}\over -2k}|_0^\infty ~] \nonumber \\
&=& {1\over K}\int \frac{d\bo}{2\pi} ~|\bo| \phi^2
\eea
using $k = |\bo|/v$. 
Notice the singular dependence on the Matsubara frequency $|\bo|$.
The reason for its appearance is the following. In real space, even for
a quadratic action, all degrees of freedom (dof) are coupled. ( It is only
in Fourier space that every mode is decoupled). So when we integrate
out all dof except the one at the origin, the dispersion of this degree of 
freedom can change and has changed. This is why we
get the modulus factor in the effective action. When we Fourier
transform back to imaginary time, we get
\bea
\sum_{\bo} |\bo| \phi_n^2 &\rightarrow& i\int \frac{d\omega}{2\pi} \int d\tau
\int d\tau' ~e^{i\omega(\tau-\tau')} |\omega_n| 
\phi^*(\tau) \phi(\tau') \nonumber \\
&=& - \int d\tau d\tau' {2\over (\tau-\tau')^2} \phi^*(\tau) \phi(\tau') ~,
\eea
$i.e.$, an explicitly non-local interaction in imaginary time.

So now, we have an action solely in terms of the variables at the
origin with the action given by
\beq
S= {1\over 2K}\int \frac{d\bo}{2\pi} ~|\bo|\phi(\bo)^2 + \lambda\int 
\frac{d\bo}{2\pi} ~\cos [2 \sqrt{\pi}\phi(\bo)] ~. 
\eeq
The RG analysis now involves finding out how the coefficient $\lambda$
behaves as we go to lower and lower energies. 
Before we perform the RG analysis, we may ask why would we want to go
to lower energy scales? The general idea is that in spite of the fact
that in different physical problems or models, the parameter $\lambda$
may be slightly different, qualitatively many such models may have
the same behavior. This is because they are all governed by the same
fixed point Hamiltonian with the fixed point Hamiltonian being defined
as the Hamiltonian one gets when the RG flow stops. So the aim is to
keep reducing the energy scale till the RG flow stops so that we can
find out the appropriate fixed point Hamiltonian for this model.

In this problem, we want to find out whether the fixed point
Hamiltonian has a large barrier or a small barrier. To find that out,
let us perform the three steps of the renormalization group
transformation. We choose a high frequency cutoff $\Lambda$, which is
the real physical cutoff of the theory. Then we
rescale $\Lambda \rightarrow \Lambda/s$ with $s>1$ and then 
divide $\phi(\bo)$ into $\phi_<(\bo)$ (slow modes) 
and $\phi_>(\bo)$ (fast modes) for
the modes with frequencies less than or greater than the cutoff $\Lambda/s$
respectively. Finally, we integrate out the fast modes, which are the modes
between $\Lambda/s$ and $\Lambda$ and rescale $\bo\rightarrow \bo' =
s\bo$ or $\tau \rightarrow \tau'=\tau/s$ to get back to the original
range of integrations. To lowest order, (tree level contribution) we find that
\beq
\lambda \int d\tau \cos\sqfp\phi_<(x=0,\tau) \rightarrow 
\lambda s^{1-d}\int d\tau\cos\sqfp\phi_<(x=0,\tau) ~,
\label{oned}
\eeq
where $d$ is the dimension of the cosine operator. This was
explicitly computed earlier and we found that $d=K$. The RG equation is 
now easily obtained by taking $s=1+dl$, for infinitesimal $dl$. We find 
that the new $\lambda'$ after rescaling is given by
\bea
&&\lambda' = \lambda(1+dl)^{1-K} \nonumber\\
\Rightarrow &&\lambda'-\lambda = (1-K)\lambda dl \nonumber \\
\Rightarrow &&{d\lambda\over dl} = (1-K)\lambda ~.
\eea
Normally, one would have had coupled RG equations for $\lambda$ and
$K$. But here since $1/K$ is the coefficient of a singular operator, it
does not get renormalized to any order.

Notice that in Eq. (\ref{oned}), the coefficient of the operator gets
rescaled by a factor $s^{1-d}$ rather than $s^{2-d_i}$ as we had
mentioned earlier when we computed the dimension of the cosine
operator in the spin model. The difference is that the operator in the
spin model, in the action, required integration over both space and
time. So we rescaled both the space and time ( or equivalently both
the momentum and the energy). However, in this case, the operator
exists only at a fixed space point. So we only need to integrate over
the time coordinate. Hence, the naive scaling dimension or engineering
dimension of the operator is 1 and not 2. Such operators are called boundary
operators. You will learn more about them in the course on boundary
conformal field theory. 

The RG equation is now trivial to analyze. For any $K>1$, (which
corresponds to attractive interactions between the electrons), the
$\lambda$ renormalizes to zero and for any $K<1$, (corresponding to
repulsive interactions), it grows stronger and stronger. In other
words, for $K>1$, the fixed point Hamiltonian is just the free boson
Hamiltonian with no barrier and for $K<1$, the fixed point Hamiltonian
has two disconnected wires to the left and right of the origin. For $K=1$,
which is the limit of no interactions in the fermionic model, the
coupling is marginal. (This was expected, since we know that for free
fermions, both transmission and reflection occurs depending on the
strength of the barrier potential). Thus, for attractive interactions,
the barrier renormalizes to zero and the wire is `healed', whereas for
repulsive interactions, the barrier renormalizes to infinity and the
wire is `cut'. Note that both these answers are completely independent
of the strength of the barrier potential \cite{KF}.

\vskip 0.3cm
\noindent{\bf Strong barrier limit}

Since, we are doing perturbation theory, we cannot assume that this
result holds for arbitrary $\lambda$. It is strictly valid only for
$\lambda \simeq 0$. Once $\lambda \sim 1$, the perturbative analysis in 
$\lambda$ breaks down. So it is worthwhile to try and
see what happens in the other limit. Supposing we start with two
decoupled wires and then allow a small hopping between the two
wires. Will this hopping grow 
at low energies and heal the wire or will it renormalize to zero?

Here, we start with two semi-infinite Luttinger liquid wires and
analyze the effect of adding a small hopping term coupling the two
wires at $x=0$. The models for $x<0$ and $x>0$ are given by the action
\beq
S_E = {1\over 2}\int_0^\beta d\tau \int_0^L dx ~[{(\partial_\tau\phi_i)^2\over
K(x)v(x)} + {v(x)\over K(x)}(\partial_x\phi_i)^2]
\eeq
for $i=<$ and $i=>$ respectively. We can also write it in terms of the
dual variables as 
\beq
S_E = {1\over 2}\int_0^\beta d\tau \int_0^L dx ~[{ K(x)\over 
v(x)}(\partial_\tau\theta_i)^2 + K(x)v(x)(\partial_x\theta_i)^2] ~.
\eeq
Note that in terms of the dual variables, the action has $1/K$ in
position of $K$. This is because the roles of the fields and the
canonically conjugate momenta have interchanged. The fact that the
wire is cut implies that at the point $x=0$, there is zero density of
either $<$ or $>$ particles - $\psi_<^\dagger\psi_<(x=0)=0$ and
$\psi_>^\dagger \psi_>(x=0)=0$. In the bosonic language, this is
imposed as $\sqfp\phi_<(x=0)=\sqfp\phi_>(x=0) =\pi/2$ (and also
$\partial_x\phi(x=0)=0$ as can be seen from Eq. (\ref{imp})). Now a term 
which hops an electron from one wire to another in the Hamiltonian is just
\bea
H &=& -t~[\psi_<^{\dagger}\psi_> + h.c.] \nonumber\\
&=& -t~[\psi_{R<}^{\dagger}\psi_{R>} +\psi_{L<}^{\dagger}\psi_{L>}
+\psi_{R>}^{\dagger}\psi_{R<} +\psi_{L>}^{\dagger}\psi_{L<} \nonumber\\
&& ~~~~+\psi_{L<}^{\dagger}\psi_{R>} +\psi_{L>}^{\dagger}\psi_{R<}
+\psi_{R<}^{\dagger}\psi_{L>} +\psi_{R>}^{\dagger}\psi_{L<}] ~,
\eea
where the second equation involves the left and right moving
fields and we have already set $x=0$. 
Here, again, the terms that involve fields at one Fermi point
are low energy forward scattering terms which do not affect the
conductance. In terms of the bosonic fields too, they can be taken care
of by trivial redefinitions. But the intra-Fermi point scatterings 
which will affect the conductance 
can be bosonized and written in the action as
\bea
\delta S = -t\int d\tau ~[ && \eta_{L<}^{\dagger}\eta_{R>}
e^{-i(\phi_{L<} +\phi_{R>})} + \eta_{L>}^{\dagger}\eta_{R<}
e^{-i(\phi_{L>} +\phi_{R<})} \nonumber \\ 
&+& \eta_{R<}^{\dagger}\eta_{L>}e^{-i(\phi_{R<} +\phi_{L>})} + 
\eta_{R>}^{\dagger}\eta_{L<}e^{-i(\phi_{R>} +\phi_{L<})} ] ~.
\eea
Now, we impose the boundary condition on the bosonic fields that we
mentioned above, which constrains $\phi(0) = 
\phi_R(0)+\phi_L(0)$ to be equal to
$\pi/2$. Using this, we can express the above equation solely
in terms of the $\phi_L(0)-\phi_R(0)=\theta(0)$ fields and get 
\beq
\delta S= 4t\int d\tau ~\cos(\theta_{>}-\theta_{<}) ~,
\eeq
where once again, we have been able to drop the Klein factors after
checking that they do not lead to any extra minus signs in the correlation 
functions. (Physically, the reason why we only get the $\theta_i$ term at the
origin is because the constraint has set $\phi_i(x=0)=\pi/2$).
Computing the dimension of this operator, we see that to leading order,
the RG equations are given by
\beq
{dt\over dl} = (1-{1\over K})t ~.
\eeq
($K$ has been replaced by $1/K$ because we now have to compute the dimensions
in the dual action).
Thus, for repulsive interactions ($K<1$), the hopping term is
irrelevant and flows to zero. This confirms the weak barrier
calculation that the wire is insulating. On the other hand, for
attractive interactions, the hopping strength grows, ultimately healing 
the wire. This again is in accordance with the weak coupling analysis. 

\vskip 0.3cm
\noindent{\bf Intermediate fixed points?}

We have started from a wire with a weak barrier and shown that under
repulsive interactions, the barrier strength grows. We have also
started from two decoupled wires and shown that for repulsive
interactions, any small hopping term renormalizes to zero. Hence, it
seems plausible to conclude that for repulsive interactions in the
wire, any barrier will cut the wire and the conductance goes to
zero. However, one should keep in mind that our analysis is strictly
true only for $\lambda, t \simeq 0$. Hence, it could happen that for
intermediate values of the barrier strength, one could have a pair of
non-trivial fixed points (see Fig. 7).

\vskip 0.3cm
\noindent{\bf Conductance at finite voltage and temperature}

The earlier analysis only tells us how the barrier strength or the
tunneling amplitude grows or falls as we go to low energies. But instead of
allowing the energy scale to become arbitrarily low, 
we can cut off the energy scale of
renormalization at some finite energy scale, which could be the
temperature $T$ or the voltage $V$. Note that the energy scale at
which we want to cutoff the integral is related to the initial high
energy scale at which we start the RG as $E=E_0e^{-l}$. 
So for attractive interactions for which weak barriers are irrelevant
and for which one would expect perfect transmission at very low
energies will have power law corrections when we put the
lower energy cutoff as $E$. In that case, we have 
\beq
\int_{\Lambda_0}^{\Lambda}{d\lambda\over \lambda} = 
\int_0^{ln (E_0/E)} ~dl (1-K) ~,
\eeq 
which means that the effective barrier strength $\Lambda$ 
is proportional to $\Lambda_0 (E/E_0)^{K-1}$.
So by choosing $E=T,V$, we see that one can get power law
corrections to the naive conductance at $T\rightarrow 0, V\rightarrow
0$. In other words, if we measure the conductance at a finite
temperature $T$, rather than at $T=0$, instead of zero conductance for
$K<1$, we will get conductances which go as $T^{1-K}$ (roughly the
inverse of the barrier strength). Similarly, if instead of measuring
conductances as $V\rightarrow 0$, we measure them at finite voltages,
we find that the conductances go as $V^{1-K}$. On the other hand, for
repulsive interactions, we need to start at the strong coupling limit
with two decoupled Luttinger wires and allow for a small hopping,
which is irrelevant in the RG sense. Here, again, if we cutoff the
lower energy scale at $E$, we expect instead of zero transmission,
power law corrections of the form $I \sim V^{1-1/K}$ and $I\sim
T^{1-1/K}$. The only difference in the analysis at the strong coupling
fixed point and the weak coupling fixed point is that $K$ gets replaced by 
$1/K$ as we saw in the RG equations. This, in fact, is one way in which $K$ 
can be measured in experiments. They could explicitly make a constriction
in the quantum wire and measure conductances through it and extract $K$.
 
\section{Concluding Remarks}

Almost any interacting quantum system in one dimension
which is gapless and has a linear dispersion for the low-energy excitations
can be described as a Luttinger liquid at low energies and long wavelengths.
As we have seen, the properties of a Luttinger liquid are 
determined by the two parameters $v$ and $K$. These in turn depend 
on the various parameters which appear in the microscopic Hamiltonian of the 
system. Some examples of systems where Luttinger liquid theory and
bosonization can be applied are quantum spin chains (including some spin 
ladders), quasi-one-dimensional organic conductors and quantum wires (with
or without impurities), edge states in a fractional quantum Hall system, and 
the Kondo problem. Some of these examples have been discussed above.

Antiferromagnetic spin-$1/2$
chains have a long history going back to their exact solution by the Bethe
ansatz. In recent years, many experimental systems have been studied which
are well-described by quasi-one-dimensional half-odd-integer spin models with 
isotropic (Heisenberg) interactions. Such systems behave at low energies as a
$K = 1/2$ Luttinger liquid with an $SU(2)$ symmetry. It seems to be difficult 
to vary $K$ experimentally in spin systems. In contrast, a single-channel 
quantum wire (which is basically a system of interacting electrons which are 
constrained to move along one particular direction) typically 
has two low-energy sectors, both of which are Luttinger liquids (except at 
special densities like half-filling). One of these is the spin sector
which again has $K = 1/2$. The other one
is the charge sector whose $K$ value depends on a smooth way on the
different interactions present in the system. Finally, the edge states in
a fractional quantum Hall system behave as a chiral Luttinger liquid with 
$K$ taking certain discrete rational values; the value of $K$
can be changed by altering the electron density and the magnetic field in
the bulk of the system. For all these systems, many 
properties have been measured such as the response to external electric and 
magnetic fields (conductivity or susceptibility) and to disorder, scattering 
of neutrons or photons from these systems, and specific heat; so the 
two Luttinger parameters can be extracted from the experimental data.
The measurements clearly indicate the Luttinger liquid-like
behavior of these systems with various critical exponents depending in a 
non-universal way on the interactions in the system. 

On the theoretical side, a large number of exactly solvable models in one 
dimension have been shown to behave as Luttinger liquids at low energies
\cite{hald,kawa}. These include

\noindent (i) models with short range interactions which are solvable by the
Bethe ansatz, such as the $XXZ$ spin-$1/2$ chain (where $K$ can take a range 
of values from $1/2$ to $\infty$; this includes the $XY$ model with $K=1$ and 
the isotropic antiferromagnet with $K=1/2$ as special cases), and the 
repulsive $\delta$-function Bose gas (where $K$ can go from $1$ in the
limit of infinite repulsion to $\infty$ in the limit of zero repulsion), and

\noindent (ii) models with inverse-square interactions such as the 
Calogero-Sutherland model (where $K$ can go from $0$ to $\infty$) and the 
Haldane-Shastry spin-$1/2$ model (where $K=1/2$).

\noindent The models of type (ii) are {\it ideal} Luttinger liquids in 
the sense that they are scale invariant; the coefficients of all the 
marginal operators vanish, and therefore their correlation functions and
excitation energies contain no logarithmic corrections. This property makes
it particularly easy to study these systems numerically since the
asymptotic behaviors are reached even for fairly small system sizes.

\vskip 0.3cm
\noindent{\bf What has been left out?}

Finally, let us mention the various important things in this field
which has been left out. We have only worked with spinless fermions in
the transport analysis. When we include spin and do not destroy the
$SU(2)$ spin symmetry of the system, the results are very similar to
the spinless fermion case. For repulsive interactions, the barrier
becomes infinite and for attractive interaction, the barrier is
healed. However, when the $SU(2)$ symmetry is destroyed, there exists
possibilities of intermediate (non-trivial) fixed points where either spin 
or charge can be transmitted and the other reflected. The other thing that has
been left out is the phenomenon of resonant tunneling with two
impurities. This is an interesting result, because it says that for
repulsive interactions, a single impurity cuts the wire, but with two
impurities, one can have particular energies, where there can be
transmission. The reason, of course, is quantum mechanical
tunneling. Here, the energy levels, are the energy levels of the quantum
dot that is formed by the two impurities and one can have resonant
tunneling at these energy levels. If we include interactions between
the electrons on the island, (which is naturally included in the
bosonized formalism), we can obtain the physics of the Coulomb blockade. 
The other important thing that we have left out, from a physical point
of view, is what happens if there is a finite density of random
impurities. In general, one would expect Anderson localization and no
transport. But there are regimes of delocalization as well in the
phase diagram. Finally, a very important application where the physics
of the Luttinger liquids has actually been experimentally seen is in
the edge states of the fractional Quantum Hall fluid. Since here, the
edge states are chiral, a lot of the complications of backscattering
due to impurities are avoided and it is possible to explicitly
construct constrictions and allow tunneling through them. Here, both
at the theoretical and experimental level, there are a lot of
beautiful results that are worth understanding. 

Another important topic not covered here is non-abelian bosonization
\cite{gogo}. This is a powerful technique for studying one-dimensional 
quantum systems with a continuous global symmetry such as $SU(2)$. For 
instance, isotropic Heisenberg antiferromagnets and Kondo systems are
invariant under spin rotations, and they can be studied more efficiently
using non-abelian bosonization.

To conclude, let us just say that low dimensional systems and
mesoscopic systems have gained in importance in the last few years. Although 
currently, much of the theoretical work in mesoscopic systems has only
involved conventional Fermi liquid theories, it is clear that there
are regimes where strong interactions are very important. We expect
that bosonization will be one of the important non-perturbative tools
to analyze such problems for a few more years to come. 

\vskip .7 true cm
\noindent {\bf Acknowledgments}
\vskip .3 true cm

DS thanks J. Srivatsava for making Figures 1 to 5.

%\vskip 2 true cm
\newpage

\noindent {\bf Figure Captions}
\vskip .5 true cm

\noindent {1.} One-particle momentum distribution function. (a) shows
the finite discontinuity at the Fermi momentum $k_F$ for a system of
interacting fermions in more than one dimension. 
(b) shows the absence of a
discontinuity in an interacting system in one dimension.

\noindent {2.} Picture of the Fermi sea of a lattice model; the momentum
lies in the range $[-\pi , \pi ]$. The occupied states (filled circles) 
below the Fermi energy $E_F =0$ and the two Fermi points at momenta 
$\pm k_F$ are shown.

\noindent {3.} The one-particle states of a right-moving fermion showing 
the occupied states (filled circles) below zero energy and the unoccupied 
states above zero energy.

\noindent {4.} Two possible particle-hole excitations of a right-moving 
fermionic system showing the occupied states.

\noindent {5.} The one-particle states of a left-moving fermion showing the 
occupied states below zero energy and unoccupied states above zero energy.
Note that the momentum label $k$ increases towards the left.

\noindent {6.} The single channel quantum wire with Fermi liquid
leads on the left and the right.

\noindent {7.} Renormalization group flow diagram for a quantum wire
with repulsive interactions in the presence of an impurity or barrier. 
In the absence of any non-trivial fixed points, the stable fixed point 
is the strong coupling fixed point. But perturbative analyses at the
strong and weak coupling fixed points cannot rule out a pair of non-trivial
fixed points at intermediate strengths of the barrier potential.

\newpage

\begin{figure}
\vspace*{5cm}
\begin{center}
\epsfig{figure=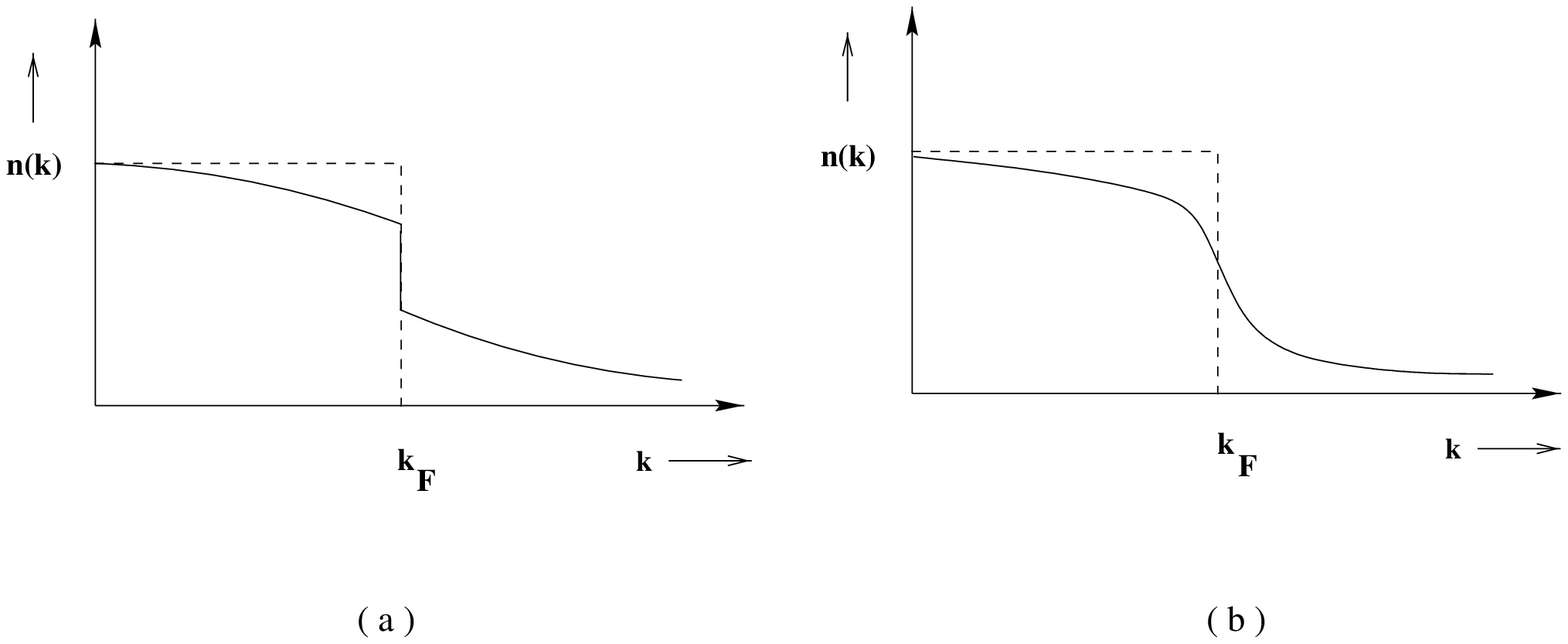,width=15cm,height=6cm}
\end{center}
\vspace*{0.5cm}
\centerline{Fig. 1}
\label{fig1}
\end{figure}

\begin{figure}
\begin{center}
\epsfig{figure=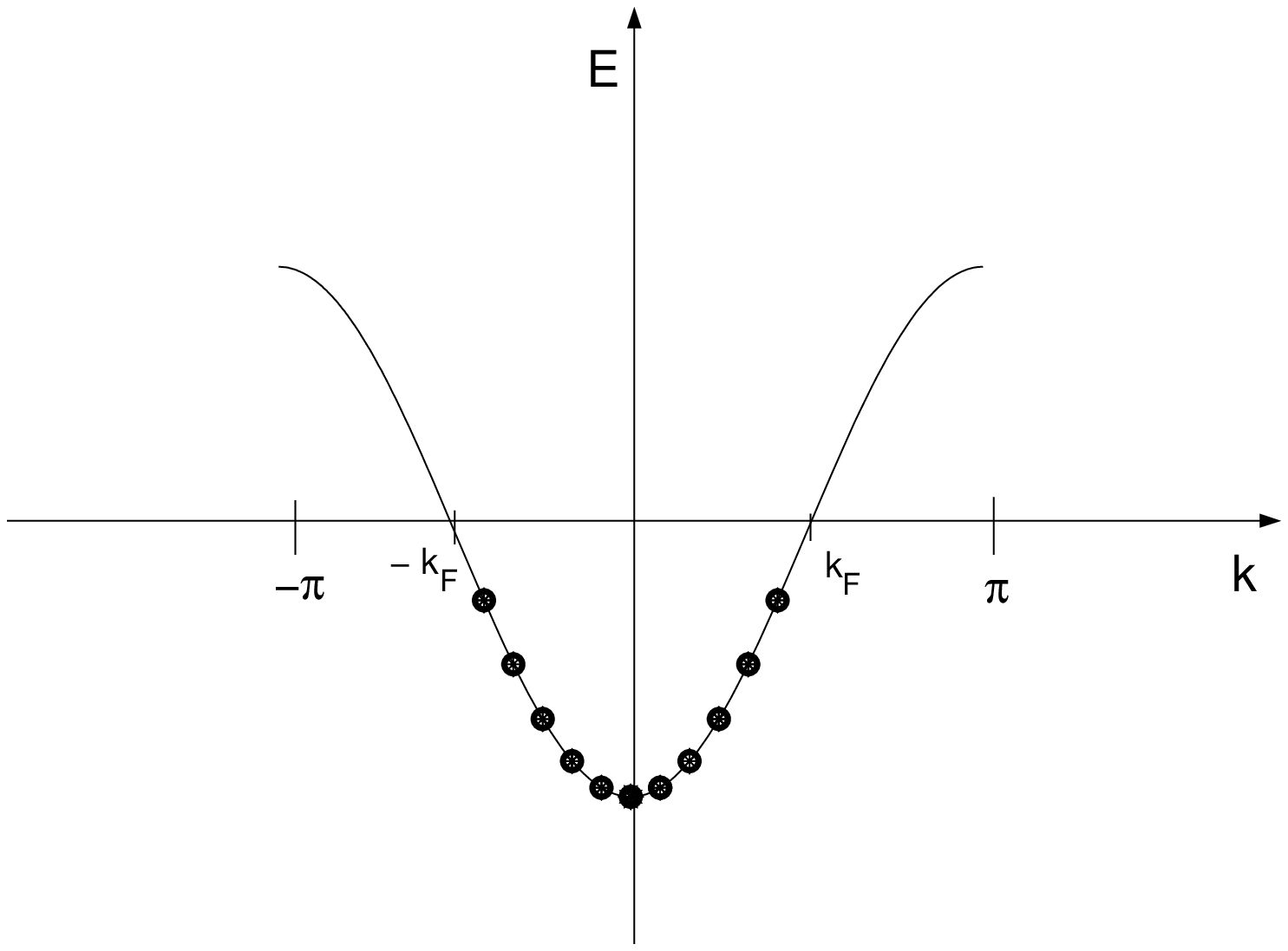,width=12cm}
\end{center}
\vspace*{0.5cm}
\centerline{Fig. 2}
\label{fig2}
\end{figure}

\begin{figure}
\begin{center}
\epsfig{figure=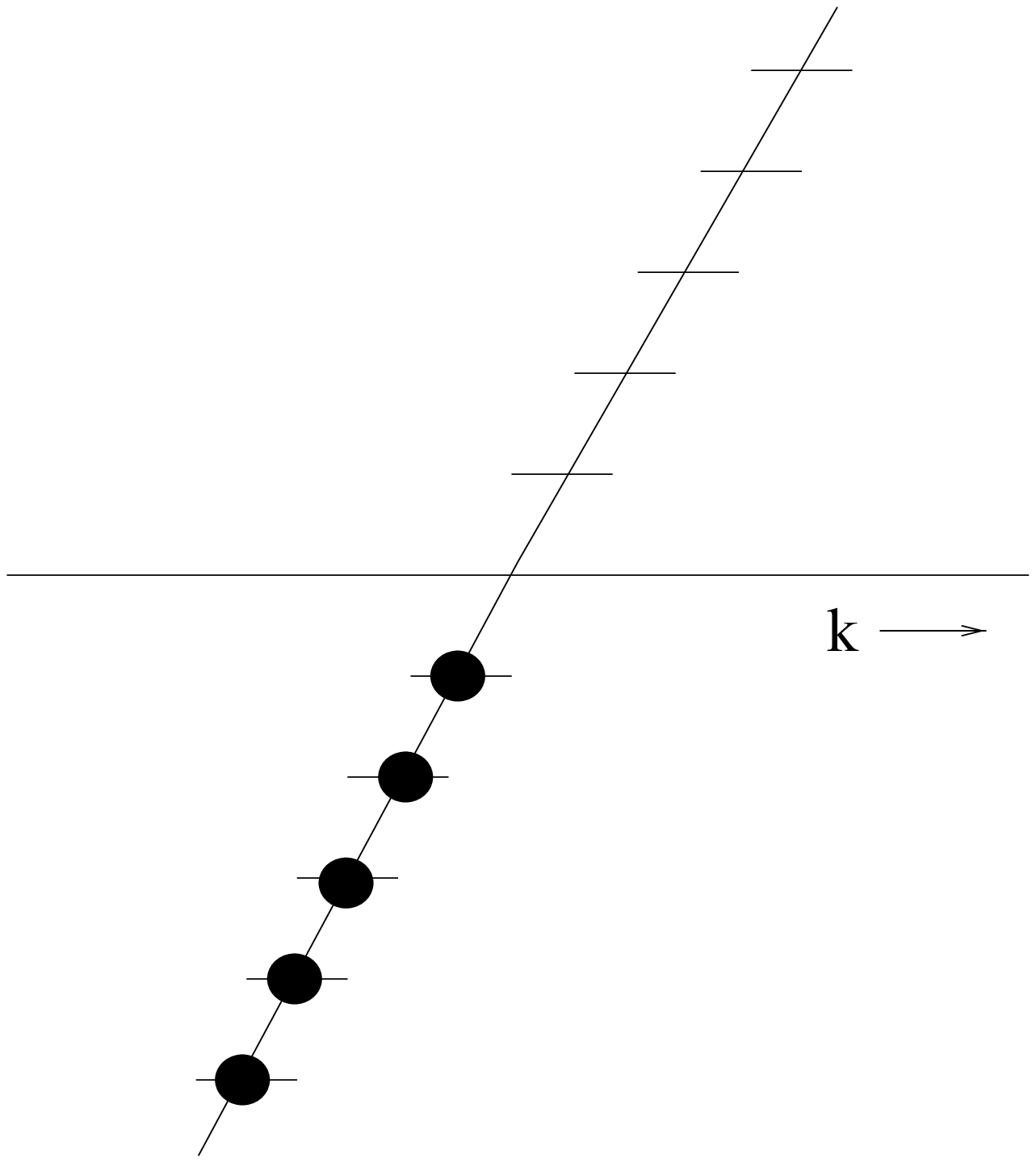,width=7cm}
\end{center}
\vspace*{0.5cm}
\centerline{Fig. 3}
\label{fig3}
\end{figure}
\vspace*{1cm}

\begin{figure}
\begin{center}
\epsfig{figure=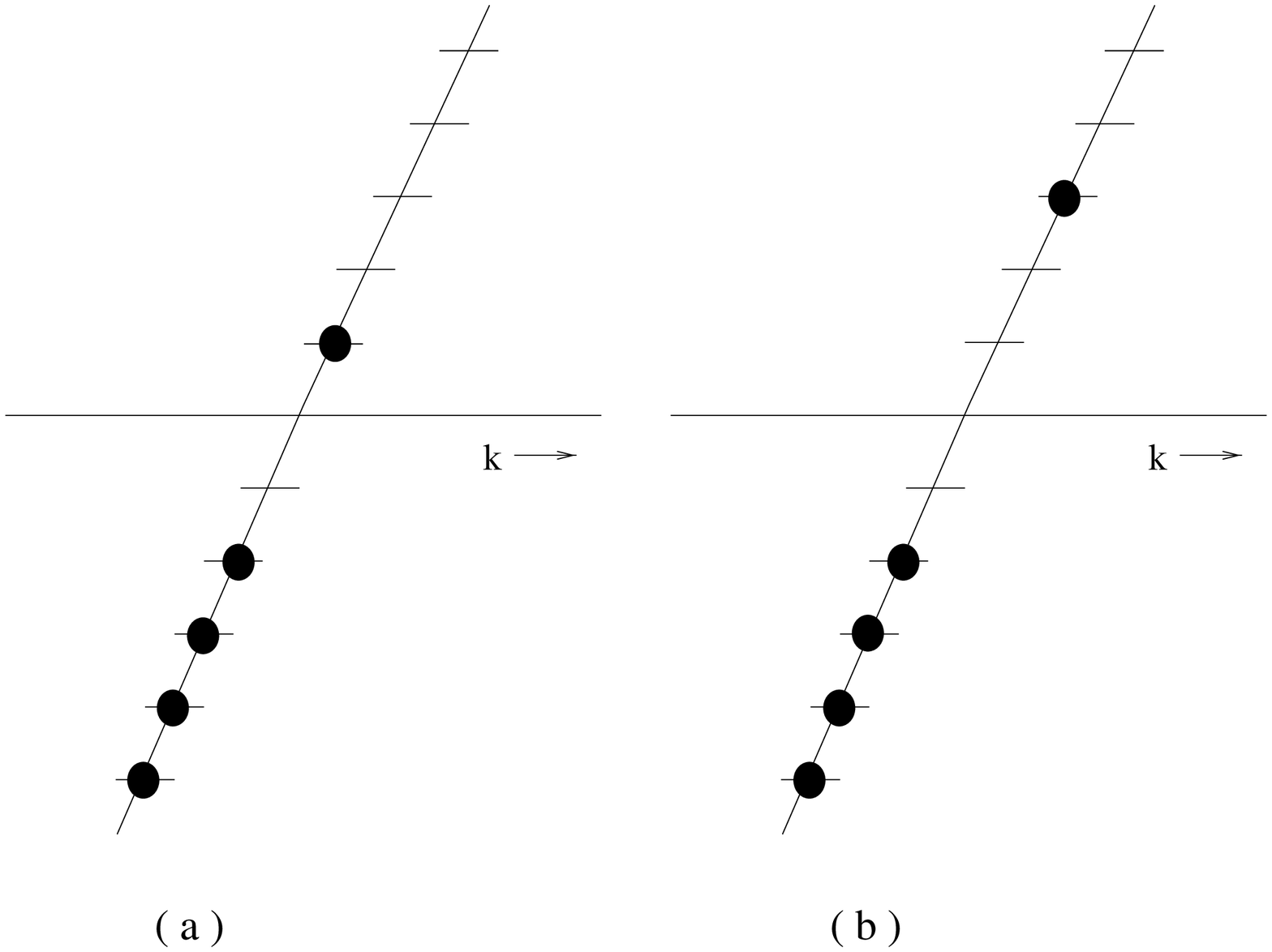,width=11cm}
\end{center}
\vspace*{0.6cm}
\centerline{Fig. 4}
\label{fig4}
\end{figure}

\begin{figure}
\begin{center}
\epsfig{figure=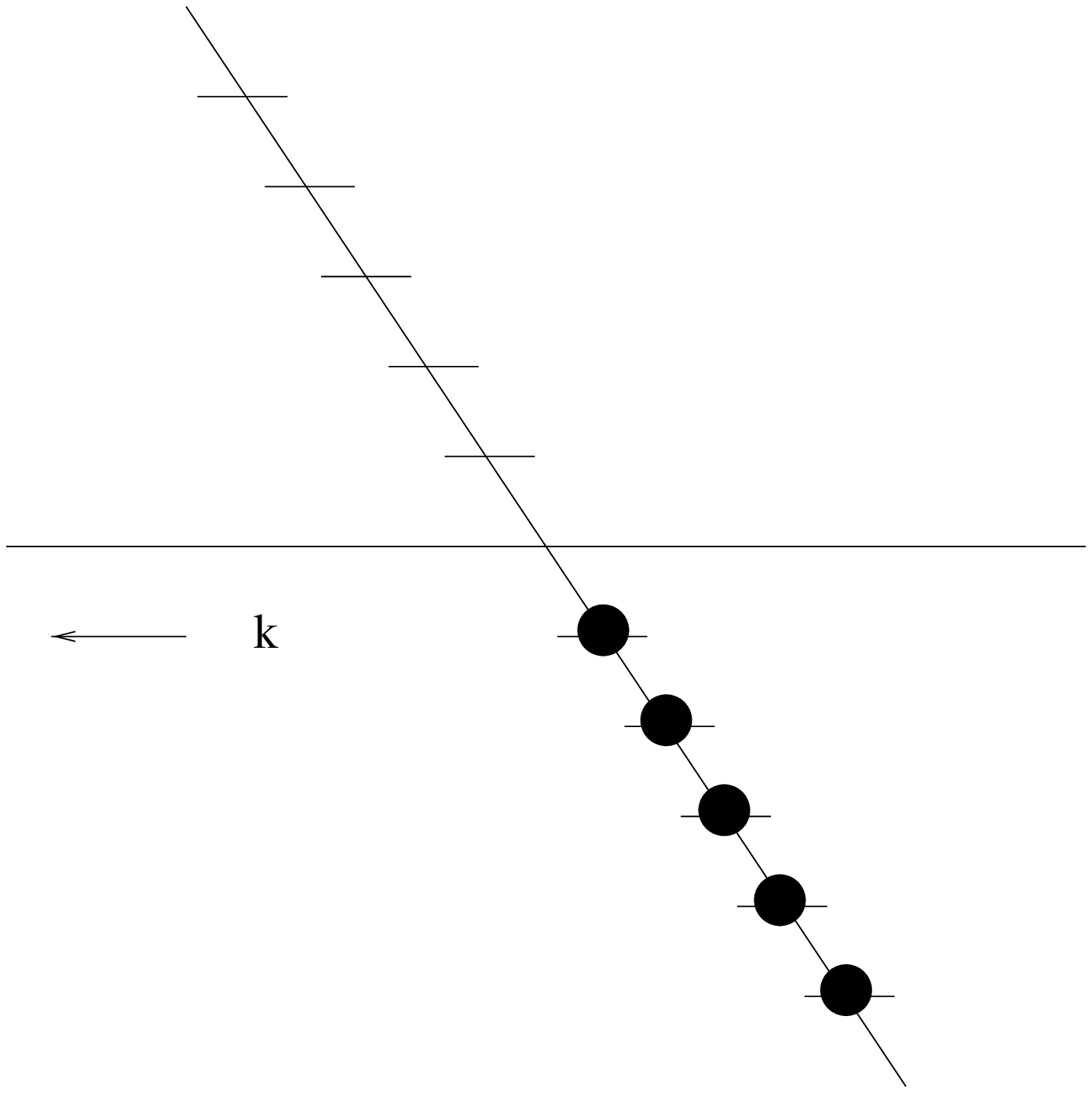,width=8cm}
\end{center}
\vspace*{0.8cm}
\centerline{Fig. 5}
\label{fig5}
\end{figure}

\begin{figure}
\begin{center}
\epsfig{figure=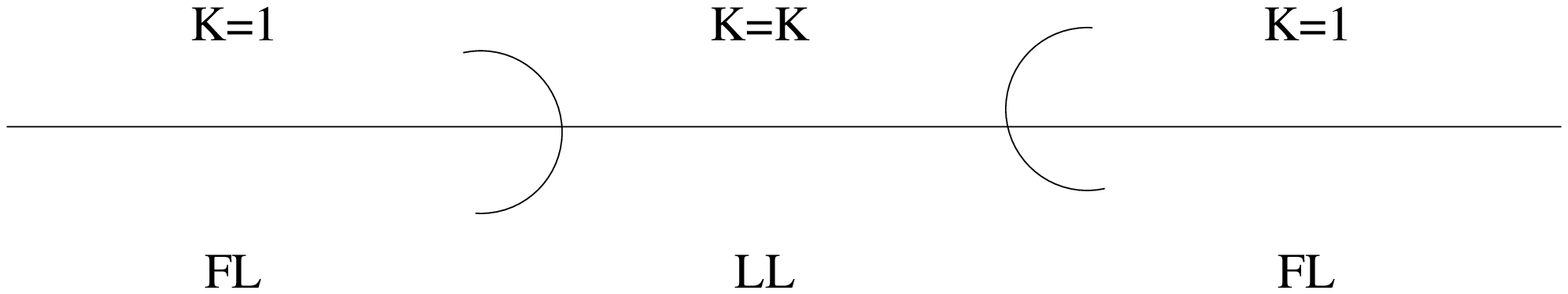,height=14cm,angle=-90}
\end{center}
\vspace*{0.8cm}
\centerline{Fig. 6}
\label{fig6}
\end{figure}
\vspace*{1cm}

\begin{figure}
\begin{center}
\epsfig{figure=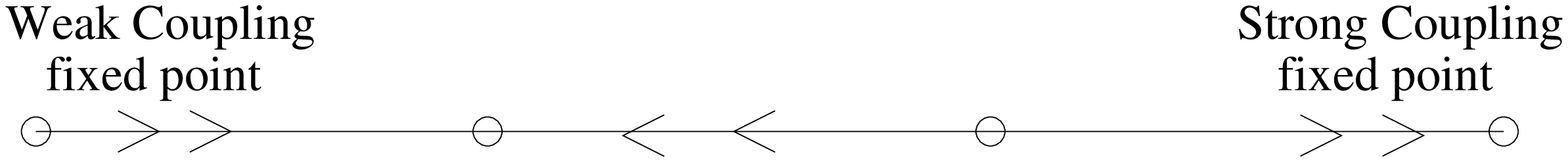,height=14cm,angle=-90}
\end{center}
\vspace*{0.8cm}
\centerline{Fig. 7}
\label{fig7}
\end{figure}


\newpage

\begin{thebibliography}{99}

\bibitem{schu} H. J. Schulz, G. Cuniberti and P. Pieri, to appear in {\it 
Field Theories for Low Dimensional Condensed Matter Systems}, edited by G. 
Morandi, A. Tagliacozzo and P. Sodano, Springer Lecture Notes in Physics 
(2000), cond-mat/9807366; H. J. Schulz, in Proceedings of Les Houches Summer 
School LXI, edited by E. Akkermans, G. Montambaux, J. Pichard and J. 
Zinn-Justin (Elsevier, Amsterdam, 1995), cond-mat/9503150.

\bibitem{hald} F. D. M. Haldane, Phys. Rev. Lett. {\bf 45}, 1358 (1980); 
Phys. Rev. Lett. {\bf 47}, 1840 (1981); J. Phys. C {\bf 14}, 2585 (1981).

\bibitem{gogo} A. O. Gogolin, A. A. Nersesyan and A. M. Tsvelik, {\it 
Bosonization and Strongly Correlated Systems} (Cambridge University Press, 
Cambridge, 1998).

\bibitem{kriv} V. Ya. Krivnov and A. A. Ovchinnikov, Sov. Phys. JETP {\bf
55}, 162 (1982); A. V. Zabrodin and A. A. Ovchinnikov, Sov. Phys. JETP {\bf
63}, 1326 (1986).

\bibitem{aff1} I. Affleck, in {\it Fields, Strings and Critical Phenomena}, 
edited by E. Brezin and J. Zinn-Justin (North-Holland, Amsterdam, 1989).

\bibitem{shan1} R. Shankar, Lectures given at the BCSPIN School, Kathmandu, 
1991, in {\it Condensed Matter and Particle Physics}, edited by Y. Lu, J. 
Pati and Q. Shafi (World Scientific, Singapore, 1993).

\bibitem{tomo} S. T. Tomonaga, Prog. Theor. Phys. {\bf 5}, 544 (1950);
J. M. Luttinger, J. Math. Phys. {\bf 4}, 1154 (1963); D. C. Mattis and E. H.
Lieb, J. Math. Phys. {\bf 6}, 304 (1965).

\bibitem{vond} J. von Delft and H. Schoeller, Ann. der Physik {\bf 4}, 225 
(1998), cond-mat/9805275.

\bibitem{scho} K. Sch\"onhammer and V. Meden, Am. J. Phys. {\bf 64},
1168 (1996).

\bibitem{aff2} I. Affleck, D. Gepner, H. J. Schulz and T. Ziman, J. Phys. 
A {\bf 22}, 511 (1989).

\bibitem{raja} R. Rajaraman, {\it Solitons and Instantons} (North-Holland,
Amsterdam, 1982).

\bibitem{kawa} N. Kawakami and S.-K. Yang, Phys. Rev. Lett. {\bf 67}, 2493
(1991).

\bibitem{SB} See lectures by S. M. Bhattacharjee and D. Kumar in this school.

\bibitem{MAHAN} G. D. Mahan, {\it Many Particle Physics} (Plenum Press, New 
York).

\bibitem{shan2} R. Shankar, Int. J. of Mod. Phys. {\bf B}, 2371 (1990).

\bibitem{KF} C. L. Kane and M. P. A. Fisher, Phys. Rev. B {\bf 46}, 15233 
(1992).

\bibitem{MS} D. L. Maslov and M. Stone, Phys. Rev. B {\bf 52}, R5539 (1995).

\bibitem{SAFI} I. Safi and H. J. Schulz, Phys. Rev. B {\bf 52}, R17040; I. 
Safi, Phys. Rev. B {\bf 55}, R7331; I. Safi and H. J. Schulz, Phys. Rev. B 
{\bf 58}.

\end{thebibliography}
\end{document}